%% file: zp3.tex
\documentclass[12pt]{article}
\usepackage{graphics,cite,epsfig}
\begin{document}
\input FEYNMAN

\begin{titlepage}

\begin{flushright}
\end{flushright}

\vspace{1.cm}

\begin{center}
{\Large \bf\boldmath
Phenomenology of supersymmetric
Z$'$ decays\\
\vspace{0.2cm}
 at the Large Hadron Collider}
\end{center}

\vspace{2mm}

\begin{center}
{\large \bf Gennaro Corcella}\\

\vspace{5mm}

{\sl INFN - Laboratori Nazionali di Frascati}\\ 
{\sl Via E.~Fermi 40, I-00044, Frascati (RM), Italy}
\vspace{3mm}

\end{center}

\par \vspace{2mm}
\begin{center}
{\large \bf Abstract}
\end{center}
\begin{quote}
  \pretolerance 10000
I study the phenomenology of heavy neutral bosons
$Z'$, predicted in GUT-inspired U(1)$'$ models, at the Large Hadron Collider.
In particular, I investigate
possible signatures due to $Z'$ decays into supersymmetric 
particles, such as chargino, neutralino and sneutrino pairs,
leading to final states with charged leptons and 
missing energy.
The analysis is carried out at $\sqrt{s}=14$~TeV,
for a few representative points of the
parameter space of the Minimal Supersymmetric Standard Model,
suitably modified to accommodate the extra $Z'$ boson and 
consistent with the discovery of
a Higgs-like boson with mass around 125 GeV.
Results are presented 
for several observables and compared with
those obtained for direct $Z'$ decays into lepton pairs,
as well as direct production of supersymmetric particles.
For the sake of comparison, $Z'$ phenomenology in an
effective supersymmetric extension of the Sequential
Standard Model is also discussed.
\end{quote}
\end{titlepage}

\section{Introduction}
Searching for heavy neutral gauge bosons $Z'$ is one of the
challenging goals of the experiments performed at the Large Hadron Collider (LHC). 
In fact, such bosons are predicted in extensions of the Standard Model involving
an extra U(1)$'$ gauge group, inspired by Grand Unification Theories (GUTs)
(see, e.g., \cite{langa,rizzo} for a review).
Furthermore, $Z'$ bosons are also present in the
so-called Sequential Standard Model (SSM), where the $Z'$ has the same couplings
to fermions as the Standard Model (SM) $Z$ boson. 
Though not being theoretically motivated, the SSM is often
used as a benchmark for the experimental searches.

The LHC experiments have so far searched for high-mass neutral gauge bosons
$Z'$ and have set exclusion limits on its mass $m_{Z'}$.
In detail, the ATLAS Collaboration \cite{atlas} set the limits in the
range $m_{Z'}>2.90$~TeV for a SSM $Z'$ and $m_{Z'}>2.51-2.62$~TeV
for GUT-inspired U(1)$'$ models. The same numbers for CMS \cite{cms} are
instead $m_{Z'}>2.90$~TeV for the SSM and $m_{Z'}>2.57$~TeV in U(1)$'$ models.
However, such analyses were carried out looking for high-mass dilepton pairs 
($e^+e^-$ or $\mu^+\mu^-$) and
assuming that the $Z'$ has only Standard Model decay modes.
Possible decays Beyond the Standard Model (BSM), e.g. in supersymmetric
particles, were investigated first in
\cite{gherghetta} and lately reconsidered in \cite{baum,chang,corgen}
within the Minimal Supersymmetric Standard Model \cite{haber,barbieri}.
Although SM decays are still dominant and the most promising
for the searches, the opening of new channels
decreases the branching ratios into electron and muon pairs and therefore
the mass exclusion limits. 
Reference~\cite{cor1}, using a representative point of the MSSM parameter space
as in \cite{corgen},
found that the LHC exclusion limits decrease by an amount
$\Delta m_{Z'}\simeq$~150-300~GeV, once accounting for BSM decay modes
at $\sqrt{s}=8$~TeV.

From the viewpoint of supersymmetry, the lack of
evidence of new particles in the LHC 
runs at 7 and 8 TeV, together with the discovery of a boson with 
mass $m_h=125.7\pm 0.4$~GeV \cite{pdg}
and properties consistent with the Standard Model Higgs boson \cite{higgs},
sets some tight constraints on the mass spectra and couplings 
of possible supersymmetric models. While awaiting the collisions
at 13 and ultimately 14 TeV, it is therefore worthwhile
thinking of scenarios, not yet excluded by the current searches
and compatible with the Higgs discovery, which may deserve
some specific analyses at high luminosity and energy.
Extending the MSSM via a U(1)$'$ group presents some features
which makes it a pretty interesting scenario, so that 
novel analyses, looking for signals of supersymmetric
$Z'$ decays by using current and future data, may be well 
justified.  
Unlike direct sparticle production in $q\bar q$ or $gg$ 
annihilation, the $Z'$ is colorless and its mass sets
a constraint on the invariant mass of the sparticle pair.
Therefore, if one had to discover a $Z'$, its decay
modes would be an ideal environment to look for supersymmetry,
as they would yield a somewhat cleaner signal, with respect to 
direct sparticle production. 
Supersymmetric $Z'$ decays would also be an excellent framework
to study electroweak interactions in regions of the phase space
which would not be accessible through other processes, such
as Drell--Yan interactions.
Moreover, possible decays into pairs of the lightest neutralinos
of the MSSM
would lead to mono-photon or mono-jet final states, like those which
are investigated when looking for Dark Matter candidates. 

The reference point of the parameter space 
chosen in Refs.~\cite{corgen,cor1} yielded substantial 
decay rates into supersymmetric particles and was
consistent with the present 
exclusion limits, but did not
take into account for the recent discovery of a Higgs-like boson.
In this paper, I shall extend the work in \cite{corgen} 
giving some useful benchmarks for possible $Z'$ searches 
within supersymmetry. First, it will be chosen a set of points in the
parameter space yielding a SM-like Higgs boson with a mass around 125 GeV. 
Then, thanks to the Monte Carlo implementation of the U(1)$'$ models
along with the MSSM, 
a phenomenological analysis will be performed and
a few final-state distributions in events with supersymmetric $Z'$ decays
will be presented. On the contrary, Ref.~\cite{corgen} only
calculated total production cross sections and branching ratios
and left the investigation of differential distributions as an open issue. 
Furthermore, I shall also account for an effective supersymmetric extension
of the Sequential Standard Model, denoted by S-SSM hereafter,
wherein the couplings of the $Z$ and $Z'$ to 
supersymmetric particles are the same.

In detail, in Section 2 I will briefly review the theoretical
framework of the investigation here undertaken, paying special attention to
the new features of the MSSM once a $Z'$ boson is included.
In Section 3 I shall discuss the practical implementation
of supersymmetric $Z'$ decays in a few computing codes
and Monte Carlo event generators.
Sections 4, 5 and 6 will deal with the phenomenology of 
the $Z'$  in three scenarios, namely the $Z'_\psi$ and 
$Z'_\eta$ models, within U(1)$'$ gauge theories, and the S-SSM, respectively.
Section 7 will contain some final remarks and comments on possible further
developments of this work.

\section{U(1)$'$ gauge group and Minimal Supersymmetric 
Standard Model}

In this section, I shall discuss 
the theoretical framework of supersymmetric $Z'$ decays, already thoroughly reviewed
in \cite{gherghetta} and, more recently, in \cite{corgen}. 
As discussed in \cite{langa,rizzo}, U(1)$'$ groups typically arise from the breaking
of a Grand Unification gauge group E$_6$ of rank 6.
The neutral boson $Z'_\psi$ is associated with U(1)$'_\psi$, coming from the
breaking into SO(10) as follows:
\begin{equation}\label{upsi}
{\rm E}_6\to {\rm SO}(10)\times {\rm U}(1)'_\psi.
\end{equation}
The $Z'_\chi$ is instead related to the subsequent breaking of SO(10) according to:
\begin{equation}
{\rm SO}(10)\to {\rm SU}(5)\times {\rm U}(1)'_\chi.
\end{equation}
The $Z'_\psi$ and $Z'_\chi$ mix into a generic $Z'(\theta)$ depending on the
mixing angle $\theta$:
\begin{equation}\label{ztheta}
Z'(\theta)=Z'_\psi\cos\theta-Z'_\chi\sin\theta.
\end{equation}
The $Z'_\psi$ and $Z'_\chi$ models correspond to $\theta=0$ and $\theta=-\pi/2$, 
respectively.
Another scenario, which is often investigated from both theoretical and experimental
viewpoints, is the one, characteristic of superstring theories, where E$_6$ 
breaks in the Standard Model (SU(2)$_{\rm L}\times$ U(1)$_{\rm Y}$) and an extra
U(1)$'_\eta$:
\begin{equation}\label{ueta}
{\rm E}_6\to {\rm SM}\times U(1)'_\eta.
\end{equation}
Equation~(\ref{ueta}) leads to a $Z'_\eta$ boson, with a mixing angle
$\theta=\arccos\sqrt{5/8}$ in Eq.~(\ref{ztheta}).
One can anticipate that the following
analysis will be performed for the $Z'_\psi$ and $Z'_\eta$ models, since 
other models like those leading to the $Z'_\chi$,
as well as the $Z_{\rm I}$, $Z'_{\rm S}$ and $Z'_{\rm N}$, 
corresponding to the mixing angle $\theta$ described in
\cite{corgen}, are less interesting, as the $Z'$ branching ratios into 
supersymmetric final states are rather low.

As far as the MSSM is concerned, 
a few relevant features are inherited by the presence of
the extra $Z'$ boson. In addition to the scalar Higgs doublets $H_d$
and $H_u$ of the MSSM, an extra neutral singlet $S$ is necessary
to break the U(1)$'$ gauge symmetry and give mass to the $Z'$.
Hereafter, the Higgs bosons will be denoted as follows:
\begin{equation}\label{phi123}
H_d=\left(\begin{array}{c}
H_d^0 \\ H_d^- \end{array}\right)\ ,\ 
H_u=\left(\begin{array}{c}H_u^+ \\ H_u^0 \end{array}\right)\ ,\ 
S=S^0\ ,
\end{equation}
and their vacuum expectation values like
$v_d$, $v_u$ and $v_s$, respectively.
The Higgs superfields will then contain a Higgsino component as well,
i.e. $\tilde H_u$, $\tilde H_d$ and $\tilde S$ fields. 

The superpotential of the MSSM, once it is extended by means of a U(1)$'$ group,
is then given by\cite{martin,chang}
\begin{equation}\label{superp}
W=u^c y_u Q H_u-d^c y_d Q H_d-e^c y_e L H_d+\lambda
H_uH_dS,
\end{equation}
where, following the notation in \cite{gherghetta},
$y_{u,d,e}$ are the Yukawa coupling matrices
for up- and down-type quarks, $Q$ and $L$ are the 
MSSM superfields containing left-handed (s)quarks and (s)leptons, 
$u^c$, $d^c$ and $e^c$  
are the singlet fields of right-handed 
up-, down-type (s)quarks and (s)leptons, respectively.
The trilinear term $\lambda H_uH_dS$
involving all three Higgs superfields, 
is a feature of the U(1)$'$ extension of the MSSM and gives rise
to the well-known $\mu$ term, which can be expressed in terms
of $\lambda$ and the vacuum expectation value of
$S$ as $\mu=\lambda v_s/\sqrt{2}$.\footnote{Without the U(1)$'$ group,
the $\mu$-term in the MSSM superpotential would just be $\mu H_uH_d$.}

For the analysis which will be hereafter undertaken, the 
soft supersymmetry-breaking Lagrangian plays a crucial role.
It is given by the following expression \cite{gherghetta,martin}:
\begin{eqnarray}\label{lsoft}
{\cal L} &=& -\frac{1}{2}(M_3\tilde g\tilde g+M_2\tilde W\tilde W+
M_1\tilde B\tilde B+M'\tilde B'\tilde B'+{\rm h. c.})\nonumber\\
&\ &-(\tilde {u^c}A_u\tilde Q H_u-\tilde {d^c} A_d\tilde QH_d-\tilde {e^c}A_e\tilde 
L H_d+{\rm h. c.})
\nonumber\\
&\ &-\tilde Q^\dagger (m^0_Q)^2\tilde Q-\tilde L^\dagger (m^0_L)^2\tilde  L
-\tilde {u^c}(m^0_{\tilde u})^2 {\tilde {u^c}}^\dagger-
\tilde {d^c} (m^0_{\tilde d})^2 {\tilde{d^c}}^\dagger
-\tilde{e^c}(m^0_e)^2 {\tilde{e^c}}^\dagger\nonumber\\
&\ & -m_{H_d}^2H_d^\dagger H_d-m_{H_u}^2{H_u}^\dagger
H_u-m_S^2 S^\dagger S
+\frac{i}{\sqrt{2}}\lambda A_\lambda(H_d^\dagger
\sigma_2H_uS+{\rm h.c.}).
\end{eqnarray}
In Eq.~(\ref{lsoft}), $M_3$, $M_2$ and $M_1$ are the
soft masses of gluino ($\tilde g$), wino ($\tilde W$) and
bino ($\tilde B$) fields of the MSSM, while $M'$
is the mass of $\tilde B'$, the supersymmetric partner of 
$B'$, the gauge boson associated with the U(1)$'$ group.
Moreover, $m_{H_u}$, $m_{H_d}$ and $m_S$ are the
soft masses of $H_u$, $H_d$ and $S$ in (\ref{phi123}),  
$m^0_Q$, $m^0_L$ and $m^0_{\tilde f}$ are the soft
masses of the left-handed
superfields $Q$ and $L$ and of the right-handed 
$\tilde f$, respectively. $A_u$, $A_d$ and $A_e$ are the soft
trilinear couplings of squarks and sleptons with the 
Higgs fields, in one-to-one correspondence with the Yukawa
couplings in the superpotential (\ref{superp}); 
one usually writes the trilinear couplings
as $A_f=m_fA_{f,0}$, where $A_{f,0}$ is dimensionless.
$A_\lambda$ is the soft Higgs trilinear
coupling, with $\sigma_2$ being one of the Pauli matrices;
the term $\sim \lambda A_\lambda  H_d^\dagger\sigma_2 H_uS$ is the counterpart 
in the soft Lagrangian of the trilinear contribution $\lambda H_uH_dS$
in the superpotential.
Also, in Eq.~(\ref{lsoft}) $\tilde Q$, $\tilde L$,
$\tilde u^c$, $\tilde d^c$ and $\tilde e^c$ are the squark/slepton components
in the left- and right-handed superfields, already introduced in
(\ref{superp}).

In the Higgs sector, after electroweak symmetry breaking 
and giving mass to $W$, $Z$ and $Z'$ bosons,
one is left with two charged $H^\pm$ and four neutral Higgs bosons,
namely one pseudoscalar $A$ and three scalars $h$, $H$ and $H'$, where
$H'$ is due to the U(1)$'$ gauge group
and is typically much heavier than the $Z'$.
Furthermore, with respect to the MSSM, 
two extra neutralinos are present, associated with the
supersymmetric partners of $Z'$ and $H'$, for 
a total of six neutralinos: in \cite{corgen}
it was nevertheless argued that these new
neutralinos are typically too heavy to be significant in
$Z'$ phenomenology. 

In order to reliably compute the sfermion masses,
one would need to perform this analysis in a specific
scenario for supersymmetry breaking, such as gauge-,
gravity- or anomaly-mediated mechanisms.
Investigations of supersymmetry-breaking models
are beyond the scopes of this paper; it is nevertheless
mandatory to recall that supersymmetry can be spontaneously
broken if the so-called D-term and/or the F-term in the scalar
potential have non-zero vacuum expectation values.
The contribution of D- and F-terms to the potential reads:
\begin{equation}
V_{D,F}(\phi,\phi^*)=F^{*i}F_i+\frac{1}{2}D^aD_a,\ F_i=\frac{\delta W}{\delta\phi_i}\ ,\ 
D_a=-g_a(\phi_i^*T^a\phi^i),
\end{equation}
where $W$ is the superpotential,
$\phi_i$ are the scalar (Higgs) fields,
$g_a$ and $T^a$ the coupling constant and the generators
of the gauge groups of the theory.
The F-terms are proportional to the particle masses,
and therefore they are mostly important for 
stop quarks; the D-terms are relevant for both light and
heavy sfermions and contain two contributions.
The first one, already present in the MSSM, is related to the
hyperfine splitting due to electroweak symmetry breaking:
for a sfermion $a$, it depends on its weak isospin $T_{3,a}$,
electric charge $Q_a$ and weak hypercharge $Y_a$, as well as 
on the vacuum expectation values of the
two MSSM Higgs doublets ($v_1$ and $v_2$):
\begin{equation}
\Delta\tilde m_a^2=
(T_{3,a}g_1^2-Y_ag_2^2)(v_1^2-v_2^2)=(T_{3,a}-Q_a\sin^2\theta_W)m_Z^2\cos 2\beta,
\label{d1}
\end{equation}
where $g_1$ and $g_2$ are the coupling constants of U(1) and
SU(2), respectively, and $\theta_W$ is the Weinberg angle.
A second contribution is due to possible 
extensions of the MSSM, such as the U(1)$'$ group, and is related to
the Higgs bosons which break the new symmetry:
\begin{equation}
\Delta{\tilde m}_a^{\prime 2}=\frac{g'^2}{2}Q'_a(Q_{H_u}'v_u^2+Q_{H_d}'v_{H_d}^2+Q_S'v_S^2),
\label{dt}
\end{equation}
where $g'$ is the U(1)$'$ coupling, 
$Q'_{H_u}$, $Q'_{H_d}$, $Q'_S$ and $Q'_a$ 
are the U(1)$'$ charges of the Higgs fields $H_u$, $H_d$ and
$S$ and of the sfermion $a$. 
As a result, the soft sfermion masses $m^0_f$ in (\ref{lsoft}) get
F- and D-term corrections: as they are not 
positive definite, one may even be driven 
to unphysical scenarios, where the sfermion squared 
mass gets negative (see few examples in Ref.~\cite{corgen}). 

In general, sfermion mass eigenstates are obtained 
by diagonalizing the following mass matrix:
\begin{equation}
{\cal M}_{\tilde f}^2=\left( \begin{array} {cc}
(M_{LL}^{\tilde f})^2 &
(M_{LR}^{\tilde f})^2\\ (M_{LR}^{\tilde f})^2 &   
(M_{RR}^{\tilde f})^2  \end{array} \right),
\label{smass}
\end{equation}
where the matrix elements are obtained by summing
the squared soft masses in (\ref{lsoft}) and the D- and F-term
corrections.
As an example, the matrix elements for down-type squarks are given by
\begin{eqnarray}
( M^{\tilde d}_{LL})^2 &=& (m^0_{\tilde d_{L}})^2 + 
m^2_d + \left(-\frac{1}{2}+\frac{1}{3}x_W \right)m_{Z'} ^2\ 
\cos 2\beta + \Delta {\tilde m}^{\prime 2}_{\tilde d_L} \label{mll}\\
(M^{\tilde d}_{RR})^2 &=& 
(m^0_{\tilde d_R})^2 + m^2_d -
\frac{1}{3} x_W m_{Z'}^2\ \cos 2\beta + 
\Delta {\tilde m}^{\prime 2}_{\tilde d_R}
\label{mrr}\\  
( M^{\tilde d}_{LR})^2&=&  m_d\left( A_d - \mu \tan \beta \right),
\label{mlr}
\end{eqnarray}
where  $x_W=\sin\theta_W^2$,  $m^0_{\tilde d_{L,R}}$ is the $\tilde u_{L,R}$ 
soft mass at the $Z'$  energy scale
and $A_d=m_dA_{d,0}$ is the coupling 
entering in the Higgs--sfermion interaction term in the soft 
supersymmetry-breaking Lagrangian.
The mixing matrix element $M^{\tilde d}_{LR}$ is due to the F-term and,
as anticipated, is proportional to the quark mass $m_d$;
the expressions for the F- and D-term corrections to the soft mass
of up-type squarks and sleptons can be found in \cite{gherghetta}.

In the following, besides GUT-inspired models, I will also account for the Sequential 
Standard Model; unlike the U(1)$'$ gauge groups, the SSM is not a real model,
but nonetheless it is used by the experimental collaborations as a benchmark for the
searches. In fact, if the $Z'$ has the same couplings to the fermions
as the $Z$, the production cross section can be straightforwardly
computed as a function of the $Z'$ mass. 
Following \cite{chang,corgen} I will consider an effective model,
named S-SSM in the following, where
the $\tilde Z'$ is too heavy to be visible at the LHC
and the couplings of the $Z'$ to
sfermions and gauginos are the same as the $Z$ in the MSSM. In principle,
a consistent SSM should be built up along the lines of \cite{mpv}, wherein
it was explained that any sequential $Z'$ must be accompanied by another 
$Z'$ and a longitudinal $W'$. However, employing this improved formulation
of the SSM goes beyond
the goals of this paper and therefore I shall stick to the approximations in 
\cite{chang,corgen}, with a $Z'_{\rm S-SSM}$ coupled to SM and
BSM particles like the Standard Model $Z$.

\section{Framework for $Z'$ supersymmetric decays}

Hereafter, I will present a phenomenological analysis of $Z'$ production
and decay at the LHC, paying special attention to supersymmetric
decay modes and comparing the results with those obtained in standard analyses,
where only Standard Model channels are allowed. 
As discussed before, the investigation will be concentrated on the
$Z'_\psi$ and $Z'_\eta$ models and, for each scenario,
it will be chosen a point in the parameter space not yet
excluded by the LHC searches and leading to an interesting
phenomenology within supersymmetry.
In all cases, I will set the $Z'$ mass to the value 
\begin{equation}
m_{Z'}=2~{\rm TeV} 
\end{equation}
and will use
the following relation between U(1)$'$ and U(1)$_{\rm Y}$ coupling constants
$g'$ and $g_1$, typical of GUTs:
\begin{equation}
g'=\sqrt{\frac{5}{3}}g_1.
\end{equation}
When dealing with the S-SSM, the $Z'$ coupling constant to fermions
will be the same as the $Z$:
\begin{equation}
g_{\rm S-SSM}=\frac{g_2}{2\cos\theta_W}.
\end{equation}

In \cite{corgen} the authors fixed the $Z'$ mass and the MSSM parameters and 
calculated, either analytically or numerically, particle masses and $Z'$ branching
ratios into SM and MSSM final states. However, the computation was carried
out at leading order (LO) in the couplings $g_1$, $g_2$ and $g'$ 
and therefore the mass of the lightest
MSSM neutral Higgs boson, which roughly plays the role of the Standard
Model Higgs, was around the value of the $Z$ mass, i.e. about 90 GeV.
In this paper, I shall include higher-order corrections, 
especially top and stop loops, in such a way to recover a light
Higgs mass about 125 GeV.
For this purpose, I will make use of the Mathematica package SARAH \cite{sarah}
which calculates the mass matrices by using the renormalization group equations at one 
loop.\footnote{The most updated SARAH version \cite{sar4} even includes
two-loop corrections to the renormalization group equations.} Among the implemented models, 
SARAH includes the so-called
UMSSM, namely the extension of the MSSM through a U(1)$'$ gauge group:
the output of SARAH is used as a source code for 
SPheno \cite{spheno}
to create a precision spectrum generator for the given scenario.
Model files in the Universal FeynRules Output (UFO) 
 format \cite{feyn} are then used by the MadGraph code \cite{madgraph}
to generate the hard-scattering process, with $Z'$ production,
i.e. $q\bar q\to Z'$, and decay according to the chosen mode.
The events are thus written in the Les Houches 
format and the HERWIG Monte Carlo event generator 
\cite{herwig} can provide them with parton showers and hadronization,
eventually leading to exclusive final states.
The analysis within the S-SSM is somewhat different, since SARAH and SPheno
do not contain this benchmark model. A straightforward implementation
can nevertheless be achieved within the package 
FeynRules itself \cite{feyn}, by simply adding to the MSSM code
a $Z'$ boson, coupled
to SM and BSM particles as the $Z$ in the Standard Model. 
FeynRules then constructs the UFO model files which can be read by MadGraph
and HERWIG to simulate full hadron-level events.

In order to perform a consistent investigation and comparison
with previous work in \cite{gherghetta,chang, corgen}, few further
changes
were implemented into SARAH and FeynRules. In SARAH, I added Dirac right-handed
neutrinos and sneutrinos, not present in its default version,
in order to allow $Z'$ decays into both left- and right-handed 
neutrino and sneutrino pairs.
When modifying SARAH,
the mass of the right-handed neutrino is set to zero by default.
In the FeynRules implementation of the effective S-SSM,
the $Z'WW$ coupling was suppressed:
in fact, if one naively  assumed that the $Z'$ couples to $WW$ pairs 
like the $Z$, on the one hand the decay $Z'\to WW$ would largely dominate,
on the other the unitarity of the theory would be in trouble,
because of the enhancement of $WW$ scattering mediated by a $Z'$.
A consistent S-SSM, possibly built up along the lines of \cite{mpv}, would not suffer
from this drawback.\footnote{Updated releases of SARAH and FeynRues including
such changes are in progress. For the time being, the computing
code to obtain the results presented in this paper can be requested 
from the author.}

In the choice of the working reference point for this investigation,
I will make use of the results in \cite{nazila1,nazila2}, wherein the
authors determined the regions of the supersymmetric phase space which
are not yet excluded by the direct searches and are consistent
with a Higgs of 125 GeV, taking care of the limits from
flavour physics and Dark Matter searches.
Strictly speaking, the results of \cite{nazila1,nazila2}
are obtained for the so-called phenomenological MSSM
(pMSSM), which makes a few simplifying assumptions in order to 
reduce the number of parameters. In detail, the
pMSSM assumes that the soft supersymmetry-breaking terms are real, 
there is no new source of CP violation, we have diagonal matrices 
for the sfermion masses and trilinear couplings, i.e.
no flavor change at tree level, and the same soft masses and trilinear
couplings at least for the first two generations
of squarks and sleptons at the electroweak scale.
The leftover parameters are then the ratio of the MSSM neutral Higgs
vacuum expectation values $\tan\beta=v_u/v_d$, the Higgs (higgsino) mass
parameter $\mu$, the soft masses of bino and wino $M_1$ and $M_2$, 
the sfermion masses and the trilinear couplings.
As in \cite{gherghetta,corgen}, because of the $Z'$, one has an extra gaugino
$\tilde B'$, whose soft mass parameter is named $M'$.

For all the scenarios which will be studied, $M_1$, $M'$, $\tan\beta$ and $\mu$ 
will be set as follows:
\begin{equation}
M_1=400~{\rm GeV},\ M'=1~{\rm TeV},\ \tan\beta=30,
\ \mu=200~{\rm GeV}.
\label{mubeta}
\end{equation}
Given $M_1$, the wino mass $M_2$ can be obtained by using the relation 
$M_2=(3/5)\cot^2\theta_W\simeq 827$~GeV.

Furthermore, in the Standard Model it is well known that 
bottom and especially top quarks play 
a fundamental role in Higgs phenomenology: in fact, 
heavy-quark loops give
the highest corrections to the Higgs mass and the largest contribution 
to the Higgs production cross section in gluon fusion.
It is therefore obvious that in the MSSM stops and sbottoms,
the supersymmetric partners of top and bottom quarks,
will deserve special attention and, although they have not
been observed, the measured mass of the Higgs boson
sets some constraints on their masses. 
In fact, they can be very heavy, i.e. their mass in
the TeV range, 
but even quite light, say of the order of a few hundred GeV, 
provided that 
the mixing is large, i.e. the mixing parameter $A_t$ is about a few TeV
(see, e.g., the discussion in \cite{carena}).
The latter case is often chosen in the supersymmetry studies, namely
the first two squark generations heavier than
sbottoms and stops.
In this paper, I shall consider both possibilities: all three 
squark generations heavy and degenerate, as well as the option
of lighter stops and sbottoms.
The authors of \cite{nazila1} define the
mixing parameter:
\begin{equation}
x_t=A_t-\mu\cot\beta,
\label{xt}
\end{equation}
which runs in the range $0<x_t<\sqrt{6}~M_S$, $M_S$ being the
geometrical average of the stop masses, i.e. 
$M_S=\sqrt{m_{\tilde t_1}m_{\tilde t_2}}$, where $m_{\tilde t_1}$
and $m_{\tilde t_2}$ are obtained after adding to 
the soft mass $m_{\tilde t}^0$
the D-term (see \cite{gherghetta}) and diagonalizing the
stop mixing matrix.

In  Eq.~(\ref{xt}), $A_t$ is a dimensionful quantity
related to the dimensionless trilinear coupling $A_{t,0}$
in \cite{corgen} via
$A_t=A_{t,0} m_t$, where 
$m_t$ is the top quark mass, fixed to $m_t=173$~GeV.
For $x_t=4$~TeV, one obtains that, using the numbers in
(\ref{mubeta}), $A_t\simeq 4$~TeV and $A_{t,0}\simeq 23.2$.
Later on, all the trilinear couplings, 
as well as $A_\lambda$, contained in the
soft supersymmetry-breaking Lagrangian (\ref{lsoft}), 
will be set the same value:\footnote{Note that SARAH requires $A_\lambda/\sqrt{2}\simeq 2.8$~TeV
as an input, rather than $A_\lambda$ in Eq.~(\ref{aaa}).}
\begin{equation}
A_q=A_\ell=A_\lambda\simeq 4~{\rm TeV}.
\label{aaa}
\end{equation}

In the following sections, I shall present the results yielded at the LHC by
the models U(1)$'_\psi$, U(1)$'_\eta$ and S-SSM. As in \cite{corgen}, 
a few decay chains will be taken into account: they all
start with a primary 
supersymmetric decay, e.g. into pairs of charged sleptons,
sneutrinos, charginos or neutralinos, and eventually yield
final states with two or four charged leptons and missing transverse
energy (MET), associated with neutrinos or light neutralinos.
For each model, I will consider a specific point in the parameter space,
with the goal of maximizing the branching ratio in at least one of
the supersymmetric modes.
Then I shall present some leptonic final-state distributions,
in the scenario which maximizes the BSM $Z'$ decay rate.
Whenever it makes sense, the results will be confronted with those
from the standard search strategies, where the $Z'$ decays into
a SM charged-lepton pair and has no BSM decay width.

\section{Phenomenology: U$(1)'_\psi$ model}

The model U(1)$'_\psi$, leading to a heavy boson 
$Z'_\psi$, corresponds to a mixing angle $\theta=0$ in
Eq.~(\ref{ztheta}).
In \cite{corgen}, it was found that, in a reference point of the
parameter space and for a $Z'_\psi$ mass between 1 and 5 TeV,
about 35\% of the $Z'_\psi$ width is 
due to the supersymmetric modes.  However, as discussed above,
that scenario was not consistent with a Higgs mass of 125 GeV
and the supersymmetric mass spectrum was computed only at tree level:
such approximations will be relaxed in the present analysis.

Hereafter, the representative points of the parameter space will be
chosen in order to satisfy the Higgs mass constraint and the supersymmetry
exclusion limits. The quantities $M_1$, $M'$, $\mu$ and $\tan\beta$
are fixed as in Eq.~(\ref{mubeta});
as for sfermions, I assume that
sleptons, as well as the first two generations of squarks,
are degenerate at the $Z'_\psi$ mass scale and have mass\footnote{Alternatively, 
one can fix the sfermion masses at a very high
scale, such as the Planck mass, and evolve them down to the
$Z'$ scale by means of renormalization group equations}:
\begin{equation}
\label{sfer1}
m_{\tilde\ell}^0=m_{\tilde\nu_\ell}^0=1.2~{\rm TeV}\ ,\ 
m^0_{\tilde q}=5.5~{\rm TeV},
\end{equation}
where $\ell=e,\mu,\tau$, $\nu=\nu_e,\nu_\mu,\nu_\tau$ and $q=u,d,c,s$.
The soft masses of stops and sbottoms are instead fixed as follows:
\begin{equation}\label{sfer2}
m^0_{\tilde t}=m^0_{\tilde b}=2.2~{\rm TeV}.
\end{equation}
The sfermion masses at the $Z'_\psi$ mass scale are obtained after summing 
to the numbers in 
(\ref{sfer1}) and (\ref{sfer2}) 
the F- and D-terms due to U(1)$'$ and 
electroweak symmetry breaking;
at leading order, the masses yielded by the SARAH code
agree with those computed by using the expressions in 
\cite{gherghetta}.
For $m_{Z'}=2$~TeV, the sfermion masses are quoted in Tables
\ref{tabmassq} and \ref{tabmassl}, for squarks and sleptons, respectively.
The notation $\tilde q_{1,2}$, $\tilde \ell_{1,2}$ and
$\tilde\nu_{1,2}$ refers to the mass eigenstates, which differ from the
gauge ones  $\tilde q_{L,R}$, $\tilde\ell_{L,R}$ and $\tilde\nu_{L,R}$ 
because of the mixing; 
such mixing terms are proportional to 
the fermion squared masses, and therefore they are 
mostly relevant in the case of the stops.
From such tables, one can learn that the impact of the D-term is
about 100 GeV on squarks and even larger than 200 GeV on sleptons;
also, in the chosen reference point, 
the D-term can be either positive or negative.

The Higgs masses, computed by SARAH to one-loop accuracy,
are reported in Table~\ref{tabmassh}:
the lightest scalar $h$
has roughly the same mass as the SM-like Higgs boson, 
$H$ is approximately as heavy as the $Z'_\psi$, whereas $H'$, $A$ and
the charged $H^\pm$ are above 4 TeV, and therefore too heavy to be
significant for $Z'_\psi$ phenomenology.
The $\lambda$ parameter, 
contained in the trilinear potential $V_\lambda$,
is related to $\mu$ and the vacuum expectation
value $v_S$ of the extra Higgs boson $S$ via 
$\lambda=\sqrt{2}\mu/v_S\simeq 5.4\times 10^{-2}$.
Table~\ref{tabmasscn} contains the masses of the two charginos ($\tilde\chi^\pm_{1,2}$)
and of the six neutralinos ($\tilde\chi^0_1,\dots \tilde\chi^0_6$):
in principle, with the exception of $\tilde\chi^0_6$, whose mass 
is even above 6 TeV, several $Z'_\psi$ decay modes 
into pairs of charginos and neutralinos are kinematically permitted.
\begin{table}[b]
\caption{Masses of squarks in the MSSM, for the chosen reference point
and accounting for the U$(1)'_\psi$ modifications. 
The masses of
$\tilde q_{1,2}$ differ from those of the gauge eigenstates
$\tilde q_{L,R}$ because of the mixing contribution, relevant especially in the
stop case. All numbers are expressed in GeV.}
\label{tabmassq}
\begin{center}
\small
\begin{tabular}{|c|c|c|c|c|c|}
\hline
$m_{\tilde d_1}$ &   $m_{\tilde u_1}$ &  $m_{\tilde s_1}$ &  $m_{\tilde c_1}$ &  
$m_{\tilde b_1}$ & $m_{\tilde t_1}$\\ 
\hline
 5609.8 & 5609.4  & 5609.9 & 5609.5 & 2321.7 & 2397.2 \\
\hline
$m_{\tilde d_2}$ &   $m_{\tilde u_2}$ &  $m_{\tilde s_2}$ &  $m_{\tilde c_2}$ &  
$m_{\tilde b_2}$ & $m_{\tilde t_2}$\\ 
\hline
 5504.9 & 5508.7  & 5504.9 & 5508.7 & 2119.6 & 2036.3 \\
\hline\end{tabular}
\end{center}
\end{table}
\begin{table}[htp]
\caption{As in Table~\ref{tabmassq}, but for charged sleptons ($\ell=e,\mu$)
and sneutrinos.}
\label{tabmassl}
\begin{center}
\small
\begin{tabular}{|c|c|c|c|c|c|c|c|}
\hline
$m_{\tilde \ell_1}$ &   $m_{\tilde \ell_2}$ &  $m_{\tilde\tau_1}$ & $m_{\tilde\tau_2}$ &
$m_{\tilde \nu_{\ell,1}}$ &  $m_{\tilde \nu_{\ell,2}}$ &
$m_{\tilde \nu_{\tau,1}}$ &  $m_{\tilde \nu_{\tau,2}}$ \\ 
\hline
 1392.4 & 953.0  & 1398.9 & 971.1 & 1389.8  & 961.5 & 1395.9 & 961.5\\ 
\hline\end{tabular}
\end{center}
\end{table}
\begin{table}[htp]
\caption{Masses of neutral and charged Higgs bosons in GeV in the 
chosen point of the MSSM extended by
means of the U$(1)'_\psi$ gauge model.}
\label{tabmassh}
\begin{center}
\small
\begin{tabular}{|c|c|c|c|c|}
\hline
$m_h$ &   $m_H$&  $m_{H'}$ &  $m_A$ & $m_{H^\pm}$\\ 
\hline
 125.0 &  1989.7 & 4225.0  & 4225.0 & 4335.6 \\ 
\hline\end{tabular}
\end{center}
\end{table}
\begin{table}[htp]
\caption{Masses of charginos and neutralinos in the reference point for
the $Z'_\psi$ model.}
\label{tabmasscn}
\begin{center}
\small
\begin{tabular}{|c|c|c|c|c|c|c|c|}
\hline
$m_{\tilde\chi_1^+}$ &   $m_{\tilde\chi_2^+}$ & $m_{\tilde\chi_1^0}$ &   $m_{\tilde\chi_2^0}$ 
& $m_{\tilde\chi_3^0}$ &   $m_{\tilde\chi_4^0}$ & $m_{\tilde\chi_5^0}$ &   $m_{\tilde\chi_6^0}$  \\ 
\hline
 204.8 & 889.1 & 197.2  & 210.7 & 408.8 & 647.9 & 889.0 & 6193.5 \\ 
\hline\end{tabular}
\end{center}
\end{table}\par
Given the numbers in Tables \ref{tabmassq}--\ref{tabmasscn},
one can calculate, by means of the SPheno program, the branching ratios of
the $Z'_\psi$ into Standard Model and supersymmetric final states.
At leading order in $g'$, i.e. ${\cal O}(g'^2)$, the main $Z'_\psi$ branching
ratios are quoted in Table~\ref{tabbrpsi}, for $m_{Z'}=2$~TeV
and omitting decay rates which are below 0.1\%.
\begin{table}[htp]
\caption{$Z'_\psi$ decay rates for $m_Z'=2$~TeV.}
\label{tabbrpsi}
\begin{center}
\small
\begin{tabular}{|c|c|}
\hline
Final State & $Z'_\psi$ Branching ratio (\%) \\
\hline
$\tilde\chi_1^+\chi_1^-$ & 10.2 \\
\hline
$\tilde\chi_1^0\tilde\chi_1^0$ & 4.9 \\
\hline
$\tilde\chi_1^0\tilde\chi_3^0$ & 0.2 \\
\hline
$\tilde\chi_2^0\tilde\chi_2^0$ & 5.1 \\
\hline
$\tilde\chi_4^0\tilde\chi_4^0$ & 8.0  \\
\hline
$hZ$ & 1.4\\
\hline
$W^+W^-$ & 2.9\\
\hline
$\sum_i d_i\bar d_i$ & 25.1\\
\hline
$\sum_i u_i\bar u_i$ & 25.0 \\
\hline
$\sum_i \nu_i\bar \nu_i$ & 8.3 \\
\hline
$\sum_i \ell^+_i\ell^-_i$ & 8.3 \\
\hline\end{tabular}
\end{center}
\end{table}
\begin{table}[htp]
\caption{Chargino $\tilde\chi_1^+$ decay rates in the reference point for
the $Z'_\psi$ model.}
\label{tabbrch}
\begin{center}
\small
\begin{tabular}{|c|c|}
\hline
Final State & $\chi_1^+$ branching ratio (\%) \\
\hline
$\tilde\chi^0_1\  u\bar d$ & 34.3 \\
\hline
$\tilde\chi^0_1\  u\bar c$ & 1.8 \\
\hline
$\tilde\chi_1^0\  c\bar d$ & 1.6 \\
\hline
$\tilde\chi_1^0\  c\bar s$ & 29.3 \\
\hline
$\tilde\chi_1^0\  e^+\nu_e$ & 12.0 \\
\hline
$\tilde\chi_1^0\  \mu^+\nu_\mu$ & 12.0 \\
\hline
$\tilde\chi_1^0\  \tau^+\nu_\tau$ & 8.9 \\
\hline\end{tabular}
\end{center}
\end{table}
The Standard Model decays are still the dominant ones, but one has an overall
28.3\% branching ratio into supersymmetric final states, which deserves
further investigation. In particular, the decay into chargino pairs 
$\tilde\chi^+_1\tilde\chi^-_1$ accounts
for about 10\%, whereas the ratios into neutralino pairs 
vary from 0.2\% ($\tilde\chi_1^0\tilde\chi_3^0$) to
8\% ($\tilde\chi_4^0\tilde\chi_4^0$).
Decays into pairs of the lightest neutralinos, i.e. 
$\tilde\chi^0_1\tilde\chi^0_1$, possibly relevant for the searches for Dark Matter
candidates, have non-negligible branching ratio, accounting for
about 5\%.

Since the highest BSM rate is the one into chargino pairs, it is
worthwhile carrying out the phenomenological analysis for final
states originated from a $Z'_\psi\to \tilde\chi_1^+\tilde\chi_1^-$
process.
As discussed in \cite{corgen}, primary decays into chargino pairs can lead to
a chain yielding charged leptons and missing energy in the final states.
To gauge the rates of the different final states, one must
compute the branching ratios of the 2- and 3-body decays
of the charginos $\tilde\chi^\pm_1$:
these numbers, calculated by means of SPheno,
are quoted in Table~\ref{tabbrch}.

\begin{figure}
\begin{center}
\begin{picture}(24000,4000)
\begin{small}
\drawline\photon[\E\REG](0,0)[4]
\drawarrow[\LDIR\ATTIP](\pmidx,\pmidy)
\global\advance\pmidy by 1000
\global\advance\pmidx by -700
\put(\pmidx,\pmidy){$Z'$}
\drawline\fermion[\SE\REG](\pbackx,\pbacky)[4000]
\drawarrow[\LDIR\ATTIP](\pmidx,\pmidy)
\global\advance\pmidy by 200
\put(\pmidx,\pmidy){\ \ \  $\tilde\chi_1^-$}
\drawline\fermion[\E\REG](\pbackx,\pbacky)[4000]
\drawarrow[\LDIR\ATTIP](\pmidx,\pmidy)
\put(\pbackx,\pbacky){\ \  $\tilde\chi_1^0$}
\drawline\photon[\SE\REG](\pfrontx,\pfronty)[4]
\drawarrow[\LDIR\ATTIP](\pmidx,\pmidy)
\global\advance\pmidy by -200
\put(\pmidx,\pmidy){\ \  \ $W^-$}
\drawline\fermion[\E\REG](\pbackx,\pbacky)[4000]
\drawarrow[\LDIR\ATTIP](\pmidx,\pmidy)
\put(\pbackx,\pbacky){\ \  $\bar\nu$}
\drawline\fermion[\SE\REG](\pfrontx,\pfronty)[4000]
\drawarrow[\LDIR\ATTIP](\pmidx,\pmidy)
\put(\pbackx,\pbacky){\ \  $\ell^-$}
\drawline\fermion[\NE\REG](4000,0)[4000]
\drawarrow[\LDIR\ATTIP](\pmidx,\pmidy)
\global\advance\pmidx by 200
\global\advance\pmidy by -200
\put(\pmidx,\pmidy){\ \ \  $\tilde\chi_1^+$}
\drawline\fermion[\E\REG](\pbackx,\pbacky)[4000]
\drawarrow[\LDIR\ATTIP](\pmidx,\pmidy)
\put(\pbackx,\pbacky){\ \  $\tilde\chi_1^0$}
\drawline\photon[\NE\REG](\pfrontx,\pfronty)[4]
\drawarrow[\LDIR\ATTIP](\pmidx,\pmidy)
\global\advance\pmidx by 200
\global\advance\pmidy by -200
\put(\pmidx,\pmidy){\ \ \  $W^+$}
\drawline\fermion[\E\REG](\pbackx,\pbacky)[4000]
\drawarrow[\LDIR\ATTIP](\pmidx,\pmidy)
\put(\pbackx,\pbacky){\ \  $\nu$}
\drawline\fermion[\NE\REG](\pfrontx,\pfronty)[4000]
\drawarrow[\LDIR\ATTIP](\pmidx,\pmidy)
\put(\pbackx,\pbacky){\ \  $\ell^+$}
\end{small}
\end{picture}\vspace{3.cm}
\caption{Final state with two charged leptons and missing energy,
due to neutrinos and neutralinos, 
through a primary decay of the $Z'$ into a chargino pair.}
\label{zch}
\end{center}
\end{figure}
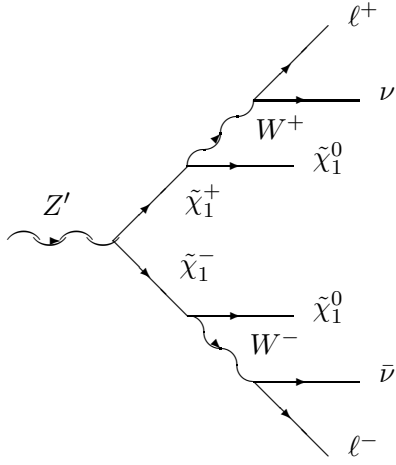

As hadronic final states are likely affected by large QCD backgrounds, 
I shall focus on the modes with neutralinos and leptons,
which will eventually lead to the following decay chain: 
\begin{equation}\label{zpc1c1}
pp\to Z'_\psi\to\tilde\chi_1^+\tilde\chi_1^-\to
(\tilde\chi_1^0\ell^+\nu_\ell)(\tilde\chi_1^0\ell^-\bar\nu_\ell),
\end{equation} 
with $\ell=\mu,e$. The neutrinos and neutralinos in (\ref{zpc1c1})
will give rise to some missing energy; the diagram of such a process is
presented in Fig.~\ref{zch}.
The U(1)$'_\psi$/MSSM masses and coupling constants,
in the UFO format, can be used by MadGraph
to generate parton-level events and then by HERWIG to simulate
showers and hadronization.
The cross section for the process $pp\to Z'_\psi$,  
computed by MadGraph at LO, by using
the CTEQL1 set \cite{cteq} for the initial-state parton distributions,
is $\sigma(pp\to Z'_\psi)\simeq 0.13$~pb.
The cross section for the decay chain
(\ref{zpc1c1}) is then given by
$\sigma (pp\to Z'_\psi\to \ell^+\ell^- +\rm{MET})\simeq 7.9\times 10^{-4}$~pb
at 14 TeV.
This means that such events can be, e.g.,
about 80 for a luminosity ${\cal L}\simeq 100~{\rm fb}^{-1}$,
almost 240 at 300 fb$^{-1}$ and so on.
Though being less likely than SM channels, 
supersymmetric decays are nevertheless pretty interesting,
since, unlike direct production of squark, slepton and 
gaugino pairs in $pp$ collisions, the final state 
with two charginos decaying in 
two charged leptons and two neutrinos has 
a fixed invariant mass.
\vspace{1.5cm}


\vspace{10.cm}
In the following, I will present some relevant leptonic distributions
and compare them with
those from direct decays $Z'_\psi\to \ell^+\ell^-$,
accounted in the LHC searches for $Z'$ bosons
carried out so far. Furthermore, the final state in process
(\ref{zpc1c1}) can even occur in events with
direct chargino production, i.e.
\begin{equation}\label{c1c1dir}
pp\to \tilde\chi_1^+\tilde\chi_1^-\to
(\tilde\chi_1^0\ell^+\nu_\ell)(\tilde\chi_1^0\ell^-\bar\nu_\ell),
\end{equation}
which represent a sort of
supersymmetric background for the events initiated by 
a $Z'_\psi$ decay.
In the chosen reference point, the LO cross section 
for direct $\tilde\chi^+_1\tilde\chi^-_1$ production 
is $\sigma(pp\to \tilde\chi^+_1\tilde\chi^-_1)\simeq 0.2$~pb; accounting
for the chargino branching ratios into neutralinos
and leptons (muons and electrons), the rate of the process (\ref{c1c1dir})
is then given by $\sigma\simeq 1.15\times 10^{-2}$~pb,
higher than in the chain (\ref{zpc1c1}).
Before presenting some distributions and comparisons, 
one can anticipate that, while in 
processes like (\ref{zpc1c1}) 
the chargino-pair invariant mass is
forced to reproduce $m_{Z'}$, in (\ref{c1c1dir}) the charginos
do not have this constraint and can therefore be very soft: 
the kinematics of the leptons produced in chargino decays
will in fact reflect this property.

\begin{figure}[htp]
\centerline{\resizebox{0.49\textwidth}{!}
{\includegraphics{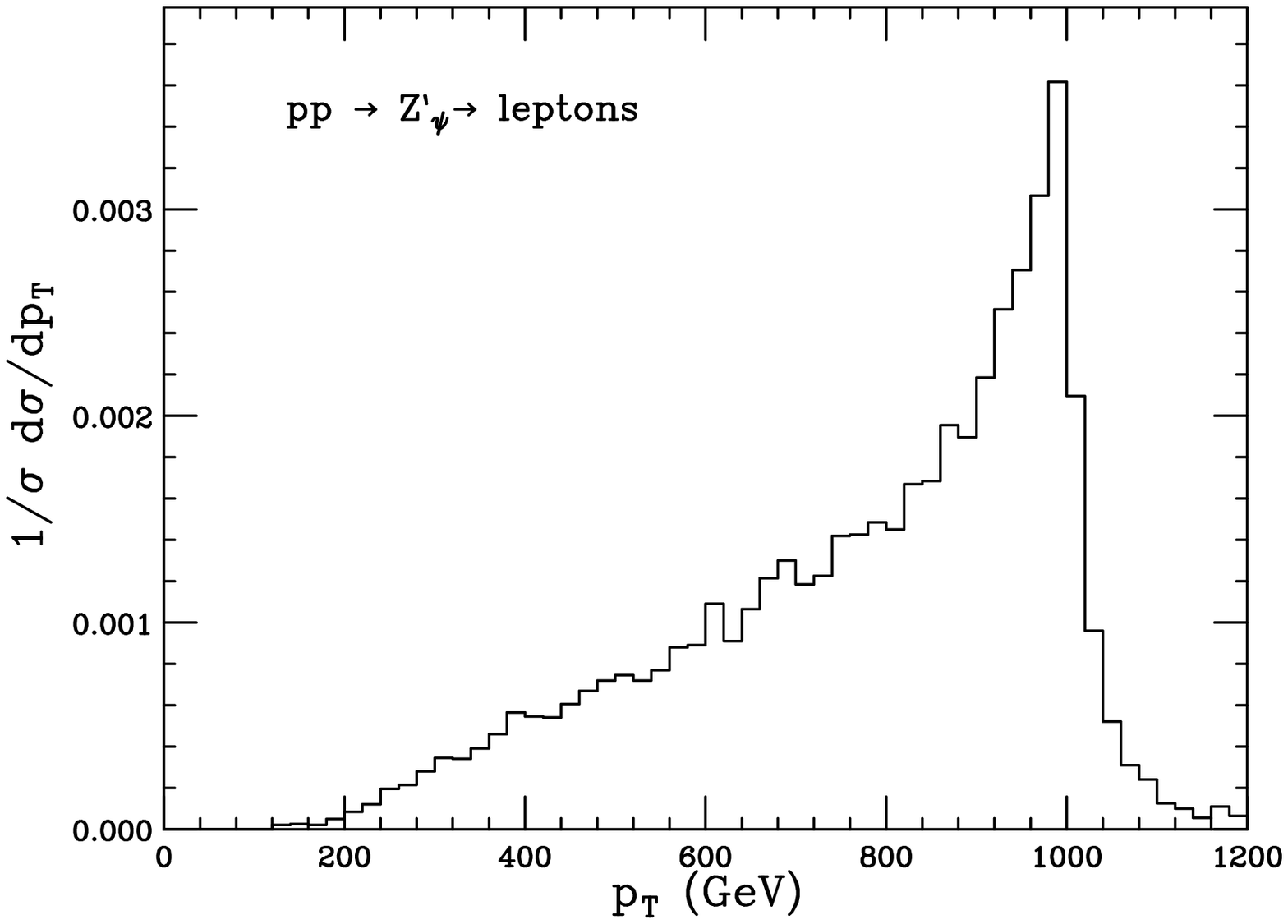}}%
\hfill%
\resizebox{0.49\textwidth}{!}{\includegraphics{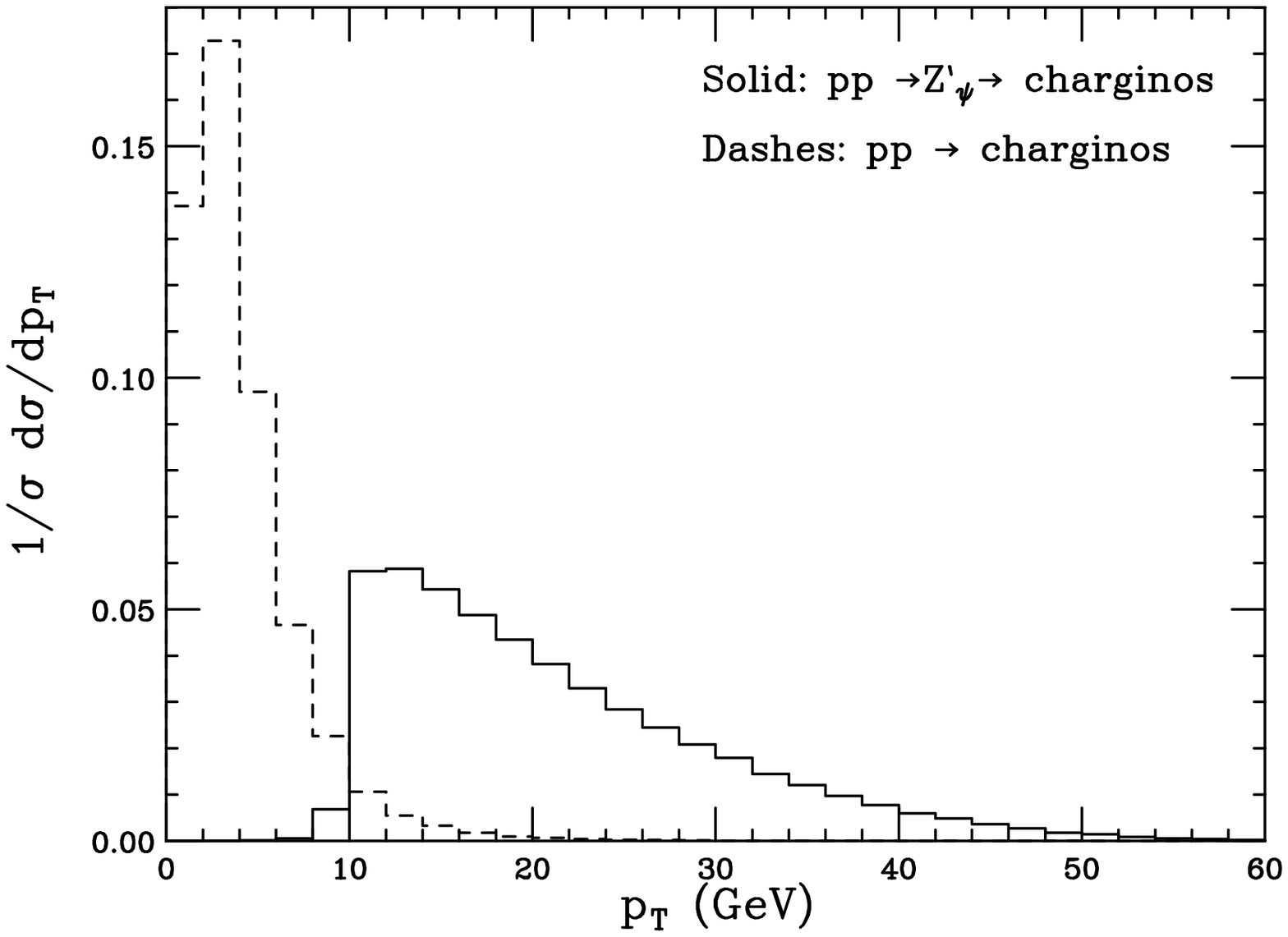}}}
\caption{Lepton transverse momentum for the $Z'_\psi$ model at $\sqrt{s}=14$~TeV
and $m_{Z'}=2$~TeV, for a direct $Z'_\psi\to \ell^+\ell^-$ decay (left) and 
chains initiated by $Z'_\psi\to\tilde\chi^+_1\chi^-_1$ or direct 
chargino production processes (right).}
\label{zpsipt}
\end{figure}

Figure~\ref{zpsipt} presents the transverse momentum spectrum of leptons
produced in all three processes: $Z'_\psi\to \ell^+\ell^-$
$Z'_\psi\to\tilde\chi^+_1\tilde\chi^-_1$, i.e.
Eq.~(\ref{zpc1c1}), and direct chargino-pair
production, like in Eq.~(\ref{c1c1dir}).
Since the kinematics of $\ell^+$ and $\ell^-$ 
is symmetric, the histograms contain the $p_T$ of both leptons.
For direct production (Figure~\ref{zpsipt}, left), 
the $p_T$ distribution starts to be non-negligible
for $p_T>200$~GeV, i.e. about $m_{Z'}/10$, then increases and reaches a peak about
$p_T\simeq m_{Z'}/2=1$~TeV; above 1 TeV the spectrum rapidly decreases.
In the case of the decay chain (\ref{zpc1c1}), 
the lepton transverse momentum has a completely different behavior: 
there are nearly no events below $p_T\simeq 8$~GeV, then the spectrum 
increases, reaches its peak at $p_T\simeq 15$~GeV and smoothly decreases,  
being negligible for $p_T>60$~GeV. 
For direct chargino production,
i.e. Eq.~(\ref{c1c1dir}), the lepton $p_T$ distribution
is mostly concentrated in the range $0<p_T<20$~GeV, exhibiting
a sharp peak about 5 GeV.
The observed $p_T$ spectra can easily be understood: 
for direct $Z'_\psi\to\ell^+\ell^-$, 
the two leptons get the full initial-state transverse momentum
and therefore the $p_T$ spectrum is substantial at high values, 
whereas, in the case of the cascades (\ref{zpc1c1}) and (\ref{c1c1dir}),
a consistent (missing) $p_T$ is lent to
neutrinos and neutralinos, which significantly decreases the $p_T$ of $\ell^+$
and $\ell^-$.  In particular, for direct charginos (\ref{c1c1dir})
there is no cutoff on the invariant mass of $\chi^+_1\chi^-_1$ pairs,
which can therefore be very soft, thus yielding 
mostly low-$p_T$ leptons. When the charginos come from $Z'$ decays,
$m_{Z'}$ is a constraint on their invariant mass, shifting
the lepton transverse momentum to higher values with respect
to those produced in (\ref{c1c1dir}).
\begin{figure}[htp]
\centerline{\resizebox{0.49\textwidth}{!}
{\includegraphics{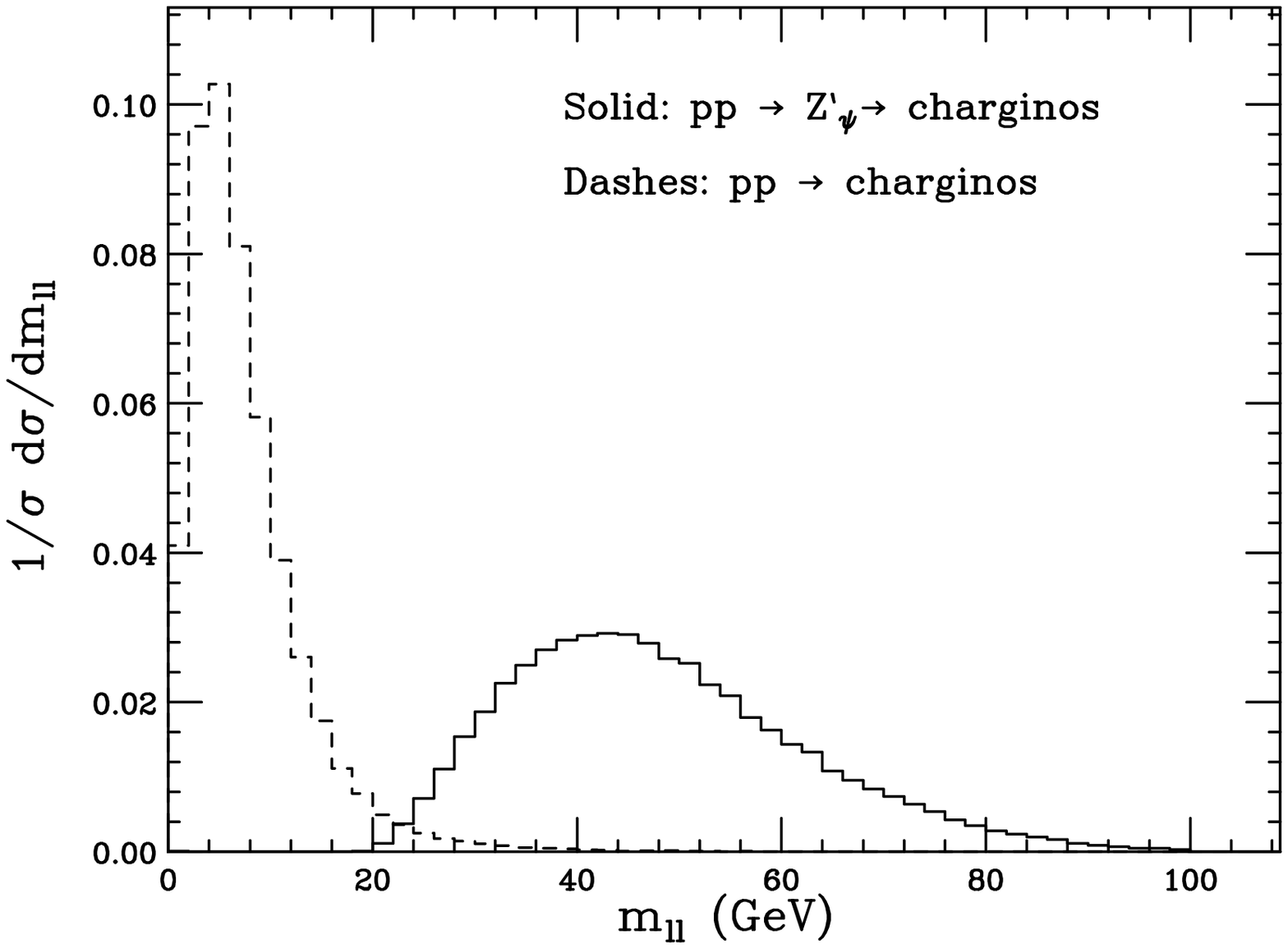}}%
\hfill%
\resizebox{0.49\textwidth}{!}{\includegraphics{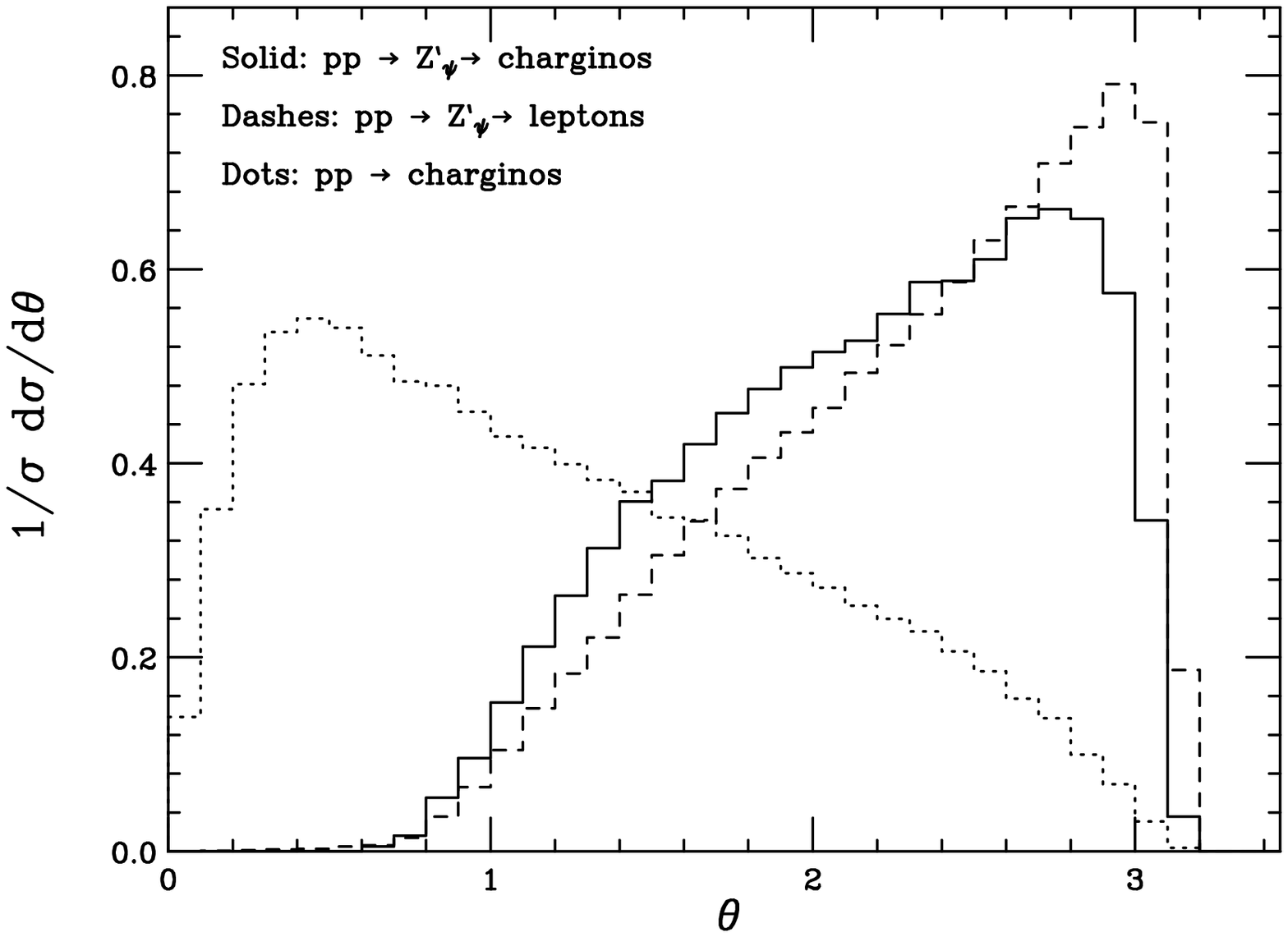}}}
\caption{Left: $\ell^+\ell^-$ inviariant-mass distribution of charged-lepton
pairs in the events (\ref{zpc1c1}) and (\ref{c1c1dir}). 
Right: angle between the two leptons $\ell^\pm$ in the laboratory frame for
direct $Z_\psi'\to\ell^+\ell^-$ production (dashes) and after the decay
chains (\ref{zpc1c1}) (solid) and (\ref{c1c1dir}) (dots).}
\label{zpsimth}
\end{figure}
\begin{figure}[htp]
\centerline{\resizebox{0.55\textwidth}{!}{\includegraphics{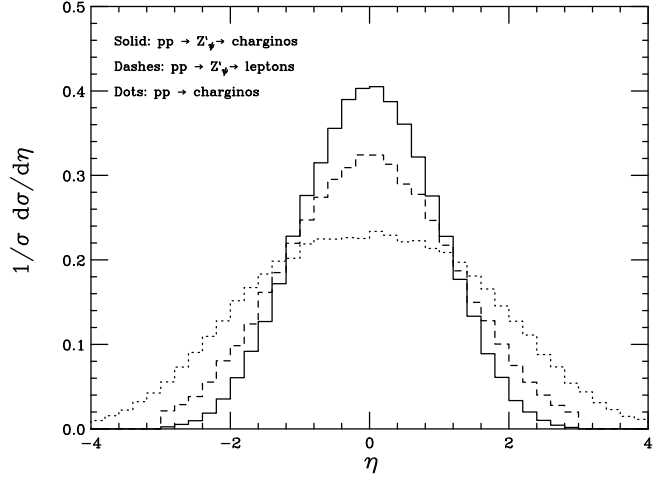}}}
\caption{Lepton rapidity distributions for standard $Z'_\psi$ decays into lepton pairs
(dashes) and in the supersymmetric cascades in Eqs.~(\ref{zpc1c1}) (solid
histogram) and
(\ref{c1c1dir}) (dots).}
\label{zpsieta}
\end{figure}
\begin{figure}[htp]
\centerline{\resizebox{0.49\textwidth}{!}
{\includegraphics{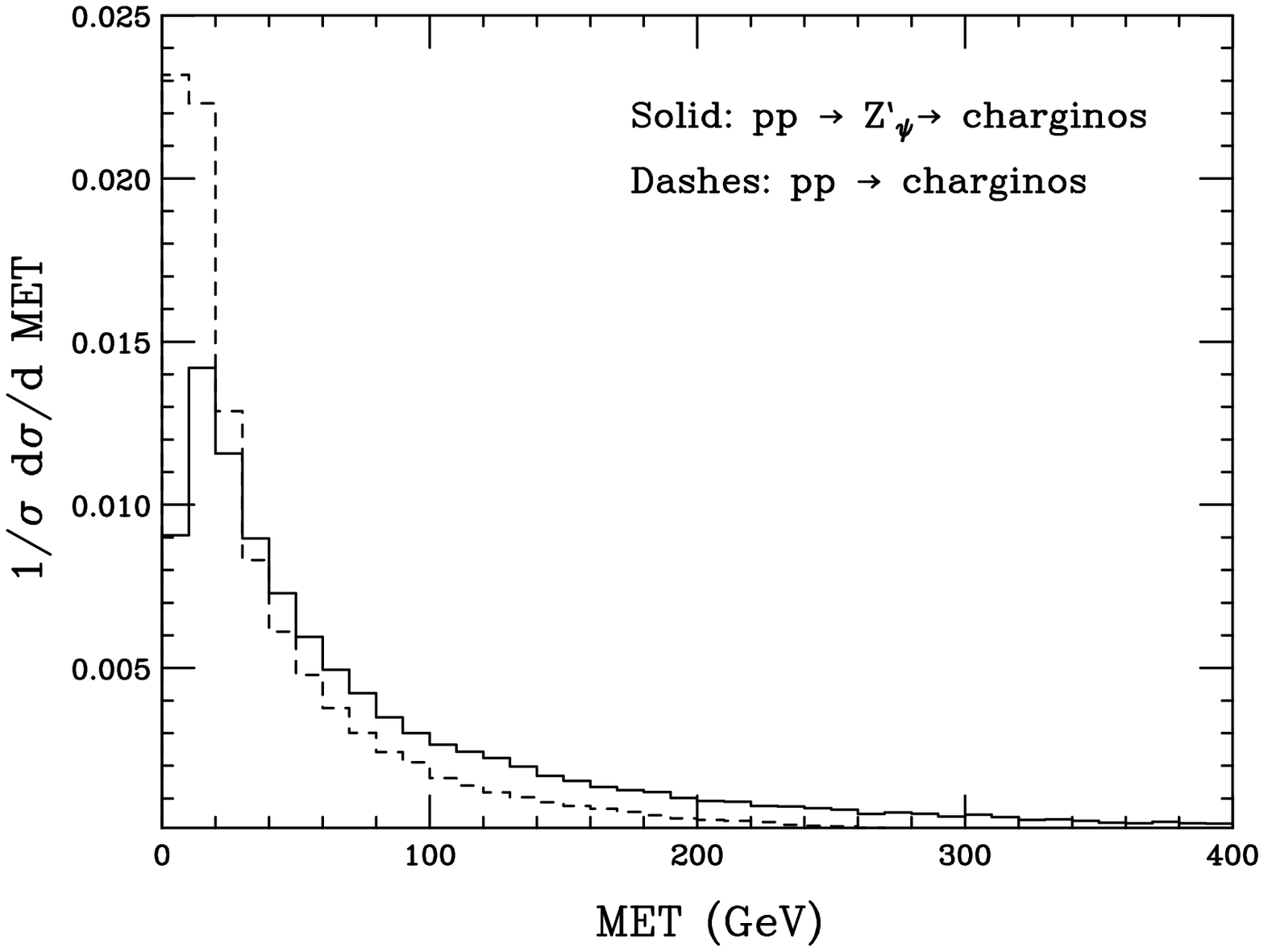}}%
\hfill%
\resizebox{0.49\textwidth}{!}{\includegraphics{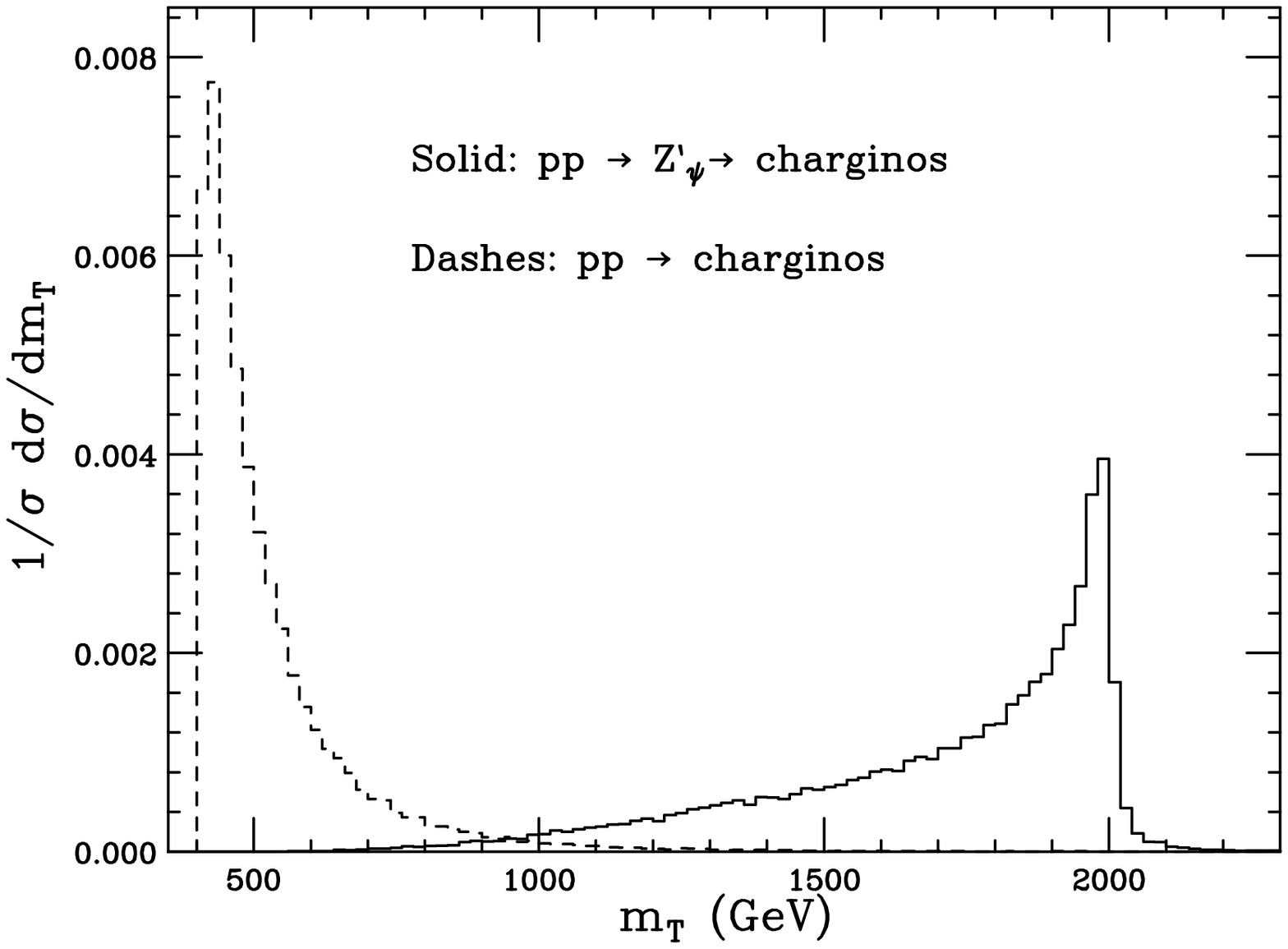}}}
\caption{
Left: missing transverse energy due to the neutrinos and neutralinos in
the cascade initiated by a primary $Z'_\psi\to\tilde\chi^+_1\tilde\chi^-_1$ decay
(solid histogram) and for direct chargino production (dashes).
Right: transverse mass for the final-state particles 
(leptons, neutrinos and neutralinos)
in the reactions (\ref{zpc1c1}) (solid) and (\ref{c1c1dir}) (dashes).}
\label{mtmetzpsi}
\end{figure}
\par In Fig.~\ref{zpsimth} one can instead find the 
$\ell^+\ell^-$ invariant mass $m_{\ell\ell}$
(left) and the angle $\theta$ between the two charged leptons in the laboratory 
frame (right). 
The invariant mass is plotted only for the
cascades (\ref{zpc1c1}) and (\ref{c1c1dir}), 
since, for direct $Z'_\psi\to\ell^+\ell^-$, it would just be 
a narrow resonance with the same mass and width as the $Z'_\psi$.
In the cascade (\ref{zpc1c1}), $m_{\ell\ell}$ 
varies essentially in the range 20 GeV~$<m_{\ell\ell}<$~100 GeV and has its
maximum value about $m_{\ell\ell}\simeq 45$~GeV.
For direct chargino production, $m_{\ell\ell}$ 
is peaked about 5 GeV and rapidly decreases, so that there are
nearly no events for $m_{\ell\ell}> 30$~GeV:
as observed before for the purpose of the $p_T$ distribution,
processes like (\ref{c1c1dir}) are dominated by soft 
charginos and therefore $\ell^\pm$ are
substantially produced at small $m_{\ell\ell}$.

As for the $\theta$ spectrum
(Fig.~\ref{zpsimth}, right),
for direct $Z'_\psi\to \ell^+\ell^-$ production it exhibits a maximum 
about $\theta\simeq 3$, a value close to back-to-back production,
i.e. $\theta=\pi$. When the leptons are accompanied by missing energy
in $Z'_\psi\to\tilde\chi^+_1\tilde\chi^-_1$ events, 
the $\theta$ distribution is broader;
it lies above the direct-production spectrum
at small and middle angles, below at high $\theta$, and is peaked
at a lower $\theta\simeq 2.75$.
The angular distribution of charged leptons in 
the chain (\ref{c1c1dir}) is instead completely different:
since there is no cutoff imposed by the $Z'$ mass,
$\ell^+$ and $\ell^-$ are essentially produced at small angles
and the $\theta$ spectrum is pretty broad, being peaked about $\theta\simeq \pi/6$
and negligible for back-to-back leptons.

Figure~\ref{zpsieta} presents the $\ell^\pm$ rapidity distributions:
the $\eta$ spectrum for leptons originated from the supersymmetric
cascade (\ref{zpc1c1}) has the highest fraction of leptons with
$\eta\sim 0$, corresponding to production
perpendicular to the beam axis, and the lowest at large $|\eta|$,
i.e. small angles with respect to the beam.
The $\eta$ distribution in direct chargino production,
i.e. process (\ref{c1c1dir}), is instead the lowest
at small $|\eta|$ and the highest at large $|\eta|$;
for direct lepton production in $Z'_\psi$ decays, it lies between
the other two distributions. 
As observed for the $\theta$ spectra, such a behaviour can
easily be understood in terms of the kinematics of the
processes which have been investigated: in $Z'\to\ell^+\ell^-$
the production is dominantly back-to-back, whereas in
Eq.~(\ref{c1c1dir}) the leptons are substantially
collinear to the beam axis.

Figure~\ref{mtmetzpsi} presents the differential distributions of
two observables which are typically studied in supersymmetry searches:
the sum of the transverse momenta
of `invisible' particles like neutrinos and neutralinos, also called
MET (missing transverse energy), and the transverse mass $m_T$ of
all final-state particles (neutrinos, neutralinos and charged leptons)
in the decay chains (\ref{zpc1c1}) and (\ref{c1c1dir}).
They are defined as follows:
\begin{eqnarray}
{\rm MET}=\sqrt{\left(\sum_i p_{x,i}\right)^2+
\left(\sum_i p_{y,i}\right)^2}\ &,&\   i=\nu,\bar \nu, \tilde\chi^0_1;
\nonumber \\
m_T=\sqrt{\left(\sum_jE_{T,j}\right)^2-\left(\sum_j\vec p_{T,j}\right)^2} &,&  
E_{T,j}=\sqrt{m_j^2+p_{T,j}^2} \ ,\ j=\ell^+,\ell^-,\nu,\bar \nu,\tilde\chi^0_1.
\label{metdef}
\end{eqnarray}
In both processes (\ref{zpc1c1}) and (\ref{c1c1dir}),
the MET spectrum is significant in the low range:
in the chain (\ref{zpc1c1}), it is 
sharply peaked at MET$\simeq 20$~GeV and smoothly decreases, vanishing 
for MET$>300$~GeV.
For direct chargino production, the MET exhibits an even sharper
peak at MET$\simeq 10$~GeV and decreases very rapidly, so that 
it is negligible above 200~GeV.

The transverse mass distribution exhibits instead a completely
different behaviour for processes (\ref{zpc1c1}) and
(\ref{c1c1dir}). In (\ref{c1c1dir}), leptons and neutralinos are
likely rather soft and collinear with respect to the beam and therefore
the transverse mass of the final state is
substantial only at small $m_T$: it is peaked around $m_T\simeq$~500 GeV 
and vanishes above 1 TeV. The chain
(\ref{zpc1c1}) is initiated by a $Z'_\psi$ with mass 2 TeV: 
the transverse mass 
is thus relevant in the range $m_{Z'}/2<m_T<m_{Z'}$
and maximum at $m_T\simeq 1.8$~TeV, just below the $Z'_\psi$ mass threshold.

Before moving to the investigation of direct decays into
light neutralinos, I wish to point out that, as a result
of the study so far carried out for a few observables, 
plotted in Figs.~\ref{zpsipt}-\ref{mtmetzpsi},
final states initiated by $Z'_\psi$ decays into
charginos can be safely discriminated from those coming from direct 
decays into lepton pairs, as well as from direct chargino production.
The last finding is not trivial, since the final states
of processes (\ref{zpc1c1}) and (\ref{c1c1dir}) are the same
and, in principle, direct chargino production would have been
a background for supersymmetric signals in $Z'_\psi$ decays. 

Since the branching ratio into neutralino pairs $\tilde\chi^0_1\tilde\chi^0_1$
is almost 5\%, even the process
\begin{equation}
pp\to Z'_\psi\to\tilde\chi^0\tilde\chi^0_1
\label{zpsin1n1}
\end{equation}
has a substantial cross section, i.e. 
$\sigma(pp\to Z'_\psi\to \tilde\chi^0_1\tilde\chi^0_1)\simeq 6.4\times 10^{-3}$~pb
at $\sqrt{s}=14$~TeV,
which yields about 640 events at ${\cal L}=100$~fb$^{-1}$ and up to almost $2\times 10^3$
at 300 fb$^{-1}$.
In fact, unlike charginos, the lightest neutralinos are stable particles
in the MSSM, and therefore the cross section of the process (\ref{zpsin1n1})
does not get any further branching fraction which possibly dilutes the
event rate. Therefore, 
the U(1)$'_\psi$ extension of the MSSM could be an interesting scenario
to search for Dark Matter candidates in the 14 TeV run of the LHC.
The typical signature is given by mono-photon or mono-jet final states,
with the photon and jet being associated with initial-state radiation
from the incoming quarks. The actual implementation
of photon isolation criteria or jet-clustering algorithms goes 
beyond the scopes of this paper and will not debated here.

\begin{figure} 
\centerline{\resizebox{0.55\textwidth}{!}{\includegraphics{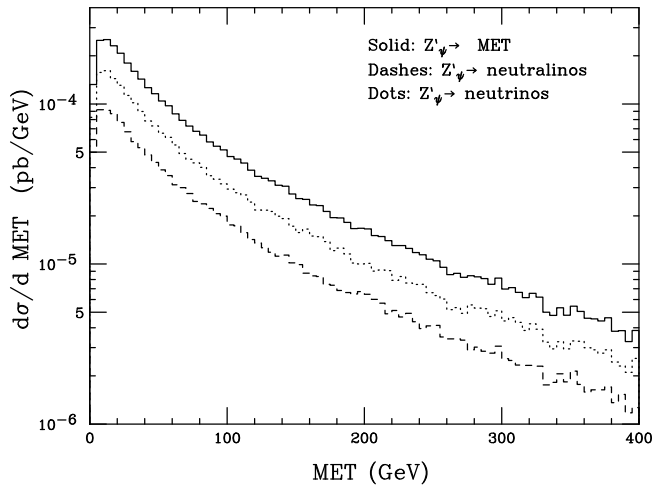}}}
\caption{Missing transverse energy in $Z'_\psi$ decays: plotted are
the neutralino (dashes), neutrino (dots) and total (solid) contributions to 
the missing transverse energy.}
\label{metneu}
\end{figure}

Competing processes, leading to final states with just
missing energy, are $Z'_\psi$ decays into neutrino pairs,
amounting to about $\sigma(pp\to Z'_\psi\to\nu\bar\nu)\simeq 
1.1\times 10^{-2}$~pb at 14 TeV, with ${\cal O}(10^3)$ 
events at 100 and 300 fb$^{-1}$.
Figure~\ref{metneu} displays the total
missing transverse energy (MET) spectrum and the contribution due to the 
neutrino and neutralino pairs in $Z'_\psi$ decays; unlike
previous distributions, they are normalized to the total LO 
cross section and not to unity, in such a way to appreciate the
discrepancy between the two subprocesses.
All plots are peaked at ${\rm MET}\simeq 10$~GeV 
and smoothly decrease, up to the point of being quite negligible
for ${\rm MET}> 300$~GeV. The shapes of both neutrino- and 
neutralino-induced spectra are very similar, which is quite reasonable since
the $\tilde\chi^0_1$ particles, though being massive, are still much
lighter than the decaying $Z'_\psi$. 
Nevertheless, the total number of events
at any MET value is substantially higher, by about 60\%,
if neutralinos contribute.

\section{Phenomenology - U$(1)'_\eta$ model}

The model U$(1)'_\eta$ corresponds to a mixing angle $\theta=\arccos{\sqrt{5/8}}$
and, even in the reference point considered in \cite{corgen},
gives rise to an interesting $Z'_\eta$ phenomenology within supersymmetry,
the BSM channels accounting for about 1/4 of the total width.
In the following, though keeping the constraints due to the Higgs mass and 
direct supersymmetry searches, I shall choose a slightly 
different representative point of the parameter space, with respect
to the previous U$(1)'_\psi$ model, 
in order to enhance supersymmetric decays. 
In particular, the $Z'_\eta$ will still have mass $m_{Z'}=2$~TeV, 
$M_1$, $M_2$, $M'$, $\tan\beta$, $\mu$, $A_q$,
$A_\ell$ and $A_\lambda$ will be set to the values in
Eqs.~(\ref{mubeta}) and (\ref{aaa}), like in the $Z'_\psi$ scenario,
whereas all three generations of squarks and sleptons will be degenerate 
at the $Z'_\eta$ scale, with masses equal to the following values: 
\begin{equation}
m_{\tilde\ell}^0=m_{\tilde\nu_\ell}^0=1.5~{\rm TeV},\ 
m^0_{\tilde q}=3~{\rm TeV},
\end{equation}
where $q=u,d,c,s,t,b$ and $\ell=e,\mu,\tau$.
After adding the D-term, the masses of squarks and sleptons are
quoted in Tables~\ref{tabmassqeta} and \ref{tabmassleta},
and exhibit a substantial impact of the D-term. 
The squark masses increase or decrease by few hundred GeV,
whereas 
$\tilde\ell_2$  and $\tilde\nu_1$ get slightly heavier,
$m_{\tilde\ell_2}$ a bit lower and $\tilde\nu_2$ 
considerably lighter, by about 850 GeV.
This is therefore an example of negative D-term; 
in fact, in \cite{corgen}, negative and
large D-terms had even led to the exclusion of a few $Z'$ models,
since some sfermion squared masses had become negative.
Table~\ref{tabmassheta} contains the masses of the Higgs bosons, which are rather
similar to those obtained for the $Z'_\psi$ case: $m_h\simeq 125$~GeV,
$m_H\simeq m_{Z'}$, with $H'$, $A$ and $H^+$
above 4 TeV. With those numbers for the Higgs boson, the $\lambda$
parameter in the trilinear potential $V_\lambda$
is now equal to $\lambda=\sqrt{2}\mu/v_3\simeq 4.3\times 10^{-2}$.
Chargino and neutralino masses are reported in 
Table~\ref{tabmasscneta}: $\tilde\chi^\pm_1$, $\tilde\chi^\pm_2$
and the first four neutralinos ($\tilde\chi^0_1\dots\tilde\chi^0_4$)
are roughly as heavy as those in the $Z'_\psi$ model previously 
considered; $\tilde\chi^0_5$ and $\tilde\chi^0_6$ have masses 
above 1.5 and 2.5 TeV, respectively, and they are therefore
negligible for $Z'_\eta$ phenomenology in this scenario.
\begin{table}[htp]
\caption{Masses in GeV of the squarks in the $Z'_\eta$ model in the representative
point of the parameter space, for a soft mass 
$m^0_{\tilde q}=3$~TeV and $m_{Z'}=2$~TeV.}
\label{tabmassqeta}
\begin{center}
\small
\begin{tabular}{|c|c|c|c|c|c|}
\hline
$m_{\tilde d_1}$ &   $m_{\tilde u_1}$ &  $m_{\tilde s_1}$ &  $m_{\tilde c_1}$ &  
$m_{\tilde b_1}$ & $m_{\tilde t_1}$\\ 
\hline
3130.8  & 3129.8  & 3130.8 & 3129.8 & 3130.8 & 3175.5 \\
\hline
$m_{\tilde d_2}$ &   $m_{\tilde u_2}$ &  $m_{\tilde s_2}$ &  $m_{\tilde c_2}$ &  
$m_{\tilde b_2}$ & $m_{\tilde t_2}$\\ 
\hline
 3065.9 & 2863.6  & 3065.9 & 2863.6 & 3065.9 &  2823.5 \\
\hline\end{tabular}
\end{center}
\end{table}
\begin{table}[htp]
\caption{Masses of sleptons in the $Z'_\eta$ scenario, with a soft term
$m^0_{\tilde\ell}=m_{\tilde\nu}=1.3$~TeV. 
All numbers are in GeV and $\ell=e,\mu$.}
\label{tabmassleta}
\begin{center}
\small
\begin{tabular}{|c|c|c|c|c|c|c|c|}
\hline
$m_{\tilde \ell_1}$ &   $m_{\tilde \ell_2}$ &  $m_{\tilde\tau_1}$
& $m_{\tilde\tau_2}$ &
$m_{\tilde \nu_{\ell,1}}$ &  $m_{\tilde \nu_{\ell,2}}$ &
$m_{\tilde \nu_{\tau,1}}$ &  $m_{\tilde \nu_{\tau,2}}$ \\ 
\hline
 1194.6 & 1364.5  & 1208.8 & 1307.7 &  1361.8  & 456.0 & 1368.0 & 456.0\\ 
\hline\end{tabular}
\end{center}
\end{table}
\begin{table}[htp]
\caption{Higgs bosons in the $Z'_\eta$ model, with masses expressed in GeV.}
\label{tabmassheta}
\begin{center}
\small
\begin{tabular}{|c|c|c|c|c|}
\hline
$m_h$ &   $m_H$ &  $m_{H'}$ &  $m_A$ & $m_{H^+}$\\ 
\hline
  124.9 &  2004.2 & 4229.4  & 4229.4 & 4230.0\\ 
\hline\end{tabular}
\end{center}
\end{table}
\begin{table}[htp]
\caption{Masses in GeV of charginos and neutralinos in the $Z'_\eta$ model. }
\label{tabmasscneta}
\begin{center}
\small
\begin{tabular}{|c|c|c|c|c|c|c|c|}
\hline
$m_{\tilde\chi_1^+}$ &   $m_{\tilde\chi_2^+}$ & $m_{\tilde\chi_1^0}$ &   $m_{\tilde\chi_2^0}$ 
& $m_{\tilde\chi_3^0}$ &   $m_{\tilde\chi_4^0}$ 
& $m_{\tilde\chi_5^0}$ &   $m_{\tilde\chi_6^0}$  \\ 
\hline
 206.5 & 882.4 & 199.3  & 212.5 & 408.2 & 882.3 & 1562.8 & 2569.2 \\ 
\hline\end{tabular}
\end{center}
\end{table}
\par Table~\ref{tabbreta} presents the branching ratios of the $Z'_\eta$ into
the most significant decay channels: the Standard Model
modes are still the most relevant, with the supersymmetric
channels accounting for about 21\% of the total width.
Among the supersymmetric channels, 
sneutrino pairs $\tilde\nu_2\tilde\nu_2^*$ exhibit the highest rate,
slightly below 10\% after adding up all three flavors;
the decay into $\tilde\chi^+_1\tilde\chi^-_1$
accounts for about 6\%, into neutralino pairs for another 5\%. 
\begin{table} 
\caption{$Z'_\eta$ decay rates in the MSSM reference point for
a mass $m_{Z'}=2$~TeV.}
\label{tabbreta}
\begin{center}
\small
\begin{tabular}{|c|c|}
\hline
Final State & $Z'$ Branching ratio (\%) \\
\hline
$\tilde\chi_1^+\chi_1^-$ & 5.6 \\
\hline
$\tilde\chi_1^0\tilde\chi_1^0$ & 1.9 \\
\hline
$\tilde\chi_2^0\tilde\chi_2^0$ & 2.1 \\
\hline
$\tilde\chi_1^0\tilde\chi_2^0$ & 1.5 \\
\hline
$\sum_\ell\tilde\nu_{\ell,2}\tilde\nu_{\ell,2}^*$ & 9.4  \\
\hline
$hZ$ & 1.5\\
\hline
$W^+W^-$ & 3.0\\
\hline
$\sum_i d_i\bar d_i$ & 16.1 \\
\hline
$\sum_i u_i\bar u_i$ & 25.5 \\
\hline
$\sum_i \nu_i\bar \nu_i$ & 27.8 \\
\hline
$\sum_i \ell^+_i\ell^-_i$ & 5.3 \\
\hline
\end{tabular}
\end{center}
\end{table}
As done for the previous model, 
the phenomenological analysis will be undertaken for
the supersymmetric mode with the highest branching ratio,
i.e. $Z'_\eta\to\tilde\nu_2\tilde\nu_2^*$. 
In the notation used in this paper, $\tilde\nu_2$ 
is the supersymmetric partner of the $\nu_2$,
which, after the mixing, is mostly a right-handed neutrino. 
The sneutrinos decay into neutrinos $\nu_2$ and neutralinos,
with branching ratios given in Table~\ref{brsnu}: the
highest rates are into  
neutralino--neutrino pairs $\tilde\chi_3^0\nu_2$
and $\tilde\chi_2^0\nu_2$.
In order to discriminate among the final states yielded
by these two decay modes, one needs to evaluate,
by using SPheno, the
rates of neutralinos $\tilde\chi^0_3$ and $\tilde\chi^0_2$:
they are quoted in Tables~\ref{brchi03} and \ref{brchi02},
respectively.
\begin{table} 
\caption{Sneutrino $\tilde\nu_2$ branching ratios, in the representative
point of the $Z'_\eta$ model, where $m_{\tilde\nu_2}\simeq 456$~GeV.}
\label{brsnu}
\begin{center}
\small
\begin{tabular}{|c|c|}
\hline
Final state & $\tilde\nu_2$ branching ratio (\%) \\
\hline
$\tilde\chi_1^0\nu_2$ & 4.0 \\
\hline
$\tilde\chi_2^0\nu_2$ & 37.3 \\
\hline
$\tilde\chi_3^0\nu_2$ & 58.7 \\
\hline\end{tabular}
\end{center}
\end{table}
\begin{table} 
\caption{Branching ratios of the neutralino $\tilde\chi^0_3$ in the representative
point of the $Z'_\eta$ model.}
\label{brchi03}
\begin{center}
\small
\begin{tabular}{|c|c|}
\hline
Final State & $\tilde\chi^0_3$ Branching ratio (\%) \\
\hline
$\tilde\chi^\pm_1W^\mp$ & 56.4 \\
\hline
$\tilde\chi^0_1h$ & 19.3 \\
\hline
$\tilde\chi^0_2h$ & 1.2 \\
\hline
$\tilde\chi^0_2Z$ & 20.2 \\
\hline
$\tilde\chi^0_1Z$ & 3.0\\
\hline\end{tabular}
\end{center}
\end{table}
\begin{table} 
\caption{As in Table~\ref{brchi03}, but for the lighter neutralino
$\tilde\chi^0_2$.}
\label{brchi02}
\begin{center}
\small
\begin{tabular}{|c|c|}
\hline
Final State & $\tilde\chi^0_2$ Branching ratio (\%) \\
\hline
$\sum_i\tilde\chi_1^0q_i\bar q_i$ & 63.3 \\
\hline
$\sum_i\tilde\chi^0_1\ell^+_i\ell^-_i$ & 13.4 \\
\hline
$\sum_i\tilde\chi_1^0\nu_i\bar\nu_i$ & 20.6 \\
\hline\end{tabular}
\end{center}
\end{table}
Because of its higher mass, the neutralino $\tilde\chi_3^0$ 
is capable of decaying
according to $\tilde\chi^0_3\to \tilde\chi_1^\pm W^\mp$,
with a branching fraction about 
56\%; the other main channels are 
$\tilde\chi^0_{1,2}h$ and $\tilde\chi^0_{1,2}Z$ pairs,
accounting for about 20 and 23\%, respectively.
As for $\tilde\chi^0_2$, 
it undergoes decays into the lightest
$\tilde\chi_1^0$ and a pair of quarks, charged leptons or
neutrinos, through an intermediate
charged slepton $\tilde\ell^\pm$, with branching ratios 
varying from about 63 to 13\%, as quoted in
Table~\ref{brchi02}.

As a result, in order to end up with a 
final state with leptons and missing energy, and
considering only electrons and muons, 
one has
\begin{eqnarray}\label{br23}&\ &
{\rm B}(\tilde\nu_2\to \tilde\chi^0_2\nu_2)\times 
{\rm B}(\tilde\chi^0_2\to \tilde\chi^0_1\ell^+\ell^-) \simeq  3.3\%,\\ 
&\ &{\rm B}(\tilde\nu_2\to \tilde\chi^0_3\nu_2)\times 
{\rm B}(\tilde\chi^0_3\to \tilde\chi_1^\pm W^\mp)\times
{\rm B}(\tilde\chi_1^\pm\to \tilde\chi^0_1\ell^\pm\nu_\ell)
\times {\rm B}(W^\mp\to \ell^\mp\nu_\ell) 
\simeq 3.0\%.\nonumber
\end{eqnarray}
From Eq.~(\ref{br23}) one learns that,
although the decay $\tilde\nu_2\to \tilde\chi^0_3\nu_2$
is more probable than $\tilde\nu_2\to \tilde\chi^0_2\nu_2$,
after accounting for all the subprocesses, the overall
branching ratios are comparable, with the one originated from
a sneutrino decay into $\tilde\chi^0_2\nu_2$ being even slightly
larger.

\begin{figure}
\begin{center}
\begin{picture}(24000,4000)
\begin{small}
\drawline\photon[\E\REG](0,0)[4]
\drawarrow[\LDIR\ATTIP](\pmidx,\pmidy)
\global\advance\pmidy by 1000
\global\advance\pmidx by -1000
\put(\pmidx,\pmidy){$Z'$}
\drawline\scalar[\SE\REG](\pbackx,\pbacky)[2]
\drawarrow[\LDIR\ATTIP](\pmidx,\pmidy)
\global\advance\pmidy by -400
\global\advance\pmidx by -2000
\put(\pmidx,\pmidy){$\tilde\nu^*_2$}
\drawline\fermion[\E\REG](\pbackx,\pbacky)[4000]
\drawarrow[\LDIR\ATTIP](\pmidx,\pmidy)
\put(\pbackx,\pbacky){\ \  $\bar\nu_2$}
\drawline\fermion[\SE\REG](\pfrontx,\pfronty)[4000]
\drawarrow[\LDIR\ATTIP](\pmidx,\pmidy)
\global\advance\pmidy by -500
\global\advance\pmidx by -3500
\put(\pmidx,\pmidy){\ \  \ $\tilde\chi^0_2$}
\drawline\scalar[\E\REG](\pbackx,\pbacky)[2]
\drawarrow[\LDIR\ATTIP](\pmidx,\pmidy)
\global\advance\pmidy by 500
\put(\pmidx,\pmidy){$\tilde\ell^-$}
\drawline\fermion[\SE\REG](\pbackx,\pbacky)[4000]
\drawarrow[\LDIR\ATTIP](\pmidx,\pmidy)
\global\advance\pbacky by -1000
\put(\pbackx,\pbacky){\ \  $\tilde\chi^0_1$}
\drawline\fermion[\NE\REG](\pfrontx,\pfronty)[4000]
\drawarrow[\LDIR\ATTIP](\pmidx,\pmidy)
\put(\pbackx,\pbacky){\ \  $\ell^-$}
\drawline\fermion[\SE\REG](9500,-5500)[4000]
\drawarrow[\LDIR\ATTIP](\pmidx,\pmidy)
\global\advance\pbacky by -1000
\put(\pbackx,\pbacky){\ \  $\ell^+$}
\drawline\scalar[\NE\REG](4000,0)[2]
\drawarrow[\LDIR\ATTIP](\pmidx,\pmidy)
\global\advance\pmidy by 400
\global\advance\pmidx by -1000
\put(\pmidx,\pmidy){$\tilde\nu_2$}
\drawline\fermion[\E\REG](\pbackx,\pbacky)[4000]
\drawarrow[\LDIR\ATTIP](\pmidx,\pmidy)
\put(\pbackx,\pbacky){\ \  $\nu_2$}
\drawline\fermion[\NE\REG](\pfrontx,\pfronty)[4000]
\drawarrow[\LDIR\ATTIP](\pmidx,\pmidy)
\global\advance\pmidy by 500
\global\advance\pmidx by -3000
\put(\pmidx,\pmidy){\ \  \ $\tilde\chi^0_2$}
\drawline\fermion[\NE\REG](\pfrontx,\pfronty)[4000]
\drawarrow[\LDIR\ATTIP](\pmidx,\pmidy)
\drawline\scalar[\E\REG](\pbackx,\pbacky)[2]
\drawarrow[\LDIR\ATTIP](\pmidx,\pmidy)
\global\advance\pmidy by 500
\put(\pmidx,\pmidy){$\tilde\ell^+$}
\drawline\fermion[\NE\REG](\pbackx,\pbacky)[4000]
\drawarrow[\LDIR\ATTIP](\pmidx,\pmidy)
\drawarrow[\LDIR\ATTIP](\pmidx,\pmidy)
\put(\pbackx,\pbacky){\ \  $\tilde\chi^0_1$}
\drawline\fermion[\SE\REG](\pfrontx,\pfronty)[4000]
\drawarrow[\LDIR\ATTIP](\pmidx,\pmidy)
\put(\pbackx,\pbacky){\ \  $\ell^+$}
\drawline\fermion[\NE\REG](9500,5500)[4000]
\drawarrow[\LDIR\ATTIP](\pmidx,\pmidy)
\global\advance\pbacky by 500
\put(\pbackx,\pbacky){\ \  $\ell^-$}
\end{small}
\end{picture}\vspace{3.2cm}
\caption{Final state with four charged leptons and missing energy, initiated
by a $Z'$ decay into a sneutrino pair.}
\label{zsnu}
\end{center}
\end{figure}
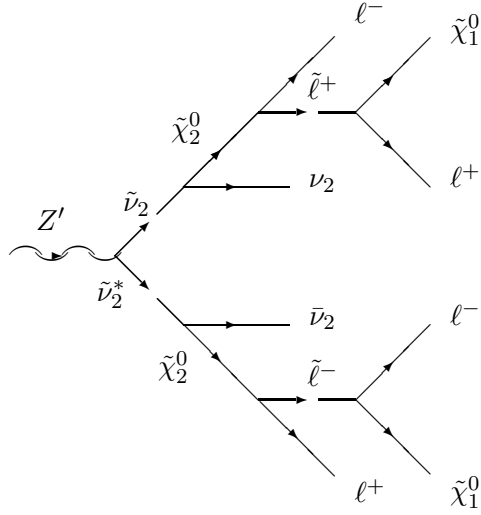

In this paper, 
I shall therefore investigate the following cascade, originating from
a $\chi_2^0\nu_2$ pair, leaving the study of 
sneutrino decays into $\tilde\chi^0_3\nu_2$ to future work:
\begin{equation}
pp\to Z'_\eta\to \tilde\nu_2\tilde\nu_2^*\to (\tilde\chi^0_2\nu_2) (\tilde\chi^0_2\bar\nu_2)
\to (\ell^+\ell^-\tilde\chi^0_1\nu_2)(\ell^+\ell^-\tilde\chi^0_1\bar\nu_2).
\label{zpsnu}
\end{equation}
The final state is thus made of four charged leptons and missing energy,
due to neutrinos $\nu_2$ and neutralinos $\tilde\chi^0_1$;
the diagram for the process (\ref{zpsnu}) is presented in
Fig.~\ref{zsnu}.
\par
The cross section for $Z'_\eta$ production in the above scenario
at 14 TeV, computed by MadGraph, is 
$\sigma(pp\to Z'_\eta)\simeq 0.18$~pb.
Given the numbers in Tables~\ref{brsnu} and \ref{brchi02},
and accounting only for $e^\pm$ and $\mu^\pm$, 
the cross section of the cascade (\ref{zpsnu}) is thus
$\sigma(pp\to Z'_\eta\to 4\ell+{\rm MET})\simeq 1.90\times 10^{-4}$~pb, which
yields about 20 events at ${\cal L}=100$~fb$^{-1}$ and 60 at 300~fb$^{-1}$.
\begin{figure}[t]
\centerline{\resizebox{0.49\textwidth}{!}
{\includegraphics{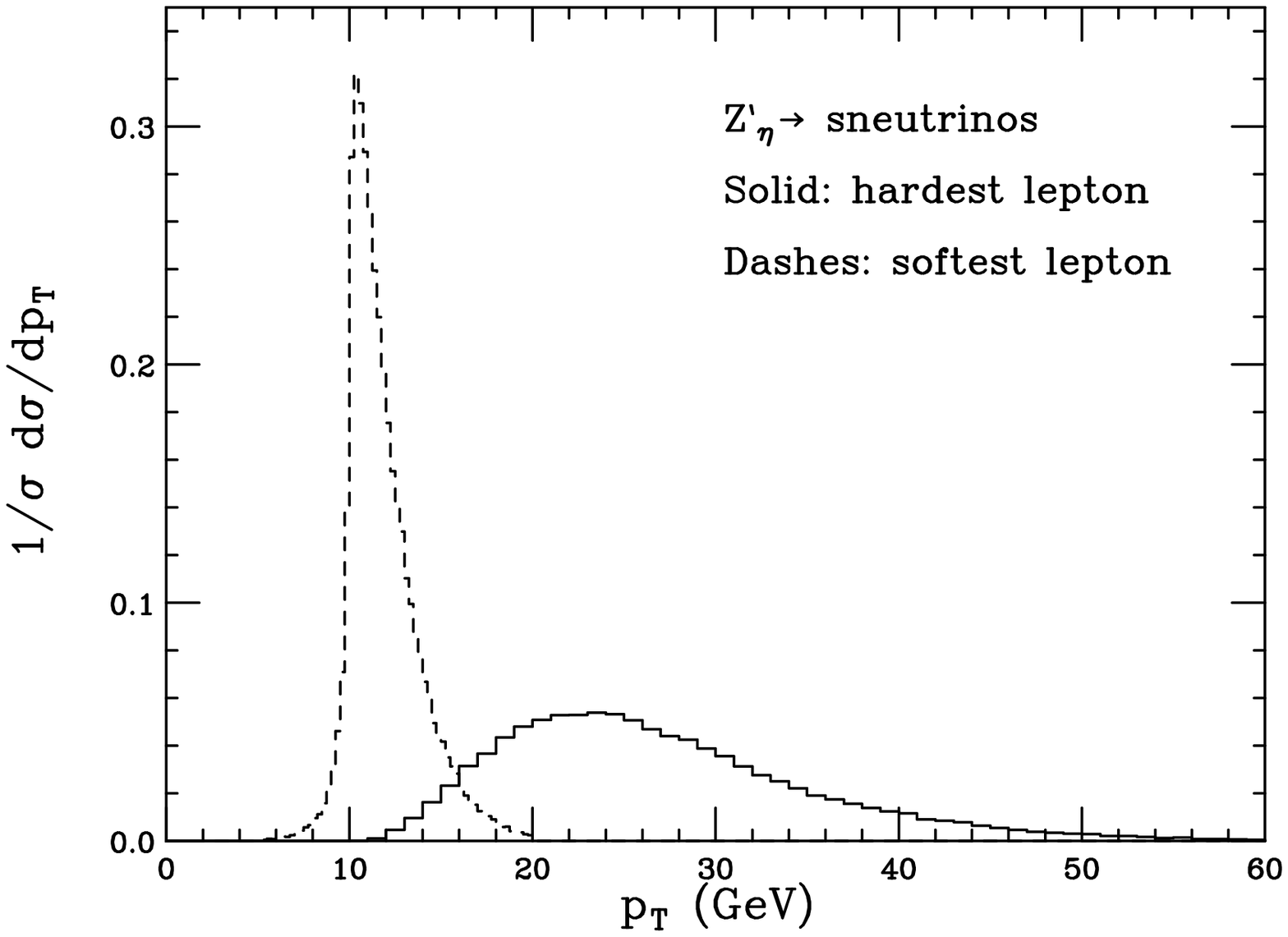}}%
\hfill%
\resizebox{0.49\textwidth}{!}{\includegraphics{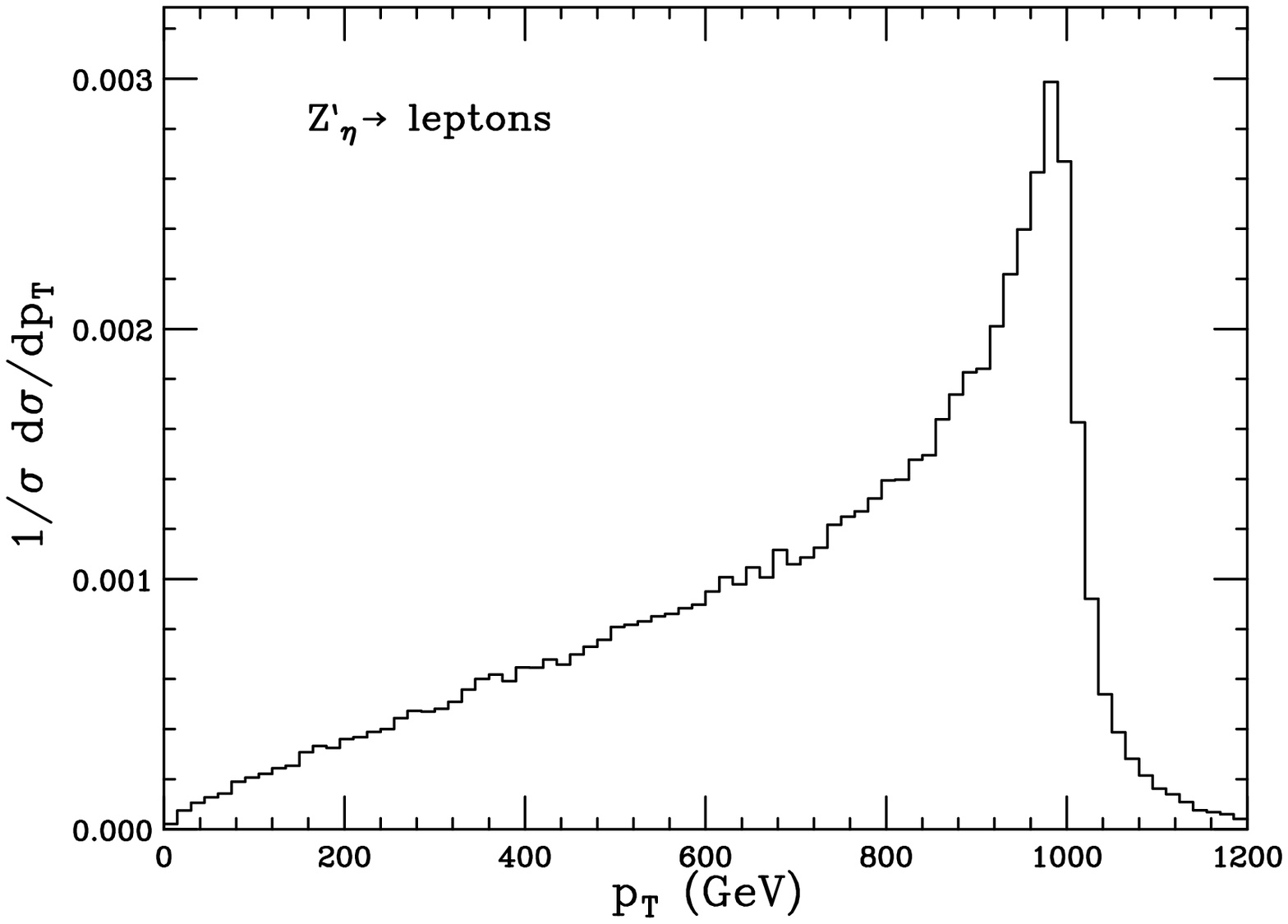}}}
\caption{Left: Transverse momentum of the hardest (solid) and softest (dashes) lepton
in the cascade (\ref{zpsnu}). Right: Lepton transverse momentum in 
$Z'_\eta\to \ell^+\ell^-$ processes.}
\label{zetapt}
\end{figure}
\begin{figure}[htp]
\centerline{\resizebox{0.49\textwidth}{!}
{\includegraphics{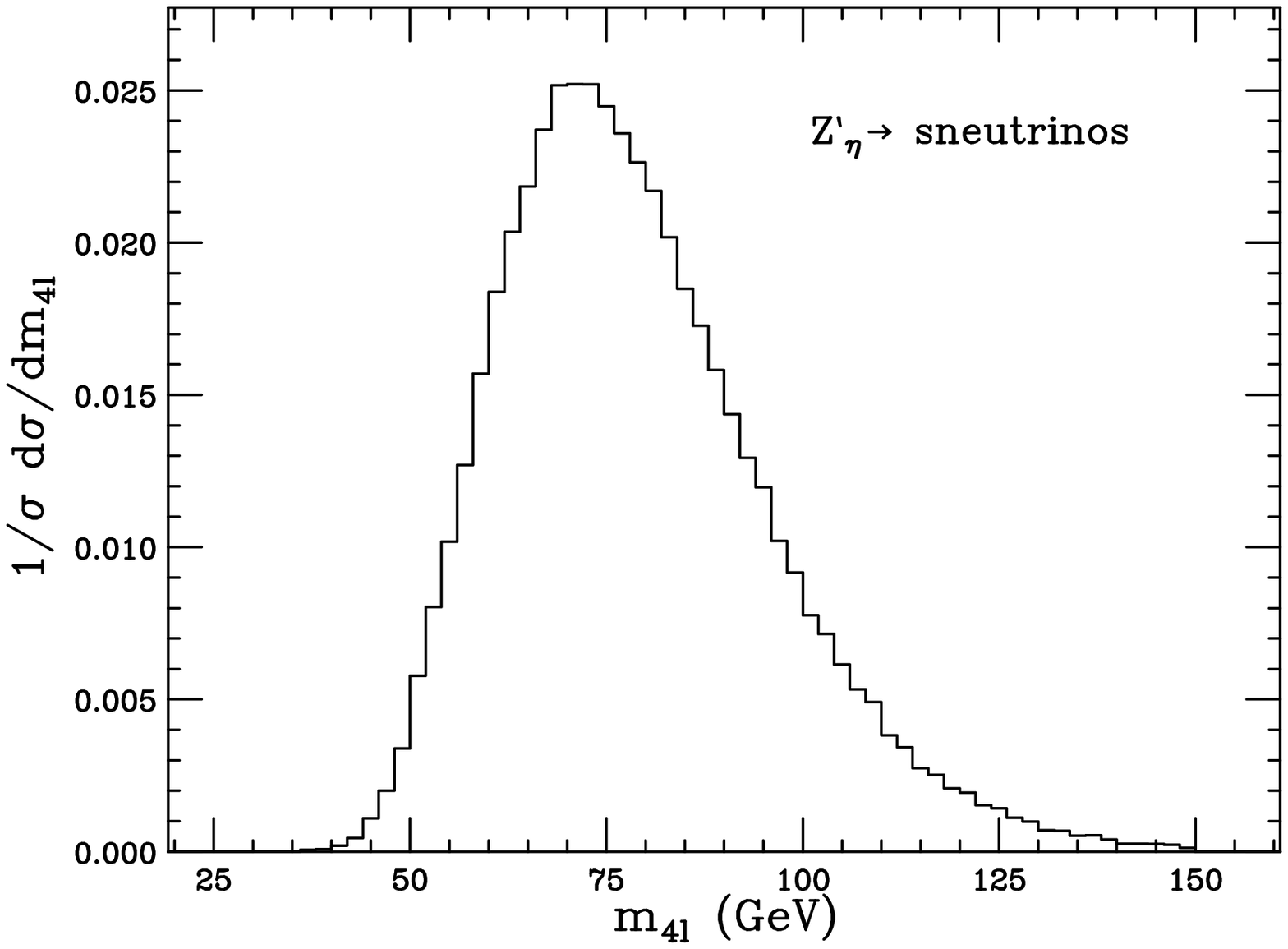}}%
\hfill%
\resizebox{0.49\textwidth}{!}{\includegraphics{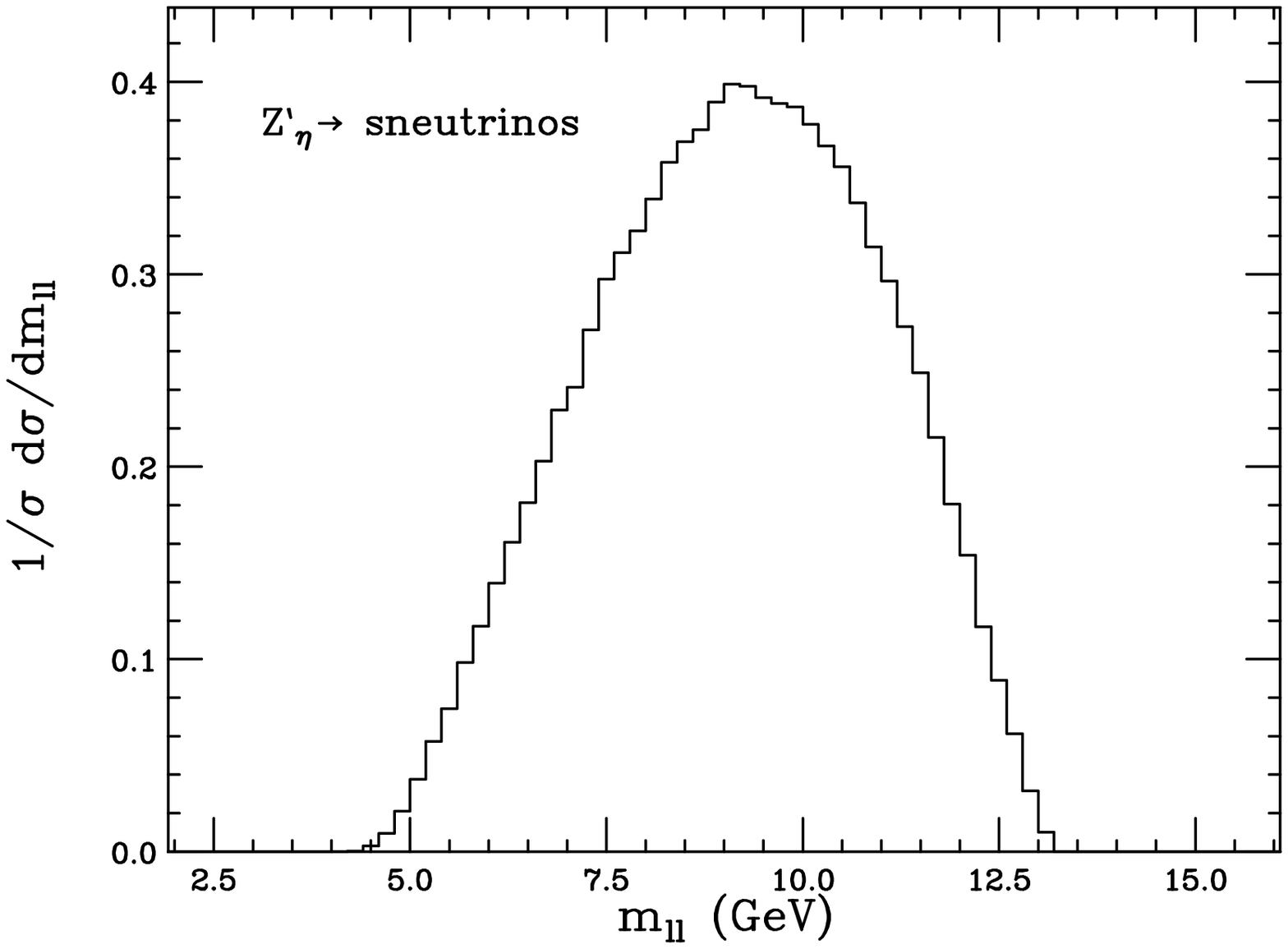}}}
\caption{Invariant mass of the four leptons in the process 
$pp\to 4\ell +~{\rm MET}$ (left) and of the $\ell^+\ell^-$ pairs coming
from each $\tilde\chi^0_2\to\tilde\chi^0_1\ell^+\ell^-$ decay in the cascade (\ref{zpsnu})
(right).}
\label{zetamll}
\end{figure}
\par In Fig.~\ref{zetapt} the lepton transverse momenta in the decay chain (\ref{zpsnu}) 
and in direct decays $Z'_\eta\to\ell^+\ell^-$ are
plotted: unlike the $Z'_\psi$ case, where we had final states with two 
charged leptons, exhibiting roughly the same kinematic properties, 
the decay chain (\ref{zpsnu}) presents four leptons, with different kinematics. 
Therefore, in Fig.~\ref{zetapt}, on the left-hand one has the spectra
in $p_T$ of the hardest (solid) and softest (dashes) lepton in the cascade  
(\ref{zpsnu}), on the right-hand side the lepton $p_T$ in 
$Z'_\eta\to\ell^+\ell^-$. 
In the cascade, the hardest lepton has a broad spectrum, relevant
in the 10 GeV$<p_T<50$~GeV range and maximum around $p_T\simeq$~20-25 GeV;
the $p_T$ of the softest $\ell^\pm$ is instead a narrow distribution, 
substantial only for 8 GeV$<p_T<20$~GeV and sharply peaked at $p_T\simeq 11$~GeV.
The spectrum in the direct production $Z'_\eta\to\ell^+\ell^-$ 
is roughly the same as in the $Z'_\psi$ case: in fact, 
using  
normalized distributions like $(1/\sigma)~d\sigma/dp_T$ minimizes
the impact of the value of the coupling. 

In Fig.~\ref{zetamll}, I have instead included two invariant-mass spectra:
$m_{4\ell}$, the invariant mass of the four charged leptons in (\ref{zpsnu}),
and $m_{\ell\ell}$, invariant mass  
of the $\ell^+\ell^-$ pairs in secondary $\chi^0_2\to\chi^0_1\ell^+\ell^-$
processes,
assuming that one is ideally able to identify
and reconstruct the leptons coming from each $\tilde\chi^0_2$ decay.
The $m_{\ell\ell}$ spectrum is significant only in the range
4 GeV~$<m_{\ell\ell}<$~13 GeV and peaked around $m_{\ell\ell}\simeq 9$~GeV;
$m_{4\ell}$ is relevant between
40 and 150 GeV and is maximum at $m_{4\ell}\simeq 70$~GeV.
\begin{figure}
\centerline{\resizebox{0.49\textwidth}{!}
{\includegraphics{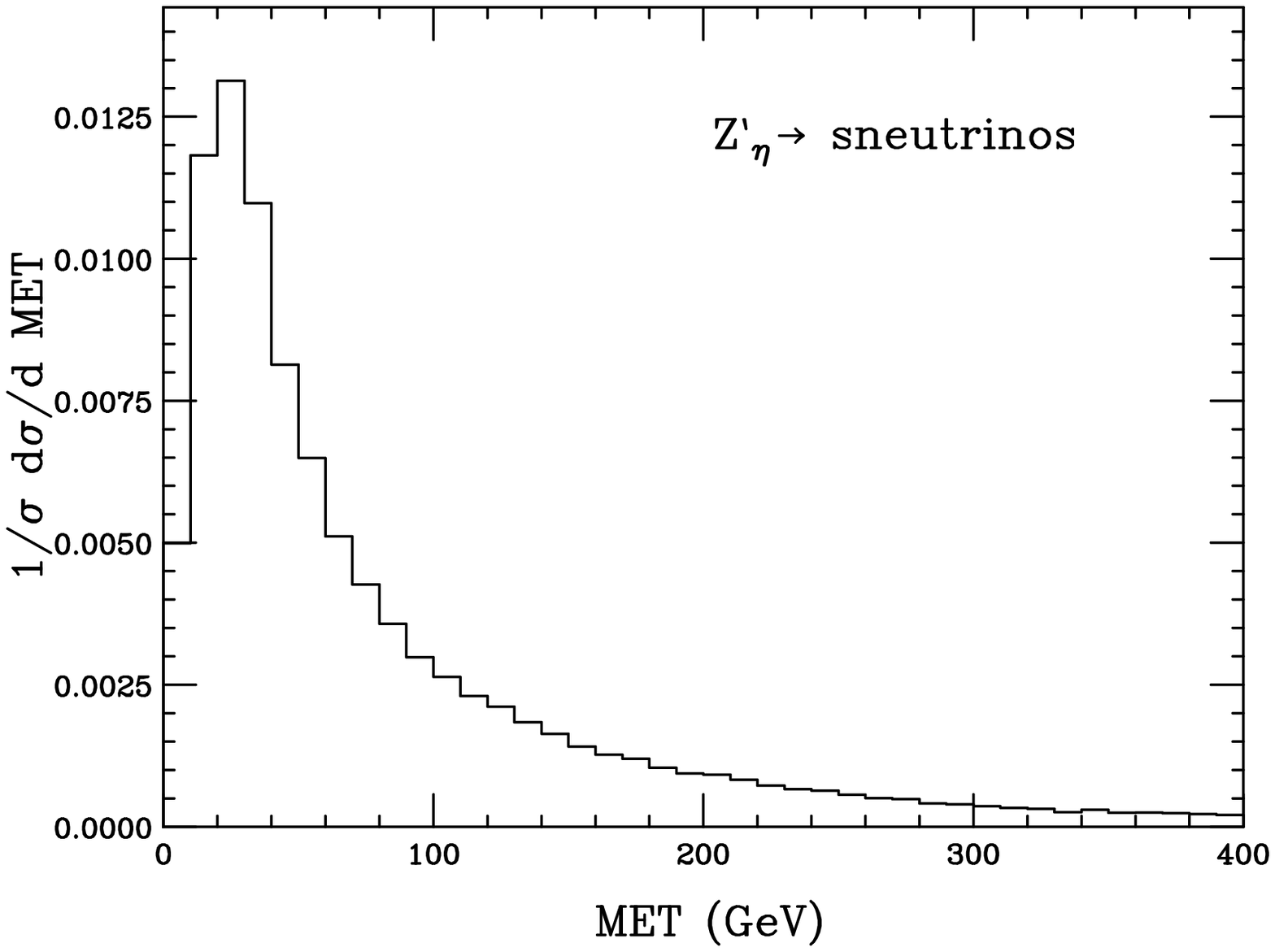}}%
\hfill%
\resizebox{0.49\textwidth}{!}{\includegraphics{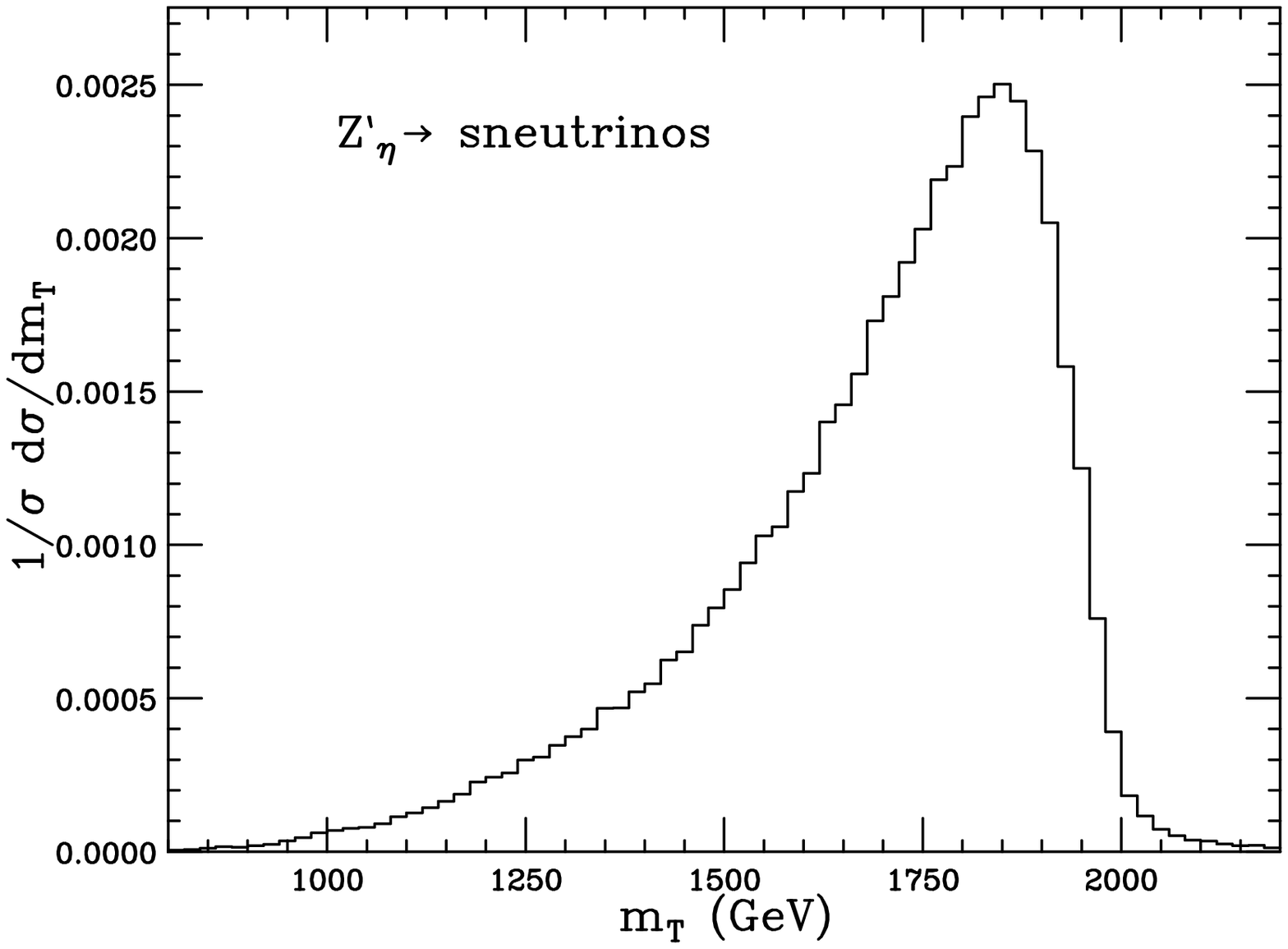}}}
\caption{Left: transverse mass for the final state in the process
described in Fig.~\ref{zsnu}.
Right: missing transverse energy due to neutrinos and neutralinos.}
\label{mtmetzeta}
\end{figure}\par
Finally, Fig.~\ref{mtmetzeta} presents the spectrum of the missing transverse energy
and transverse mass of the final states in the process in Fig.~\ref{zsnu},
defined as in Eq.~(\ref{metdef}).
The MET distribution is similar to the $Z'_\psi$ one, peaked at 20 GeV
and decreasing quite rapidly for larger MET values; the transverse mass
is relevant in the range $m_{Z'}/2<m_T<m_{Z'}$ and is overall a broader
and smoother distribution with respect to the previous model, with a peak still
around $m_T\simeq 1.8$~TeV.

\section{Phenomenology - Supersymmetric extension of the Sequential Standard Model}
In this section, I briefly discuss possible analyses within the
Sequential Standard Model (SSM):
this is the simplest extension of the Standard Model,
just containing $Z'$ and possibly $W'$ bosons, with the same couplings to fermions
as the Standard Model $Z$ and $W$. Although it does not have any strong
theoretical motivation,
as happens for GUT-inspired gauge symmetries, 
the SSM turns out to be very useful as a benchmark
model, since, once the coupling to quarks is fixed, the production 
cross section can be computed. 
In principle, in order to fairly extend the SSM to include supersymmetry,
one would also need to account for the $\tilde Z'$ and $\tilde W'$, 
the supersymmetric
partners of the $Z'$ and $W'$, for the extra Higgs fields and 
D-term corrections to the sfermion masses. 
However, following \cite{chang} and the
more recent study in \cite{corgen}, for the sake of releasing
some numbers which can be used as benchmarks in the experimental analyses, 
one can carry out the 
investigation in an effective theory, wherein the
$\tilde Z'$, the $\tilde W'$ and extra Higgs degrees of freedom
are too heavy to contribute to LHC phenomenology and
the $Z'$ couples to 
MSSM sfermions and gauginos like the Standard Model $Z$. 
In this scenario,
denoted by S-SSM, the coupling to $WW$ pairs must be suppressed,
otherwise the $WW$ scattering cross section, mediated by a $Z'$, would diverge. 
In the representative point of Ref.~\cite{corgen}, the $Z'_{\rm S-SSM}$ had
substantial branching fractions in supersymmetric channels, yielding
an overall contribution around 40\% to the total decay width.
\begin{table}[htp]
\caption{Squark masses in the S-SSM effective model, in GeV.}
\label{ssmq}
\begin{center}
\small
\begin{tabular}{|c|c|c|c|c|c|}
\hline
$m_{\tilde d_1}$ &   $m_{\tilde u_1}$ &  $m_{\tilde s_1}$ &  $m_{\tilde c_1}$ &  
$m_{\tilde b_1}$ & $m_{\tilde t_1}$\\ 
\hline
 5000.0 & 5000.0  & 5000.0 & 5000.0 & 1480.6 & 1486.8 \\
\hline
$m_{\tilde d_2}$ &   $m_{\tilde u_2}$ &  $m_{\tilde s_2}$ &  $m_{\tilde c_2}$ &  
$m_{\tilde b_2}$ & $m_{\tilde t_2}$\\ 
\hline
 5000.0 & 5000.0  & 5000.0  & 5000.0  & 1460.7 & 1390.2  \\
\hline\end{tabular}
\end{center}
\end{table}
\begin{table}[htp]
\caption{Slepton masses in the S-SSM reference point.
Numbers are in GeV; all three generations are slightly degenerate.}
\label{ssml}
\begin{center}
\small
\begin{tabular}{|c|c|c|c|}
\hline
$m_{\tilde \ell_1}$ &   $m_{\tilde \ell_2}$ &  $m_{\tilde \nu_{1,\ell}}$ &  $m_{\tilde \nu_{2,\ell}}$ 
\\ 
\hline
 502.0 & 502.0  & 495.0  & 495.0 \\ 
\hline\end{tabular}
\end{center}
\end{table}
\begin{table}[htp] 
\caption{Masses of the Higgs bosons in the $Z'_{\rm S-SSM}$ model, for a $Z'$ mass equal to
2 TeV.}
\label{ssmh}
\begin{center}
\small
\begin{tabular}{|c|c|c|c|c|}
\hline
$m_h$ &   $m_H$ &  $m_A$ & $m_{H^+}$\\ 
\hline
  125.8 &  638.7 & 632.8 & 637.8\\ 
\hline\end{tabular}
\end{center}
\end{table}
\begin{table}[htp]
\caption{Masses of charginos and neutralinos in the reference point
of the S-SSM.}
\label{ssmchi}
\begin{center}
\small
\begin{tabular}{|c|c|c|c|c|c|}
\hline
$m_{\tilde\chi_1^+}$ &   $m_{\tilde\chi_2^+}$ & $m_{\tilde\chi_1^0}$ &   $m_{\tilde\chi_2^0}$ 
& $m_{\tilde\chi_3^0}$ &   $m_{\tilde\chi_4^0}$  \\ 
\hline
198.6  & 835.8 &  193.5 & 197.7 & 413.6  & 836.0  \\ 
\hline\end{tabular}
\end{center}
\end{table}

In this effective model, I still set the $Z'_{\rm S-SSM}$ mass to
$m_{Z'}=2$~TeV and, in order to obtain
a light Higgs consistent with the recent LHC observations, 
choose a reference point yielding
squark, slepton, Higgs and gaugino masses as in 
Tables~\ref{ssmq}--\ref{ssmchi}.
The first three squark generations are degenerate,
while stops and sbottoms are considerably lighter. 
With the exception of $\tilde\chi^\pm_2$ and $\tilde\chi^0_4$,
the gaugino masses are of the order of a few hundred GeV and 
therefore they are light enough to be capable of contributing to the width
of a 2 TeV $Z'_{\rm S-SSM}$.
The branching ratios of the $Z'_{\rm S-SSM}$ into Standard Model and supersymmetric
channels are given in Table~\ref{brssm}:
the total rate into BSM final states is 27\%, with the highest
fraction being into charginos $\tilde\chi^+_1\tilde\chi^-_1$, about 17\%; 
decays into neutralinos and sneutrinos account for about 10\%, whereas 
SM modes are the remaining 73\%.
\begin{table}[htp]
\caption{$Z'_{\rm S-SSM}$ decay rates for $m_Z'=2$~TeV}
\label{brssm}
\begin{center}
\small
\begin{tabular}{|c|c|}
\hline
Final State & $Z'$ Branching ratio (\%) \\
\hline
$\tilde\chi_1^+\chi_1^-$ & 16.6 \\
\hline
$\tilde\chi_3^0\tilde\chi_4^0$ & 3.4 \\
\hline
$\sum_i\tilde\nu_i\tilde\nu^*_i$ & 4.0 \\
\hline
$\tilde\chi_2^+\tilde\chi_2^-$ & 2.5 \\
\hline
$hZ$ & 2.0\\
\hline
$\sum_i d_i\bar d_i$ & 27.1 \\
\hline
$\sum_i u_i\bar u_i$ & 20.7 \\
\hline
$\sum_i \nu_i\bar \nu_i$ & 12.2 \\
\hline
$\sum_i \ell^+_i\ell^-_i$ & 6.1 \\
\hline\end{tabular}
\end{center}
\end{table}
\begin{table} 
\caption{Chargino $\tilde\chi_1^+$ decay rates in the reference point for
the $Z'_{\rm S-SSM}$ model.}
\label{brchssm}
\begin{center}
\small
\begin{tabular}{|c|c|}
\hline
Final State & $\tilde\chi_1^+$ branching ratio (\%) \\
\hline
$\tilde\chi^0_1\  u\bar d$ & 38.9 \\
\hline
$\tilde\chi^0_1\  c\bar s$ & 28.9 \\
\hline
$\tilde\chi_1^0\  e^+\nu_e$ & 12.3 \\
\hline
$\tilde\chi_1^0\  \mu^+\nu_\mu$ & 12.1 \\
\hline
$\tilde\chi_1^0\  \tau^+\nu_\tau$ & 6.5 \\
\hline\end{tabular}
\end{center}
\end{table}
\par 
Within this effective S-SSM, $Z'$ decays
into chargino pairs, possibly leading to final states with leptons and 
missing energy, like those in Fig.~\ref{zch},
are worthwhile to be investigated.
The main chargino branching ratios are quoted in Table~\ref{brchssm}:
the Cabibbo-favored decays into $\tilde\chi^0_1u\bar d$ and 
$\tilde\chi^0_1c\bar s$ are largely
dominant, but even the decays into electrons and muons, i.e. $\tilde\chi^0_1\ e^+\nu_e$ 
and $\tilde\chi^0_1\ \mu^+\nu_\mu$ final states, 
are quite relevant, accounting for about 1/4
of the total $\tilde\chi^+_1$ rate. 
Even in the S-SSM, an interesting decay chain is (\ref{zpc1c1}),
with the obvious replacement of the $Z'_\psi$ with the $Z'_{\rm S-SSM}$,
leading to final states with two charged leptons and missing energy.
At $\sqrt{s}=14$~TeV, the inclusive cross section reads
$\sigma(pp\to Z'_{\rm S-SSM})\simeq 0.63$~pb and the one of the
chain (\ref{zpc1c1}) 
is $\sigma(pp\to Z'_{\rm S-SSM}\to\ell^+\ell^-+{\rm MET})
\simeq 6.18\times 10^{-3}$~pb, implying about 600 final states with $e^+e^-$ or
$\mu^+\mu^-$ and missing energy in the phase ${\cal L}=100$~fb$^{-1}$
and even few thousands at 300 fb$^{-1}$.
It is thus confirmed the finding of Ref.~\cite{corgen},
where it was observed that the S-SSM is the scenario which enhances both
production cross section and rates into supersymmetric final states.

In principle, even in the S-SSM one may study leptonic observables
like those investigated for the U(1)$'_\psi$ and U(1)$'_\eta$ models.
Nevertheless, it was found that, especially if one plots
normalized distributions like $(1/\sigma) d\sigma/dp_T$, the impact
of the coupling is mild and therefore the spectra 
are very similar to those obtained in Section 4
for the $Z'_\psi$.
I shall not present such observables in the S-SSM for the sake of brevity.

\section{Conclusions} 
I presented a phenomenological analysis of supersymmetric $Z'$ decays at the LHC,
for $\sqrt{s}=14$~TeV and a few models,
based on GUT-inspired U(1)$'$ symmetries.
The MSSM was suitably extended, in order to accommodate the new features due
to the U(1)$'$ group and the extra $Z'$ boson, and the reference points in
the parameter space were chosen in such a way to recover
a light Higgs with mass of 125 GeV and obtain substantial $Z'$ branching ratios
in the supersymmetric channels.   

The analysis was carried out for the
so-called U(1)$'_\psi$ and U(1)$'_\eta$ groups, since, even in previous
work on supersymmetric $Z'$ decays, they were the theoretical
scenarios enhancing the supersymmetric signal.
When fixing the soft squark masses, two options were considered, namely 
degenerate squarks
for all three generations as well as lighter stops and sbottoms with respect
to the first two generations. It was found that, in both U(1)$'$ models,
for the chosen parameters and $m_{Z'}=2$~TeV, supersymmetric modes account for about
25-30\% of the $Z'$ width, with the decays into chargino and sneutrino pairs
yielding the highest supersymmetric branching ratios for 
 the $Z'_\psi$ and $Z'_\eta$ models, respectively. The $Z'_\psi$ scenario 
had also a visible rate into the lightest neutralinos $\tilde\chi^0_1$, which could be
a useful channel to search for Dark Matter candidates.

In the $Z'_\psi$ case, it was then considered a decay chain, initiated
by a $Z'_\psi\to\tilde\chi^+_1\tilde\chi^-_1$ process, leading to a final state
with two charged leptons (electrons or muons) and missing transverse energy, due to the
production of neutrinos and neutralinos.
About ${\cal O}(100)$ of such events, for luminosities of 100 or 300~fb$^{-1}$,
are expected in the 14 TeV LHC run. 
The decay into neutralinos $(\tilde\chi^0_1\tilde\chi^0_1)$
yields an even larger number of events, about ${\cal O}(10^3)$ at
14 TeV, and, although it will have to compete with the
decay into neutrino pairs, it 
may deserve an appropriate analysis
when looking for Dark Matter particles at the LHC.
In the $Z'_\eta$ scenario, the $Z'_\eta\to\tilde\nu_2\tilde\nu^*_2$ process,
where $\tilde\nu_2$ is mostly a right-handed sneutrino, 
can give rise to a chain yielding four charged leptons and
missing energy. The expected rate of such events 
is lower than the $Z'_\psi$ scenario, but still a few dozens of
events are expected for $pp$ collisions at 14 TeV.
In both U(1)$'$ models, observables like lepton transverse momentum,
rapidity, opening angle and invariant mass, as well as 
missing transverse energy and transverse mass, 
are peculiar of supersymmetric
decays; the spectra are rather different from those
obtained in direct $Z'\to\ell^+\ell^-$ processes and, because
of the $Z'$-mass constraint, even from supersymmetric backgrounds,
such as direct chargino production.
For the sake of comparison, even the Sequential Standard Model
was extended to include supersymmetric particles, assuming
that the $\tilde Z'$ is outside the LHC reach. 
In the chosen point of the parameter space,
it is still the decay into charginos, leading to final states with
two charged leptons and missing transverse energy, the most promising
supersymmetric channel.
Several hundreds of events are in fact foreseen in the high-energy LHC
run and even ${\cal O}(10^3)$ for a luminosity of 300~fb$^{-1}$.

In summary, 
the expected rates and final-state observables make supersymmetric
$Z'$ decays a rather interesting investigation
to search for supersymmetry, once the $Z'$ mass were to be known.
For the time being, opening the supersymmetric decay channels up will
result in lowering the $Z'$ mass exclusion limits, since the expected
rates in Standard Model dilepton pairs decrease.
Therefore, although the presented analysis will be useful to
search for supersymmetry only after the possible discovery of the $Z'$,
it should be possibly 
taken into account when determining the $Z'$ mass exclusion limits.
Once the data on high-mass leptons
are available at $\sqrt{s}=14$~TeV, it will be very interesting comparing the data 
with the theory results on the product $\sigma(pp\to Z')\times {\rm BR}(Z'\to\ell^+\ell^-)$, 
as done in \cite{cor1} for the analysis at 8 TeV,
and determine the exclusion limits accounting for
supersymmetric decays. 
However, a complete analysis should necessarily compare possible supersymmetric
signals in $Z'$ decays with the backgrounds coming from the 
SM and other supersymmetric processes, 
as well as from non-supersymmetric $Z'$ decays, and include the detector simulation. 
The systematic computation of the backgrounds and the implementation 
of detector effects is presently in progress.

Another possible improvement of the analysis here presented
consists of relaxing the approximation of neglecting the interference
between $Z$ and $Z'$ bosons. In fact, Ref.~\cite{klasen}
compared a NLO + NLL resummed calculation, accounting for such an interference,
with the standard analyses, which employ
the PYTHIA \cite{pythia} event generator and rescale the
cross section to account for higher-order corrections
\cite{fewz}.
The finding of \cite{klasen} is that, after including resummed
as well as interference effects,
the $Z'$ mass exclusion bound may well vary by a few hundred GeV
in both U(1)$'$ and SSM.
It will be therefore worthwhile carrying out the
study on supersymmetric $Z'$ decays along the lines of \cite{klasen},
especially once the first high-energy LHC data are available.

Furthermore, beyond the models here studied, which are among
those accounted for in the experimental analyses,
it may be worthwhile studying in the near future other scenarios,
such as the leptophobic models (see, e.g., the pioneering
work in \cite{zwirner} or late studies in \cite{china}), wherein
the $Z'$ does couple to quarks, thus allowing production
via $q\bar q\to Z'$, but the coupling to leptons is suppressed.
Within supersymmetry, the very fact that the $Z'$ is
leptophobic necessarily decreases the SM rate and enhances the
branching ratio in supersymmetric particles.
Besides, since $Z'$ decays into charginos and neutralinos played a major role
in the present work, a possible application of this work 
can also be achieved in the context of split supersymmetry
\cite{split}, wherein the scalar particles 
are much heavier than the
gauginos, which are therefore the only supersymmetric
particles accessible at present colliders.
Investigations of leptophobic $Z'$ models as well as of
$Z'$ bosons in the framework of split supersymmetry
are in progress as well.

\section*{Acknowledgements}
I am indebted to Simonetta Gentile, coauthor of Ref.~\cite{corgen},
who contributed in the early stages of this work.
I am especially grateful to Florian Staub and Benjamin Fuks for their
unvaluable help in using the SARAH/SPheno and FeynRules/MadGraph
codes, respectively,
and to Tony Gherghetta for discussions on the results of 
Ref.~\cite{gherghetta}.
I also acknowledge conversations with
Nazila Mahmoudi, Andrea Romanino, 
Barbara Clerbaux, Hwidong Yoo and Marianna Testa 
on these and related topics.

\end{document}

%% file: FEYNMAN.tex
%
%
%
%
%
%
\message{FEYNMAN:  For generating Feynman Diagrams in LaTex}
\message{Mark 1.0 Last Altered by MJSL 2/89}
\setlength{\unitlength}{0.01pt}
\gdef\Feynmanlength{\setlength{\unitlength}{0.01pt}}  
\gdef\unlock{\catcode`\@=11}
\gdef\lock{\catcode`\@=12}
\global\newcount\LINETYPE
\global\newcount\LINEDIRECTION
\global\newcount\LINECONFIGURATION
\newcommand{\LTYPE}{\LINETYPE}
\newcommand{\LDIR}{\LINEDIRECTION}
\newcommand{\LCONFIG}{\LINECONFIGURATION}
\global\LINETYPE=1  \global\LINEDIRECTION=0  \global\LINECONFIGURATION=0
\global\newcount\fermion    \fermion=1
\global\newcount\scalar     \scalar=2
\global\newcount\photon     \photon=3
\global\newcount\gluon      \gluon=4
\global\newcount\SPECIAL    \SPECIAL=5
\gdef\N{0}  \gdef\NE{1}  \gdef\E{2}   \gdef\SE{3}
\gdef\S{4}  \gdef\SW{5}  \gdef\W{6}   \gdef\NW{7}
\global\newcount\REG            \global\REG=0
\global\newcount\FLIPPED        \global\FLIPPED=1
\global\newcount\CURLY          \global\CURLY=2
\global\newcount\FLIPPEDCURLY   \global\FLIPPEDCURLY=3
\global\newcount\FLAT           \global\FLAT=4
\global\newcount\FLIPPEDFLAT    \global\FLIPPEDFLAT=5
\global\newcount\CENTRAL        \global\CENTRAL=6
\global\newcount\FLIPPEDCENTRAL \global\FLIPPEDCENTRAL=7
\gdef\LONGPHOTON{6}             \gdef\FLIPPEDLONG{7}
\global\newcount\SQUASHEDGLUON  \global\SQUASHEDGLUON=8
\gdef\SQUASHED{\SQUASHEDGLUON}
%
\newcount\adjx \adjx=0
\newcount\adjy \adjy=0
\global\newdimen\BIGPHOTONS     \BIGPHOTONS=0pt  
\gdef\bigphotons{\global\BIGPHOTONS=12pt}
\global\newdimen\THICKPHOTONS     \THICKPHOTONS=0pt  
\global\newdimen\THICKPHOTONSWITCH    \THICKPHOTONSWITCH=0pt
\gdef\THICKPHOTONTEST{
\THICKPHOTONSWITCH=0pt
\ifdim\THICKPHOTONS=0pt \relax
  \else \ifnum\LTYPE=3
           \ifnum\LDIR=2 \THICKPHOTONSWITCH=1pt \fi 
           \ifnum\LDIR=6 \THICKPHOTONSWITCH=1pt \fi 
        \fi
\fi
}  
\gdef\THICKLINES{\thicklines  \THICKPHOTONS=1pt}
\gdef\THINLINES{\thinlines  \THICKPHOTONS=0pt}
\global\newcount\phantomswitch   \global\phantomswitch=0
\global\newcount\stemlength   \global\stemlength=275   
\global\newcount\absstemlength        
\global\newcount\stemlengthx          
\global\newcount\stemlengthy          
\newdimen\FRONTSTEM  \FRONTSTEM=0pt   
\newdimen\BACKSTEM   \BACKSTEM=0pt    
\newdimen\EITHERSTEM \EITHERSTEM=0pt  
\gdef\frontstemmed{\FRONTSTEM=1pt}            
\gdef\backstemmed{\BACKSTEM=1pt}              
\gdef\stemmed{\FRONTSTEM=1pt  \BACKSTEM=1pt}    
\global\newcount\arrowlength                
\global\newdimen\ATTIP   \global\ATTIP=0pt  
\global\newdimen\ATBASE  \global\ATBASE=1pt 
\global\newcount\unitboxnumber  
\global\newcount\unitboxnumberpo  
\global\newcount\particlelengthx  
\gdef\plengthx{\particlelengthx}
\global\newcount\particlelengthy  
\gdef\plengthy{\particlelengthy}
\global\newcount\boxlengthx  
\global\newcount\boxlengthy  
\global\newcount\particleadjustx  
\global\newcount\particleadjusty  
\global\newcount\particlelength   
\global\newcount\particlefrontx
\gdef\pfrontx{\particlefrontx}
\global\newcount\PFRONTx
\global\newcount\particlefronty
\gdef\pfronty{\particlefronty}
\global\newcount\PFRONTy
\global\newcount\particlebackx
\gdef\pbackx{\particlebackx}
\global\newcount\particlebacky
\gdef\pbacky{\particlebacky}
\global\newcount\particlemidx
\gdef\pmidx{\particlemidx}
\global\newcount\particlemidy
\gdef\pmidy{\particlemidy}
\global\newcount\seglength  \global\newcount\gaplength
\global\gaplength=850  
\global\seglength=1416  
\global\newcount\Xone    \global\newcount\Yone    
\global\newcount\Xtwo    \global\newcount\Ytwo    
\global\newcount\Xthree  \global\newcount\Ythree  
\global\newcount\Xfour   \global\newcount\Yfour   
\global\newcount\Xfive   \global\newcount\Yfive   
\global\newcount\Xsix    \global\newcount\Ysix    
\global\newcount\Xseven  \global\newcount\Yseven  
\global\newcount\Xeight  \global\newcount\Yeight  
%
%
\newsavebox{\lastline}  
\global\newcount\numlineparts   
\global\newcount\upperlineadjx  \upperlineadjx=0  
\global\newcount\upperlineadjy  \upperlineadjy=0  
\global\newcount\lowerlineadjx  \lowerlineadjx=0  
\global\newcount\lowerlineadjy  \lowerlineadjy=0  
\global\newcount\thirdlineadjx  \thirdlineadjx=0  
\global\newcount\thirdlineadjy  \thirdlineadjy=0  
\global\newcount\fourthlineadjx \fourthlineadjx=0  
\global\newcount\fourthlineadjy \fourthlineadjy=0  
\global\newcount\unitboxwidth   \unitboxwidth=1000
\global\newcount\unitboxheight  \unitboxheight=0  
\global\newcount\numupperunits  \numupperunits=8  
\global\newcount\numlowerunits  \numlowerunits=8  
\global\newcount\numthirdunits  \numthirdunits=8  
\global\newcount\numfourthunits \numfourthunits=8  
\global\newcount\fermioncount   \global\fermioncount=0
\global\newcount\scalarcount    \global\scalarcount=0
\global\newcount\photoncount    \global\photoncount=0
\global\newcount\gluoncount     \global\gluoncount=0
\global\newcount\SPECIALcount   \global\SPECIALcount=0
\global\newcount\vertexcount    \global\vertexcount=-1
%
\global\newcount\XDIR
\global\newcount\YDIR
\gdef\SETDIR{  
\ifcase\LDIR
     \global\XDIR=0  \global\YDIR=1   
\or  \global\XDIR=1  \global\YDIR=1   
\or  \global\XDIR=1  \global\YDIR=0   
\or  \global\XDIR=1  \global\YDIR=-1  
\or  \global\XDIR=0  \global\YDIR=-1  
\or  \global\XDIR=-1 \global\YDIR=-1  
\or  \global\XDIR=-1 \global\YDIR=0   
\or  \global\XDIR=-1 \global\YDIR=1   
\else\DIRECTERROR
\fi}  
\gdef\moduloeight#1{
\ifnum#1>7 \global\advance #1 by -8
\relax
\moduloeight#1
\relax
\else \relax
\fi}
\gdef\multroothalf#1{\global\multiply #1 by 7071 \global\divide #1 by 10000}
\gdef\negate#1{\global\multiply #1 by -1}
\gdef\double#1{\global\multiply #1 by 2}
\gdef\slanttest(#1,#2){
\ifodd\LDIR
\multiply #1 by 7071  \divide #1 by 10000
\multiply #2 by 7071  \divide #2 by 10000
\fi
}
\gdef\gslanttest(#1,#2){
\ifodd\LDIR
\multroothalf#1
\multroothalf#2
\fi
}
%
%
\gdef\setplength{ 
\global\particlelengthx=\unitboxwidth
\global\particlelengthy=\unitboxheight
\global\multiply \particlelengthx by \unitboxnumber
\global\multiply \particlelengthy by \unitboxnumber
\global\advance \particlelengthx by \particleadjustx
\global\advance \particlelengthy by \particleadjusty
}
\gdef\boxlengthdefault{  
\global\boxlengthx=\plengthx
\global\boxlengthy=\plengthy
\ifnum\plengthx<0 \global\multiply\boxlengthx by -1 \fi
\ifnum\plengthy<0 \global\multiply\boxlengthy by -1 \fi
}
\gdef\rearcoords{  
\global\particlebacky=\particlefronty
\global\particlebackx=\particlefrontx
\global\advance \particlebackx by \particlelengthx
\global\advance \particlebacky by \particlelengthy
}
\gdef\midcoords{  
\global\particlemidy=\particlefronty
\global\particlemidx=\particlefrontx
\global\stemlengthx=\particlelengthx  
\global\stemlengthy=\particlelengthy
\global\divide\stemlengthx by 2
\global\divide\stemlengthy by 2
\global\advance \particlemidx by \stemlengthx
\global\advance \particlemidy by \stemlengthy
}
\gdef\setparticle{\setplength\rearcoords\midcoords\boxlengthdefault}  
%
\gdef\setcoords(#1,#2,#3)(#4,#5,#6)[#7,#8]{
\global\upperlineadjx=#1
\global\lowerlineadjx=#2
\global\thirdlineadjx=#3
\global\upperlineadjy=#4
\global\lowerlineadjy=#5
\global\thirdlineadjy=#6
\global\unitboxwidth=#7
\global\unitboxheight=#8
}
%
%
%
\gdef\drawoldpic#1(#2,#3){  
\global\particlefrontx=#2
\global\particlefronty=#3
\rearcoords
\midcoords
\put(#2,#3){\usebox{#1}}
}
\gdef\drawsavedline`#1' as #2[#3#4](#5,#6)[#7]{
\global\LINETYPE=#2
\global\LINEDIRECTION=#3
\global\LINECONFIGURATION=#4
\global\particlefrontx=#5
\global\particlefronty=#6
\global\unitboxnumber=#7
\selectcase
\rearcoords
\midcoords
\ifnum\phantomswitch=0 \drawas{#1}\fi
}

\gdef\startphantom{\phantomswitch=1} 
\gdef\stopphantom{\phantomswitch=0}  

\gdef\drawas#1{
\global\savebox{#1}(\boxlengthx,\boxlengthy){
\setlength{\unitlength}{0.01pt}
\begin{picture}(\boxlengthx,\boxlengthy)
\multiput(\upperlineadjx,\upperlineadjy)(\unitboxwidth,\unitboxheight)
{\numupperunits}{\upperunitbox}
\ifnum\numlineparts > 1  
\multiput(\lowerlineadjx,\lowerlineadjy)(\unitboxwidth,\unitboxheight)
{\numlowerunits}{\lowerunitbox}
\fi
\ifnum\numlineparts > 2  
\multiput(\thirdlineadjx,\thirdlineadjy)(\unitboxwidth,\unitboxheight)
{\numthirdunits}{\thirdunitbox}
\fi
\ifnum\numlineparts > 3  
\multiput(\fourthlineadjx,\fourthlineadjy)(\unitboxwidth,\unitboxheight)
{\numfourthunits}{\lowerunitbox}
\fi
\end{picture} }
\global\PFRONTx=\pfrontx  \global\PFRONTy=\pfronty   
\SETFRONTSTEM
\THICKPHOTONTEST
\ifdim\THICKPHOTONSWITCH=1pt\global\advance\PFRONTy by 20  \fi
\put(\PFRONTx,\PFRONTy) {\usebox{#1}}   
\ifdim\THICKPHOTONSWITCH=1pt
\global\advance\PFRONTy by -40
\put(\PFRONTx,\PFRONTy) {\usebox{#1}}   
\global\advance \PFRONTy by 20  
\fi  
\SETBACKSTEM
\seglength=1416   \gaplength=850   
}
%
%

\gdef\drawandsaveline`#1' as #2[#3#4](#5,#6)[#7]{
\global\newsavebox{#1}
\drawsavedline`#1' as #2[#3#4](#5,#6)[#7]
}

\gdef\drawline#1[#2#3](#4,#5)[#6]{   
\drawsavedline`\lastline' as #1[#2#3](#4,#5)[#6]}

\gdef\saveas#1{  
\global\newsavebox#1
\drawas#1}
%
%
%
\gdef\TYPEERROR{\message{*** ERROR IN PARTICLE TYPE SELECTION ***}
\message{+++ Try with line type \fermion,\scalar,\photon,\gluon
(see manual) +++}\SETERR}
\gdef\DIRECTERROR{\SETERR\message{*** ERROR IN PARTICLE DIRECTION SELECTION
***}
\message{+++ Try again with direction N, NE, E, SE  etc. or see manual +++}}
\gdef\UNIMPERROR{\message{*** ERROR IN PARTICLE OPTIONS SELECTION ***}
\message{
+++ The requested options combination has not yet been implemented +++}\SETERR}
\gdef\SETERR{\gdef\upperunitbox{{\tiny Error}}  
\gdef\lowerunitbox{\relax}
\gdef\thirdunitbox{\relax}
}
\gdef\neglengthcheck{\ifnum\unitboxnumber < 1
\message{   *** ERROR:  PARTICLE OF NEGATIVE OR ZERO LENGTH REQUESTED. ***   }
\message{   ***         TAKING ABSOLUTE VALUE. ***   }\negate\unitboxnumber
\fi}
\gdef\selectcase{
\neglengthcheck   
\SETDIR
\ifcase\LINETYPE
\TYPEERROR  
\or \selectfermion  
\or \selectscalar   
\or \selectphoton   
\or \selectgluon    
\or \selectspecial  
\else \TYPEERROR \fi  }
\gdef\selectfermion{
\ifnum\fermioncount=0 \input FERMIONSETUP \fi
\global\advance\fermioncount by 1  
\ALLfermion
}
\gdef\selectscalar{
\ifnum\scalarcount=0 \input SCALARSETUP \fi
\global\advance\scalarcount by 1  
\ALLscalar
}
\gdef\selectphoton{   
\ifnum\photoncount=0 \input PHOTONSETUP  \fi
\selectphoton
}
\gdef\selectgluon{   
\ifnum\gluoncount=0 \input GLUONSETUP  \fi
\selectgluon
}
\gdef\selectspecial{\UNIMPERROR}
%
%
\gdef\checkvertex{ 
\ifnum\vertexcount=-1   \input VERTEX  \fi}
\gdef\drawvertex#1[#2#3](#4,#5)[#6]{\checkvertex\drawvertex#1[#2#3](#4,#5)[#6]}
\gdef\vertexcap#1{\checkvertex\vertexcap#1}
\gdef\vertexcaps{\checkvertex\vertexcaps}
\gdef\vertexlink#1{\checkvertex\vertexlink#1}
\gdef\vertexlinks{\checkvertex\vertexlinks}
\gdef\stemvertex#1{\checkvertex\stemvertex#1}
\gdef\stemvertices{\checkvertex\stemvertices}
\gdef\flipvertex{\checkvertex\flipvertex}
%
%
\global\arrowlength=349  
\gdef\drawarrow[#1#2](#3,#4){
\global\LDIR=#1
\SETDIR
\global\boxlengthx=#3  
\global\boxlengthy=#4  
\ifdim#2=1pt  
\adjx=\arrowlength      \adjy=\arrowlength
\multiply\adjx by \XDIR \multiply\adjy by \YDIR  
\slanttest(\adjx,\adjy)
\global\advance\boxlengthx by \adjx    \global\advance\boxlengthy by \adjy
\fi
\ifnum\phantomswitch=0\put(\boxlengthx,\boxlengthy){\vector(\XDIR,\YDIR){0}}\fi
}  
%
%
\gdef\SETFRONTSTEM{
\EITHERSTEM=\FRONTSTEM   \advance\EITHERSTEM by \BACKSTEM
\ifdim\EITHERSTEM>0pt
\global\stemlengthx=\stemlength   \global\stemlengthy=\stemlength
\global\absstemlength=\stemlength
\SETDIR
\gslanttest(\stemlengthx,\stemlengthy)
\gslanttest(\absstemlength,\REG)  
\ifnum\XDIR=0 \stemlengthx=0 \fi
\ifnum\YDIR=0 \stemlengthy=0 \fi
\global\multiply\stemlengthx by \XDIR
\global\multiply\stemlengthy by \YDIR
\ifdim\FRONTSTEM=1pt
\ifnum\phantomswitch=0
          \put(\pfrontx,\pfronty){\line(\XDIR,\YDIR){\absstemlength}}\fi
\global\advance\plengthx by \stemlengthx
\global\advance\plengthy by \stemlengthy
\global\advance\PFRONTx by \stemlengthx
\global\advance\PFRONTy by \stemlengthy
\global\advance\pmidx by \stemlengthx
\global\advance\pmidy by \stemlengthy
\global\advance\pbackx by \stemlengthx
\global\advance\pbacky by \stemlengthy
\ifnum\LTYPE=3
\global\photonfrontx=\PFRONTx  \global\photonfronty=\PFRONTy
\global\photonbackx=\pbackx    \global\photonbacky=\pbacky
\fi  
\ifnum\LTYPE=4
\global\gluonfrontx=\PFRONTx  \global\gluonfronty=\PFRONTy
\global\gluonbackx=\pbackx    \global\gluonbacky=\pbacky
\fi  
\fi  
\fi  
}    
\gdef\SETBACKSTEM{
\ifdim\BACKSTEM=1pt
\ifnum\phantomswitch=0
       \put(\pbackx,\pbacky){\line(\XDIR,\YDIR){\absstemlength}}\fi
\global\advance\plengthx by \stemlengthx
\global\advance\plengthy by \stemlengthy
\global\advance\pbackx by \stemlengthx
\global\advance\pbacky by \stemlengthy
\fi  
\global\stemlength=275  \FRONTSTEM=0pt  \BACKSTEM=0pt 
}    
\gdef\drawloop#1[#2#3](#4,#5){  
\input LOOPS  
\drawloop#1[#2#3](#4,#5)}
\Feynmanlength  

%% file: FERMIONSETUP.tex
\global\newcount\fermionlength  
\global\newcount\fermionlengthx
\global\newcount\fermionlengthy
\global\newcount\fermionfrontx  
\global\newcount\fermionfronty  
\global\newcount\fermionbackx
\global\newcount\fermionbacky
\gdef\ALLfermion{  
\global\fermionfrontx=\particlefrontx \global\fermionfronty=\particlefronty
\ifnum\unitboxnumber > 50000
\message{   *** WARNING *** Fermion of length
\the\unitboxnumber\space requested ***   }
\ifnum\unitboxnumber > 80000
\message{   *** Reducing fermion length to 30000 (max 80000) ***   }
\global\unitboxnumber=30000 \fi \fi  
\global\fermionlength=\unitboxnumber 
\global\particleadjustx=0   \global\particleadjusty=0 
\global\numlineparts = 1    \global\numupperunits=1
\global\upperlineadjx=-200  \global\upperlineadjy=0
\global\fermionlengthx=\fermionlength    \global\fermionlengthy=\fermionlength
\gslanttest(\fermionlengthx,\fermionlengthy)  
\global\multiply\fermionlengthx by \XDIR  
\global\multiply\fermionlengthy by \YDIR  
\global\unitboxheight=\fermionlengthy   \global\unitboxwidth=\fermionlengthx
\global\advance \fermionlengthx by \particleadjustx
\global\advance \fermionlengthy by \particleadjusty
\global\particlelengthx=\fermionlengthx
\global\particlelengthy=\fermionlengthy
\boxlengthdefault    \rearcoords    \midcoords
\global\fermionbackx=\particlebackx     \global\fermionbacky=\particlebacky
\ifcase\LINECONFIGURATION  
\ifnum\XDIR=0
\gdef\upperunitbox{\line(\XDIR,\YDIR){\boxlengthy}} 
\else
\gdef\upperunitbox{\line(\XDIR,\YDIR){\boxlengthx}}
\fi
\else \UNIMPERROR
\fi
}

%% file: SCALARSETUP.tex
\newcount\scalarlength
\newcount\scalarlengthx
\newcount\scalarlengthy
\newcount\scalarfrontx  
\newcount\scalarfronty  
\newcount\scalarbackx
\newcount\scalarbacky
\gdef\ALLscalar{
\global\scalarfrontx=\particlefrontx   
\global\scalarfronty=\particlefronty   
\numlineparts = 1      \numupperunits=\unitboxnumber
\ifcase\LINECONFIGURATION
\global\upperlineadjx=-200     \global\upperlineadjy=0
\slanttest(\seglength,\gaplength)   
\gdef\upperunitbox{\line(\XDIR,\YDIR){\seglength}}
\else \UNIMPERROR 
\fi
\global\unitboxwidth=\seglength  \global\advance\unitboxwidth by \gaplength
\global\multiply \unitboxwidth by \XDIR
\global\unitboxheight=\seglength  \global\advance\unitboxheight by \gaplength
\global\multiply \unitboxheight by \YDIR
\global\particleadjustx=\gaplength \global\multiply\particleadjustx by \XDIR
\global\particleadjusty=\gaplength \global\multiply\particleadjusty by \YDIR
\negate\particleadjustx   \negate\particleadjusty   
\setparticle  
\global\scalarlengthx=\particlelengthx  
\global\scalarlengthy=\particlelengthy  
\ifnum\boxlengthx > 50000
\message{   *** WARNING *** Scalar of length in excess of 50000cp
requested!}\fi
\ifnum\boxlengthy > 50000
\message{   *** WARNING *** Scalar of length in excess of 50000cp
requested!}\fi
\global\scalarbackx=\pbackx      \global\scalarbacky=\pbacky   
}

%% file: PHOTONSETUP.tex
\newcount\numwiggles    \newcount\numwigglespo
\global\newcount\photonlengthx
\global\newcount\photonlengthy
\global\newcount\photonfrontx  
\global\newcount\photonfronty  
\global\newcount\photonbackx
\global\newcount\photonbacky
\newcount\halfwigglelength
\global\font\Twelverom=cmr12
\global\font\Tenrom=cmr10
\gdef\Lbr{{\Twelverom(}}   \gdef\Rbr{{\Twelverom)}}
\gdef\SLbr{{\Tenrom(}}     \gdef\SRbr{{\Tenrom)}}
\gdef\Smile{{\large$\smile$}}  
\gdef\Frown{{\large$\frown$}}  
\ifdim\BIGPHOTONS>0pt  \gdef\Smile{$\smile$} \gdef\Frown{$\frown$} \fi
%
\gdef\selectphoton{   
\global\advance\photoncount by 1  
\global\photonfrontx=\particlefrontx   
\global\photonfronty=\particlefronty   
\ifnum\unitboxnumber > 50
\message{   *** WARNING *** Photon with
\the\unitboxnumber\space half-wiggles requested ***   }
\ifnum\unitboxnumber > 150
\message{   *** Reducing photon length to 10 half-wiggles (max 150) ***   }
\ifnum\unitboxnumber > 1000
\message{   *** Probable Cause:  Photon selected instead of Fermion ***   }
\fi \global\unitboxnumber=10 \fi \fi  
\numwiggles=\unitboxnumber
\divide\numwiggles by 2
\global\unitboxnumberpo=\numwiggles 
\global\multiply \unitboxnumberpo by -1
\numwigglespo=\unitboxnumber
\advance\numwigglespo by \unitboxnumberpo 
\global\numlineparts = 2  
\global\numupperunits=\numwigglespo  
\global\numlowerunits=\numwiggles  
\particleadjustx=0  
\particleadjusty=0  
\ifcase\LINEDIRECTION
     \Nphoton    
\or  \NEphoton   
\or  \Ephoton    
\or  \SEphoton   
\or  \Sphoton    
\or  \SWphoton   
\or  \Wphoton    
\or  \NWphoton   
\else\DIRECTERROR \fi
\setplength
\global\divide\plengthx by 2  \global\divide\plengthy by 2
\rearcoords  \boxlengthdefault   \midcoords
\global\photonbackx=\pbackx  
\global\photonbacky=\pbacky  
\global\photonlengthx=\plengthx  
\global\photonlengthy=\plengthy  
}
\gdef\SETUNITBOX(#1)[#2][#3]{ 
\gdef\upperunitbox{\oval(#1,#1)[#2]}
\gdef\lowerunitbox{\oval(#1,#1)[#3]}
}
\gdef\Nphoton{  
\ifcase\LINECONFIGURATION  
\setcoords(-490,-250,0)(260,1250,0)[0,2000]
\gdef\upperunitbox{\SLbr}   \gdef\lowerunitbox{\SRbr}
\particleadjusty=10
\or 
\setcoords(-271,-501,0)(250,1250,0)[0,2000]
\gdef\upperunitbox{\SRbr}   \gdef\lowerunitbox{\SLbr}
\or 
\particleadjusty=0
\setcoords(-501,-351,0)(300,1400,0)[0,2200]
\gdef\upperunitbox{\Lbr}   \gdef\lowerunitbox{\Rbr}
\or 
\setcoords(-353,-499,0)(300,1400,0)[0,2200]
\gdef\upperunitbox{\Rbr}   \gdef\lowerunitbox{\Lbr}
\or 
\setcoords(-481,-371,0)(280,1300,0)[0,2000]
\gdef\upperunitbox{\Lbr}   \gdef\lowerunitbox{\Rbr}
\particleadjusty=150
\ifnum\numwiggles=\number\numwigglespo \particleadjustx=-50 \fi
\or 
\setcoords(-321,-391,0)(280,1300,0)[0,2000]
\gdef\upperunitbox{\Rbr}   \gdef\lowerunitbox{\Lbr}
\particleadjusty=150
\ifnum\numwiggles=\number\numwigglespo \particleadjustx=80 \fi
\or 
\setcoords(-490,-260,0)(300,1500,0)[0,2400]
\gdef\upperunitbox{\Lbr}   \gdef\lowerunitbox{\Rbr}
\or 
\setcoords(-301,-531,0)(300,1500,0)[0,2400]
\gdef\upperunitbox{\Rbr}   \gdef\lowerunitbox{\Lbr}
\else \UNIMPERROR
\fi
}
\gdef\NEphoton{    
\ifcase\LINECONFIGURATION  
\setcoords(425,425,0)(1250,0,0)[1250,1250]       \SETUNITBOX(1250)[br][tl]
\ifnum\numwigglespo > \number \numwiggles \particleadjustx=15 \fi
\or 
\setcoords(1050,-200,0)(625,625,0)[1250,1250]    \SETUNITBOX(1250)[tl][br]
\ifnum\numwigglespo > \number \numwiggles \particleadjustx=25 \fi
\or 
\setcoords(500,500,0)(1400,0,0)[1400,1400]       \SETUNITBOX(1400)[br][tl]
\or 
\setcoords(1200,-200,0)(700,700,0)[1400,1400]    \SETUNITBOX(1400)[tl][br]
\or 
\setcoords(400,400,0)(1200,0,0)[1200,1200]       \SETUNITBOX(1200)[br][tl]
\or 
\setcoords(1000,-200,0)(600,600,0)[1200,1200]    \SETUNITBOX(1200)[tl][br]
\else \UNIMPERROR
\fi
\numupperunits=\numwiggles   \numlowerunits=\numwigglespo
}
\gdef\Ephoton{    
\ifcase\LINECONFIGURATION  
\setcoords(-285,715,0)(-150,-400,0)[2005,0]
\gdef\upperunitbox{\Frown}   \gdef\lowerunitbox{\Smile}
\or  
\setcoords(-285,715,0)(-420,-170,0)[2005,0]
\gdef\upperunitbox{\Smile}   \gdef\lowerunitbox{\Frown}
\else \UNIMPERROR
\fi
\particleadjustx=-15 
}
\gdef\SEphoton{   
\ifcase\LINECONFIGURATION  
\setcoords(-200,1050,0)(-625,-625,0)[1250,-1250] \SETUNITBOX(1250)[tr][bl]
\ifnum\numwigglespo > \number \numwiggles \particleadjustx=25 \fi
\or 
\setcoords(425,425,0)(0,-1250,0)[1250,-1250]     \SETUNITBOX(1250)[bl][tr]
\ifnum\numwigglespo > \number \numwiggles \particleadjustx=15 \fi
\or 
\setcoords(-200,1200,0)(-700,-700,0)[1400,-1400] \SETUNITBOX(1400)[tr][bl]
\or 
\setcoords(500,500,0)(0,-1400,0)[1400,-1400]     \SETUNITBOX(1400)[bl][tr]
\or 
\setcoords(-200,1000,0)(-600,-600,0)[1200,-1200] \SETUNITBOX(1200)[tr][bl]
\particleadjustx=-20
\or 
\setcoords(420,420,0)(0,-1200,0)[1200,-1200]     \SETUNITBOX(1200)[bl][tr]
\particleadjustx=40
\else \UNIMPERROR
\fi
}
\gdef\Sphoton{  
\ifcase\LINECONFIGURATION  
\setcoords(-252,-490,0)(-740,-1740,0)[0,-2000]
\gdef\upperunitbox{\SRbr}   \gdef\lowerunitbox{\SLbr}
\or 
\setcoords(-490,-260,0)(-740,-1740,0)[0,-2002]
\gdef\upperunitbox{\SLbr}   \gdef\lowerunitbox{\SRbr}
\or 
\setcoords(-299,-449,0)(-870,-1970,0)[0,-2200]
\gdef\upperunitbox{\Rbr}    \gdef\lowerunitbox{\Lbr}
\particleadjusty=-95
\or 
\setcoords(-517,-371,0)(-900,-2000,0)[0,-2200]
\gdef\upperunitbox{\Lbr}    \gdef\lowerunitbox{\Rbr}
\particleadjusty=-165
\or 
\setcoords(-299,-409,0)(-885,-1905,0)[0,-2000]
\gdef\upperunitbox{\Rbr}   \gdef\lowerunitbox{\Lbr}
\particleadjustx=50     \particleadjusty=-380
\ifodd\unitboxnumber\relax\else\particleadjustx=250 \particleadjusty=-400 \fi
\or 
\setcoords(-519,-449,0)(-900,-1920,0)[0,-2000]
\gdef\upperunitbox{\Lbr}   \gdef\lowerunitbox{\Rbr}
\particleadjusty=-370
\ifodd\unitboxnumber\relax\else\particleadjustx=-240 \particleadjusty=-400 \fi
\or 
\gdef\upperunitbox{\Rbr}   \gdef\lowerunitbox{\Lbr}
\setcoords(-325,-555,0)(-900,-2100,0)[0,-2400]
\particleadjusty=-40
\or 
\setcoords(-505,-275,0)(-900,-2100,0)[0,-2400]
\gdef\upperunitbox{\Lbr}   \gdef\lowerunitbox{\Rbr}
\particleadjusty=-30  
\else \UNIMPERROR
\fi
}
\gdef\SWphoton{  
\ifcase\LINECONFIGURATION  
\setcoords(-825,-825,0)(0,-1250,0)[-1250,-1250]     \SETUNITBOX(1250)[br][tl]
\or 
\setcoords(-175,-1425,0)(-625,-625,0)[-1250,-1250]  \SETUNITBOX(1250)[tl][br]
\or 
\setcoords(-900,-900,0)(0,-1410,0)[-1400,-1400]     \SETUNITBOX(1400)[br][tl]
\or 
\setcoords(-200,-1600,0)(-700,-700,0)[-1400,-1400]  \SETUNITBOX(1400)[tl][br]
\or 
\setcoords(-800,-800,0)(0,-1200,0)[-1200,-1200]     \SETUNITBOX(1200)[br][tl]
\or 
\setcoords(-200,-1400,0)(-600,-600,0)[-1200,-1200]  \SETUNITBOX(1200)[tl][br]
\else \UNIMPERROR
\fi
}
\gdef\Wphoton{
\ifcase\LINECONFIGURATION 
\setcoords(-2245,-1245,0)(-150,-400,0)[-2005,0]
\gdef\upperunitbox{\Frown}   \gdef\lowerunitbox{\Smile}
\or 
\setcoords(-2245,-1245,0)(-400,-150,0)[-2005,0]
\gdef\upperunitbox{\Smile}   \gdef\lowerunitbox{\Frown}
\else \UNIMPERROR
\fi
\particleadjustx=57 
\ifnum\numwigglespo=\number\numwiggles \particleadjustx=0  \fi
\numlowerunits=\numwigglespo   \numupperunits=\numwiggles
}
\gdef\NWphoton{  
\ifcase\LINECONFIGURATION  
\setcoords(-200,-1425,0)(625,625,0)[-1250,1250]   \SETUNITBOX(1250)[bl][tr]
\or 
\setcoords(-825,-825,0)(0,1250,0)[-1250,1250]     \SETUNITBOX(1250)[tr][bl]
\ifnum\numwigglespo > \number \numwiggles \particleadjusty=-15 \fi
\or 
\setcoords(-200,-1600,0)(700,700,0)[-1400,1400]   \SETUNITBOX(1400)[bl][tr]
\or 
\setcoords(-900,-900,0)(0,1400,0)[-1400,1400]     \SETUNITBOX(1400)[tr][bl]
\or 
\setcoords(-200,-1400,0)(600,600,0)[-1200,1200]   \SETUNITBOX(1200)[bl][tr]
\or 
\setcoords(-800,-800,0)(0,1200,0)[-1200,1200]     \SETUNITBOX(1200)[tr][bl]
\else \UNIMPERROR
\fi
}

%% file: GLUONSETUP.tex
\global\newcount\gluonlength
\global\newcount\gluonlengthx
\global\newcount\gluonlengthy
\global\newcount\gluonfrontx  
\global\newcount\gluonfronty  
\global\newcount\gluonbackx
\global\newcount\gluonbacky
%
\gdef\setunitbox(#1)[#2][#3](#4)[#5]{
\gdef\upperunitbox{\oval(#1,#1)[#2]}
\gdef\lowerunitbox{\oval(401,401)[#3]}
\gdef\thirdunitbox{\oval(#4,#4)[#5]}
}
\gdef\selectgluon{  
\global\advance\gluoncount by 1  
\global\gluonfrontx=\particlefrontx   
\global\gluonfronty=\particlefronty   
\global\particleadjustx=0     \global\particleadjusty=0
\ifnum\unitboxnumber > 40
\message{   *** WARNING *** Gluon with
\the\unitboxnumber\space loops requested ***   }
\ifnum\unitboxnumber > 85
\message{   *** Reducing gluon length to 6 loops (max 85) ***   }
\ifnum\unitboxnumber > 1000
\message{   *** Probable Cause:  Gluon selected instead of Fermion ***   }
\fi \global\unitboxnumber=6 \fi \fi  
\global\unitboxnumberpo=\unitboxnumber  
\global\advance\unitboxnumberpo by 1 
\global\numlineparts = 3
\global\numupperunits=\unitboxnumber
\global\numlowerunits=\unitboxnumber
\global\numthirdunits=\unitboxnumber
\ifcase\LINEDIRECTION
\Ngluon    
\or  \NEgluon  
\or  \Egluon   
\or  \SEgluon
\or  \Sgluon
\or  \SWgluon
\or  \Wgluon
\or  \NWgluon
\else\DIRECTERROR \fi
\setparticle
\global\gluonlengthx=\particlelengthx  \global\gluonlengthy=\particlelengthy
\global\gluonbackx=\particlebackx      \global\gluonbacky=\particlebacky
}
\gdef\Ngluon{   
\ifcase\LINECONFIGURATION   
\setcoords(600,540,600)(20,620,1220)[0,1050]
\setunitbox(1600)[tl][r](1600)[bl]
\particleadjusty=195
\or 
\setcoords(-990,-930,-990)(12,615,1215)[0,1050]
\setunitbox(1600)[tr][l](1600)[br]
\particleadjusty=195
\or 
\setcoords(440,390,440)(-10,415,840)[0,850]
\setunitbox(1250)[tl][r](1250)[bl]
\particleadjustx=0
\particleadjusty=-10
\or 
\setcoords(-820,-770,-820)(-25,400,825)[0,850]  
\particleadjusty=-10  
\setunitbox(1250)[tr][l](1250)[br]
\or \UNIMPERROR  
\or \UNIMPERROR  
\or 
\numupperunits=\unitboxnumberpo
\numlowerunits=\unitboxnumber
\numthirdunits=\unitboxnumberpo
\setcoords(-200,-200,-200)(616,1041,616)[0,850]
\setunitbox(1250)[tl][r](1250)[bl]
\particleadjusty=1238
\particleadjusty=1233
\or 
\numupperunits=\unitboxnumberpo
\numlowerunits=\unitboxnumber
\numthirdunits=\unitboxnumberpo
\setcoords(-200,-200,-200)(620,1045,620)[0,850]
\setunitbox(1250)[tr][l](1250)[br]
\particleadjusty=1245
\else \UNIMPERROR 
\fi
}
\gdef\NEgluon{
\numupperunits=\unitboxnumberpo
\numlowerunits=\unitboxnumber
\numthirdunits=\unitboxnumber
\ifcase\LINECONFIGURATION
\setcoords(900,900,900)(0,900,900)[900,900]
\setunitbox(2200)[tl][tr](401)[b]
\particleadjustx=1100     \particleadjusty=1100
\or 
\setcoords(-180,720,720)(1090,1091,1091)[900,900]
\setunitbox(2200)[br][tr](401)[l]
\particleadjustx=1110     \particleadjusty=1050
\else \UNIMPERROR 
\fi
}
\gdef\Egluon{     
\ifcase\LINECONFIGURATION
\setcoords(-210,390,990)(-800,-745,-800)[1050,0]  
\setunitbox(1600)[tr][b](1600)[tl]
\particleadjustx=130  
\or 
\setcoords(-210,390,990)(800,745,800)[1050,0]  
\setunitbox(1600)[br][t](1600)[bl]
\particleadjustx=130
\or 
\setcoords(-200,225,650)(-625,-575,-625)[850,0]
\setunitbox(1250)[tr][b](1250)[tl]
\or 
\setcoords(-200,225,650)(625,575,625)[850,0]
\setunitbox(1250)[br][t](1250)[bl]
\or 
\setcoords(-200,430,1060)(-830,-780,-830)[1260,0]
\setunitbox(1660)[tr][b](1660)[tl]
\or 
\setcoords(-200,430,1060)(830,780,830)[1260,0]
\setunitbox(1660)[br][t](1660)[bl]
\or 
\numupperunits=\unitboxnumberpo
\numlowerunits=\unitboxnumber
\numthirdunits=\unitboxnumberpo
\setcoords(440,865,440)(0,50,0)[850,0]
\setunitbox(1250)[tr][b](1250)[tl]
\particleadjustx=1260
\or 
\numupperunits=\unitboxnumberpo
\numlowerunits=\unitboxnumber
\numthirdunits=\unitboxnumberpo
\setcoords(430,855,430)(0,-50,0)[850,0]
\setunitbox(1250)[br][t](1250)[bl]
\particleadjustx=1250
\or 
\setcoords(-160,440,1040)(-600,-550,-600)[1200,0]
\gdef\upperunitbox{\oval(1600,1200)[tr]}
\gdef\thirdunitbox{\oval(1600,1200)[tl]}
\gdef\lowerunitbox{\oval(401,401)[b]}
\else \UNIMPERROR
\fi
}
\gdef\SEgluon{
\numupperunits=\unitboxnumberpo
\numlowerunits=\unitboxnumber
\numthirdunits=\unitboxnumber
\ifcase\LINECONFIGURATION
\setcoords(-200,700,700)(-1100,-1100,-1100)[900,-900]
\setunitbox(2200)[tr][br](401)[l]
\particleadjustx=1100     \particleadjusty=-1100
\or 
\setcoords(890,890,890)(0,-900,-900)[900,-900]
\setunitbox(2200)[bl][br](401)[t]
\particleadjustx=1050     \particleadjusty=-1100
\else \UNIMPERROR 
\fi
}
\gdef\Sgluon{   
\ifcase\LINECONFIGURATION  
\setcoords(-1000,-940,-1000)(0,-595,-1195)[0,-1050]
\setunitbox(1600)[br][l](1600)[tr]
\particleadjusty=-150
\or 
\setcoords(605,545,605)(-20,-615,-1215)[0,-1050]
\setunitbox(1600)[bl][r](1600)[tl]
\particleadjusty=-150
\or 
\setcoords(-820,-770,-820)(0,-425,-850)[0,-850]
\setunitbox(1250)[br][l](1250)[tr]
\or 
\setcoords(440,390,440)(0,-425,-850)[0,-850]
\setunitbox(1250)[bl][r](1250)[tl]
\or \UNIMPERROR 
\or \UNIMPERROR
\or 
\numupperunits=\unitboxnumberpo
\numlowerunits=\unitboxnumber
\numthirdunits=\unitboxnumberpo
\setcoords(-180,-180,-180)(-635,-1060,-635)[0,-850]
\setunitbox(1250)[br][l](1250)[tr]
\particleadjusty=-1290
\or 
\numupperunits=\unitboxnumberpo
\numlowerunits=\unitboxnumber
\numthirdunits=\unitboxnumberpo
\setcoords(-180,-180,-180)(-635,-1060,-635)[0,-850]
\setunitbox(1250)[bl][r](1250)[tl]
\particleadjusty=-1290
\else \UNIMPERROR 
\fi
}
\gdef\SWgluon{
\numupperunits=\unitboxnumberpo
\numlowerunits=\unitboxnumber
\numthirdunits=\unitboxnumber
\ifcase\LINECONFIGURATION
\setcoords(-1300,-1300,-1300)(0,-900,-900)[-900,-900]
\setunitbox(2200)[br][bl](401)[t]
\particleadjustx=-1100     \particleadjusty=-1100
\or 
\setcoords(-215,-1115,-1115)(-1107,-1107,-1107)[-900,-900]
\setunitbox(2200)[tl][bl](401)[r]
\particleadjustx=-1120     \particleadjusty=-1120
\else \UNIMPERROR 
\fi
}
\gdef\Wgluon{   
\ifcase\LINECONFIGURATION
\setcoords(-190,-790,-1390)(800,745,800)[-1050,0]
\setunitbox(1600)[bl][t](1600)[br]
\particleadjustx=-150  
\or 
\setcoords(-190,-790,-1390)(-800,-745,-800)[-1050,0]
\setunitbox(1600)[tl][b](1600)[tr]
\particleadjustx=-150  
\or 
\setcoords(-200,-625,-1050)(625,575,625)[-850,0]
\setunitbox(1250)[bl][t](1250)[br]
\or 
\setcoords(-200,-625,-1050)(-625,-575,-625)[-850,0]
\setunitbox(1250)[tl][b](1250)[tr]
\or 
\setcoords(-230,-860,-1490)(830,780,830)[-1260,0]
\setunitbox(1660)[bl][t](1660)[br]
\or 
\setcoords(-230,-860,-1490)(-830,-780,-830)[-1260,0]
\setunitbox(1660)[tl][b](1660)[tr]
\or 
\numupperunits=\unitboxnumberpo
\numlowerunits=\unitboxnumber
\numthirdunits=\unitboxnumberpo
\setcoords(-825,-1250,-825)(0,-50,0)[-850,0]
\setunitbox(1250)[bl][t](1250)[br]
\particleadjustx=-1250
\or  
\numupperunits=\unitboxnumberpo
\numlowerunits=\unitboxnumber
\numthirdunits=\unitboxnumberpo
\setcoords(-825,-1250,-825)(0,50,0)[-850,0]
\setunitbox(1250)[tl][b](1250)[tr]
\particleadjustx=-1250
\else \UNIMPERROR 
\fi
}
\gdef\NWgluon{
\numupperunits=\unitboxnumberpo
\numlowerunits=\unitboxnumber
\numthirdunits=\unitboxnumber
\ifcase\LINECONFIGURATION
\setcoords(-200,-1100,-1100)(1100,1100,1100)[-900,900]
\setunitbox(2200)[bl][tl](401)[r]
\particleadjustx=-1110   \particleadjusty=1100
\or  
\setcoords(-1309,-1309,-1309)(-15,885,885)[-900,900]
\setunitbox(2200)[tr][tl](401)[b]
\particleadjustx=-1120   \particleadjusty=1065
\else \UNIMPERROR 
\fi
}
%
%
%
\gdef\gluonlink{    
\input GLUONLINKS   
\gluonlink}  
\gdef\gluoncap{    
\input GLUONLINKS   
\gluoncap}  

%% file: VERTEX.tex
%
\global\advance\vertexcount by 1 
\newsavebox{\vertexbox}
\global\newcount\LDIRcount      
\global\newcount\VERTEXNUMBER   
\global\newdimen\VERTEXLINKONE  
\global\newdimen\VERTEXLINKTWO
\global\newdimen\VERTEXLINKTHREE
\global\newdimen\VERTEXLINKFOUR 
\global\newdimen\VERTEXCAPONE   
\global\newdimen\VERTEXCAPTWO
\global\newdimen\VERTEXCAPTHREE
\global\newdimen\VERTEXCAPFOUR  
\global\newdimen\STEMVERTEXONE
\global\newdimen\STEMVERTEXTWO
\global\newdimen\STEMVERTEXTHREE
\global\newdimen\STEMVERTEXFOUR 
\global\newcount\stemlengthcopy
\global\newcount\vertexonex    \global\newcount\vertexoney
\global\newcount\vertextwox    \global\newcount\vertextwoy
\global\newcount\vertexthreex  \global\newcount\vertexthreey
\global\newcount\vertexfourx   \global\newcount\vertexfoury
\global\newcount\vertexmidx    \global\newcount\vertexmidy
\global\newcount\VERTEXLINE
\global\newcount\FLIPVERTEX    \global\FLIPVERTEX=0
\gdef\flipvertex{\global\FLIPVERTEX=1}  
\newcount\vertadj              \newcount\negvertadj
\gdef\drawvertex#1[#2#3](#4,#5)[#6]{
\global\advance\vertexcount by 1 
\global\LINETYPE=#1           
\global\LINEDIRECTION=#2
\global\VERTEXNUMBER=#3  
\global\vertexonex=#4
\global\vertexoney=#5
\global\unitboxnumber=#6
\global\stemlengthcopy=\stemlength   
%
%
\ifnum\LINETYPE<3 \LINEERROR \fi
\ifnum\LINETYPE=3   
\ifnum\gluoncount=0 \def\gluonlink{\relax} \def\gluoncap{\relax} \fi
  \ifnum\VERTEXNUMBER<3 \UNIMPERROR \fi
  \ifnum\VERTEXNUMBER=3 \THREEPHOTON\fi 
  \ifnum\VERTEXNUMBER=4 \FOURPHOTON \fi 
  \ifnum\VERTEXNUMBER>4 \UNIMPERROR \fi
\fi
\ifnum\LINETYPE=4   
  \ifnum\VERTEXNUMBER<3 \UNIMPERROR \fi
  \ifnum\VERTEXNUMBER=3 \THREEGLUON \fi 
  \ifnum\VERTEXNUMBER=4 \FOURGLUON  \fi 
  \ifnum\VERTEXNUMBER>4 \UNIMPERROR \fi
\fi
\ifnum\LINETYPE>4 \LINEERROR \fi
\clearvertex   
}  
%
\gdef\advtwomodeight#1{
\global\advance\LDIRcount by2\moduloeight\LDIRcount
\diagFOURVERT#1[\LDIRcount]}
%
\gdef\THREEPHOTON{
\ifcase\LDIR  
\setvertexA[\S\REG]
\setvertexB[\NW\CURLY](0,0)[2]  \setvertexB[\NE\FLIPPEDCURLY](0,0)[3]
\or  
\setvertexA[\SW\CURLY]
\setvertexB[\N\REG](70,-100)[2] \setvertexB[\E\FLIPPED](0,0)[3]
\or  
\setvertexA[\W0]
\setvertexB[\NE\CURLY](20,0)[2] \setvertexB[\SE\FLIPPEDCURLY](20,0)[3]
\or  
\setvertexA[\NW\CURLY]
\setvertexB[\E\REG](0,0)[2] \setvertexB[\S1](20,100)[3]
\or  
\setvertexA[\N\REG]
\setvertexB[\SE\CURLY](0,0)[2]  \setvertexB[\SW\FLIPPEDCURLY](0,0)[3]
\or  
\setvertexA[\NE\CURLY]
\setvertexB[\S\REG](-20,100)[2]
\setvertexB[\W\FLIPPED](0,0)[3]
\or  
\setvertexA[\E\REG]
\setvertexB[\SW\CURLY](0,0)[2]  \setvertexB[\NW\FLIPPEDCURLY](0,0)[3]
\or  
\setvertexA[\SE\CURLY]
\setvertexB[\W\REG](0,0)[2]
\setvertexB[\N\FLIPPED](-40,-100)[3]
\else \DIRECTERROR
\fi
} 
%
%
\gdef\FOURPHOTON{
\ifnum\LDIR>-1
\global\LDIRcount=\LDIR  \global\advance\LDIRcount by 4 \moduloeight\LDIRcount
\setvertexA[\LDIRcount\REG] \advtwomodeight2 \advtwomodeight3 \advtwomodeight4
\else \UNIMPERROR \fi
\global\FLIPVERTEX=0
}
%
%
\gdef\THREEGLUON{
\vertadj=0   \adjvert  
\ifcase\LDIR  
\setvertexA[\S\CENTRAL]
\setvertexB[\NW\REG](0,0)[2]  \setvertexB[\NE\FLIPPED](0,0)[3]
\or  
\setvertexA[\SW\REG]
\setvertexB[\N\CURLY](-170,442)[2] \setvertexB[\E 3](420,-183)[3]
\setvertexC(442,442)[bl]
\or  
\setvertexA[\W6]
\setvertexB[\NE\REG](0,0)[2] \setvertexB[\SE\FLIPPED](0,0)[3]
\or  
\setvertexA[\NW\REG]
\setvertexB[\E\CURLY](420,183)[2]  \setvertexB[\S3](-183,-442)[3]
\setvertexC(442,-442)[tl]
\or  
\setvertexA[\N\CENTRAL]
\setvertexB[\SE\REG](0,0)[2]  \setvertexB[\SW\FLIPPED](0,0)[3]
\or  
\setvertexA[\NE\REG]
\setvertexB[\S\CURLY](170,-442)[2]
\setvertexB[\W\FLIPPEDCURLY](-420,183)[3]
\setvertexC(-442,-442)[tr]
\or  
\setvertexA[\E\CENTRAL]
\setvertexB[\SW\REG](0,0)[2]  \setvertexB[\NW\FLIPPED](0,0)[3]
\or  
\setvertexA[\SE\REG]
\setvertexB[\W\CURLY](-420,-183)[2]
\setvertexB[\N\FLIPPEDCURLY](170,442)[3]
\setvertexC(-442,442)[br]
\else \DIRECTERROR
\fi
} 
%
%
%
%
\gdef\FOURGLUON{
\ifodd\LDIR\vertadj=0 \else  \vertadj=412 \fi    \adjvert
\ifcase\LDIR  
\setvertexA[\S\CURLY]         \MIDADJUST(\negvertadj,\vertadj)
\WFOURVERT2    \NFOURVERT3    \EFOURVERT4
\or \FOURPHOTON  
\or  
\setvertexA[\W\CURLY]         \MIDADJUST(\vertadj,\vertadj)
\NFOURVERT2  \EFOURVERT3  \SFOURVERT4
\or \FOURPHOTON    
\or  
\setvertexA[\N\CURLY]         \MIDADJUST(\vertadj,\negvertadj)
\EFOURVERT2   \SFOURVERT3   \WFOURVERT4
\or \FOURPHOTON    
\or  
\setvertexA[\E\CURLY]         \MIDADJUST(\negvertadj,\negvertadj)
\SFOURVERT2  \WFOURVERT3  \NFOURVERT4
\or \FOURPHOTON    
\else \DIRECTERROR
\fi
\global\FLIPVERTEX=0   
} 
%
%
\gdef\MIDADJUST(#1,#2){
\ifnum\FLIPVERTEX=0
\global\advance\vertexmidx by #1   \global\advance\vertexmidy by #2
\setvertexC(0,\vertadj)[bl]        \setvertexC(0,\negvertadj)[tr]
\setvertexC(\vertadj,0)[tl]        \setvertexC(\negvertadj,0)[br]
\else  
\global\particleadjustx=#1  \global\particleadjusty=#2
\ifnum\LDIR=0 \global\multiply\particleadjustx by -1 \fi
\ifnum\LDIR=2 \global\multiply\particleadjusty by -1 \fi
\ifnum\LDIR=4 \global\multiply\particleadjustx by -1 \fi
\ifnum\LDIR=6 \global\multiply\particleadjusty by -1 \fi
\global\advance\vertexmidx by \particleadjustx
\global\advance\vertexmidy by \particleadjusty
\setvertexC(0,\negvertadj)[tl]        \setvertexC(0,\vertadj)[br]
\setvertexC(\negvertadj,0)[tr]        \setvertexC(\vertadj,0)[bl]
\fi
}
\gdef\NFOURVERT#1{
  \ifnum\FLIPVERTEX=0 \setvertexB[\N\CURLY](\negvertadj,\vertadj)[#1]
  \else \setvertexB[\N\FLIPPEDCURLY](\vertadj,\vertadj)[#1] \fi}
\gdef\SFOURVERT#1{
  \ifnum\FLIPVERTEX=0 \setvertexB[\S\CURLY](\vertadj,\negvertadj)[#1]
  \else \setvertexB[\S\FLIPPEDCURLY](\negvertadj,\negvertadj)[#1] \fi}
\gdef\EFOURVERT#1{
  \ifnum\FLIPVERTEX=0 \setvertexB[\E\CURLY](\vertadj,\vertadj)[#1]
  \else \setvertexB[\E\FLIPPEDCURLY](\vertadj,\negvertadj)[#1] \fi}
\gdef\WFOURVERT#1{
  \ifnum\FLIPVERTEX=0 \setvertexB[\W\CURLY](\negvertadj,\negvertadj)[#1]
  \else \setvertexB[\W\FLIPPEDCURLY](\negvertadj,\vertadj)[#1] \fi}
\gdef\diagFOURVERT#1[#2]{\setvertexB[#2\FLIPVERTEX](0,0)[#1]}
%
%
%
\gdef\setvertexA[#1#2]{
\global\savebox\vertexbox(0,0){
\begin{picture}(0,0)
\global\adjx=#2
\ifnum\FLIPVERTEX=1 \global\advance\adjx by 1
      \ifnum\VERTEXNUMBER=3 \global\FLIPVERTEX=0 \fi
  \fi  
\ifdim\STEMVERTEXONE=1pt\backstemmed \fi
\drawline\LINETYPE[#1\adjx](0,0)[\unitboxnumber]
\global\stemlength=\stemlengthcopy  
\ifdim\VERTEXLINKONE=1pt\gluonlink \fi 
\ifdim\VERTEXCAPONE=1pt\gluoncap \fi 
\global\vertexmidx=\particlebackx  \global\vertexmidy=\particlebacky
\end{picture}
}
\global\multiply\vertexmidx by -1  \global\multiply\vertexmidy by -1
\global\advance\vertexmidx by \vertexonex
\global\advance\vertexmidy by \vertexoney
\ifdim\STEMVERTEXONE=1pt\backstemmed \fi
\drawline\LINETYPE[#1\adjx](\vertexmidx,\vertexmidy)[\unitboxnumber]
\global\stemlength=\stemlengthcopy  
\ifdim\VERTEXLINKONE=1pt\gluonlink \fi 
\ifdim\VERTEXCAPONE=1pt\gluoncap \fi 
} 
\gdef\setvertexB[#1#2](#3,#4)[#5]{
\global\adjx=\vertexmidx   \global\adjy=\vertexmidy
\global\advance\adjx by #3   \global\advance\adjy by #4
\VERTEXLINE=#5
\ifcase\VERTEXLINE\UNIMPERROR  
\or \UNIMPERROR  
\or \ifdim\STEMVERTEXTWO=1pt\backstemmed \fi  
\or \ifdim\STEMVERTEXTHREE=1pt\backstemmed\fi 
\or \ifdim\STEMVERTEXFOUR=1pt\backstemmed\fi  
\else \UNIMPERROR                             
\fi
\drawline\LINETYPE[#1#2](\adjx,\adjy)[\unitboxnumber]
\global\stemlength=\stemlengthcopy  
\ifcase\VERTEXLINE\UNIMPERROR  
\or \UNIMPERROR  
\or \ifdim\VERTEXLINKTWO=1pt\gluonlink \fi 
    \ifdim\VERTEXCAPTWO=1pt\gluoncap \fi 
    \global\vertextwox=\particlebackx  \global\vertextwoy=\particlebacky
\or \ifdim\VERTEXLINKTHREE=1pt\gluonlink\fi 
    \ifdim\VERTEXCAPTHREE=1pt\gluoncap\fi 
    \global\vertexthreex=\particlebackx  \global\vertexthreey=\particlebacky
\or \ifdim\VERTEXLINKFOUR=1pt\gluonlink\fi 
    \ifdim\VERTEXCAPFOUR=1pt\gluoncap\fi 
    \global\vertexfourx=\particlebackx  \global\vertexfoury=\particlebacky
\else \UNIMPERROR
\fi
}
\gdef\setvertexC(#1,#2)[#3#4]{
\global\adjx=\vertexmidx   \global\adjy=\vertexmidy
\global\advance\adjx by #1   \global\advance\adjy by #2
\absstemlength=1250  
\ifnum \VERTEXNUMBER=4 \absstemlength=\vertadj\double\absstemlength \fi
\ifnum\phantomswitch=0
   \put(\adjx,\adjy) {\oval(\absstemlength,\absstemlength)[#3#4]}\fi
}
\gdef\adjvert{\negvertadj=\vertadj  \multiply\negvertadj by -1}
%
%
\gdef\vertexlink#1{
\global\VERTEXLINE=#1
\ifcase\VERTEXLINE\UNIMPERROR
\or \global\VERTEXLINKONE=1pt    \or \global\VERTEXLINKTWO=1pt
\or \global\VERTEXLINKTHREE=1pt  \or \global\VERTEXLINKFOUR=1pt
\else\UNIMPERROR\fi}
\gdef\vertexlinks{
\global\VERTEXLINKONE=1pt     \global\VERTEXLINKTWO=1pt
\global\VERTEXLINKTHREE=1pt  \global\VERTEXLINKFOUR=1pt  }
%
%
%
\gdef\vertexcap#1{
\global\VERTEXLINE=#1
\ifcase\VERTEXLINE\UNIMPERROR
\or \global\VERTEXCAPONE=1pt    \or \global\VERTEXCAPTWO=1pt
\or \global\VERTEXCAPTHREE=1pt  \or \global\VERTEXCAPFOUR=1pt
\else\UNIMPERROR\fi}
\gdef\vertexcaps{
\global\VERTEXCAPONE=1pt     \global\VERTEXCAPTWO=1pt
\global\VERTEXCAPTHREE=1pt  \global\VERTEXCAPFOUR=1pt  }
%
%
%
\gdef\stemvertex#1{
\global\VERTEXLINE=#1
\ifcase\VERTEXLINE\UNIMPERROR
\or \global\STEMVERTEXONE=1pt    \or \global\STEMVERTEXTWO=1pt
\or \global\STEMVERTEXTHREE=1pt  \or \global\STEMVERTEXFOUR=1pt
\else\UNIMPERROR\fi}
\gdef\stemvertices{
\global\STEMVERTEXONE=1pt     \global\STEMVERTEXTWO=1pt
\global\STEMVERTEXTHREE=1pt  \global\STEMVERTEXFOUR=1pt  }
\gdef\clearvertex{
\global\STEMVERTEXONE=0pt    \global\STEMVERTEXTWO=0pt
\global\STEMVERTEXTHREE=0pt  \global\STEMVERTEXFOUR=0pt
\global\VERTEXLINKONE=0pt    \global\VERTEXLINKTWO=0pt
\global\VERTEXLINKTHREE=0pt  \global\VERTEXLINKFOUR=0pt
\global\VERTEXCAPONE=0pt    \global\VERTEXCAPTWO=0pt
\global\VERTEXCAPTHREE=0pt  \global\VERTEXCAPFOUR=0pt
\global\stemlength=275}  
\global\stemlengthcopy=\stemlength
\clearvertex  
\global\stemlength=\stemlengthcopy

%% file: LOOPS.tex
\global\newcount\loopfrontx    \global\newcount\loopfronty
\global\newcount\loopbackx    \global\newcount\loopbacky
\global\newcount\loopmidx    \global\newcount\loopmidy
\global\newdimen\CENTRALLOOP
\gdef\drawloop#1[#2#3](#4,#5){
\global\CENTRALLOOP=0pt  
\global\LINETYPE=#1
\ifnum\LTYPE=\gluon\relax\else\UNIMPERROR\LTYPE=1\message{Reverting to Gluons}
\fi
\global\LINEDIRECTION=#2  
\global\fourthlineadjx=#3 
\ifnum\fourthlineadjx=0 
  \global\CENTRALLOOP=1pt  
  \global\fourthlineadjx=8
  \global\LDIR=0
\fi
\global\fourthlineadjy=\fourthlineadjx  
\global\advance\fourthlineadjy by -4
\global\loopfrontx=#4   \global\loopfronty=#5
\ifdim\CENTRALLOOP=1pt
  \global\advance\loopfrontx by -2413  \global\advance\loopfronty by -425
\fi                          
\global\unitboxnumber=1  
\ifnum\LINETYPE=\photon \unitboxnumber=2 \fi
\checkdir
\drawline\LINETYPE[\LDIR\LCONFIG](\loopfrontx,\loopfronty)[\unitboxnumber]
\DRAWLOOP
\ifnum\fourthlineadjy>-1 
\global\loopmidx=\loopfrontx   \global\loopmidy=\loopfronty
\global\advance\loopmidx by \loopbackx  \global\advance\loopmidy by \loopbacky
\divide\loopmidx  by 2 \divide\loopmidy by 2  
\ifdim\CENTRALLOOP=1pt
  \global\advance\loopfrontx by 200    \global\advance\loopfronty by 425
  \global\advance\loopbackx by -200    \global\advance\loopbacky by -425
\fi
\fi 
}
\gdef\DRAWLOOP{
\global\advance\fourthlineadjx by -1
\ifnum\fourthlineadjx=0\relax  
\else
\ifnum\fourthlineadjx=\fourthlineadjy 
   \global\loopbackx=\pbackx   \global\loopbacky=\pbacky
\fi
\global\advance\LDIR by 1
\moduloeight\LDIR
\checkdir
\drawline\LINETYPE[\LDIR\LCONFIG](\pbackx,\pbacky)[\unitboxnumber]
\fi 
\ifnum\fourthlineadjx>1 \DRAWLOOP  \fi  
}
\gdef\checkdir{
\ifnum\LTYPE=\gluon
\ifodd\LDIR \global\LCONFIG=0 \else \global\LCONFIG=2 \fi
\fi 
}